\documentclass{jfm}

\usepackage{amsmath}
\usepackage{amssymb}
\usepackage{graphicx}
\usepackage{epsfig}
\usepackage{subfig}
\newcommand{\mi}{\mathrm{i}}

\usepackage[normalem]{ulem}
\usepackage{xcolor}
\definecolor{pgreen}{RGB}{0,127,0}
\definecolor{pmagenta}{RGB}{191,0,191}

\usepackage{tikz,siunitx}
\newcommand{\solidline}{\protect\tikz[baseline]{\protect\draw[solid,line width=0.7pt](0.0mm,0.5ex)--(5mm,0.5ex)}}
\newcommand{\dashline}{\protect\tikz[baseline]{\protect\draw[dashed,line width=0.7pt](0.0mm,0.5ex)--(5mm,0.5ex)}}
\newcommand{\dotline}{\protect\tikz[baseline]{\protect\draw[dotted,line width=0.7pt](0.0mm,0.5ex)--(5mm,0.5ex)}}
\newcommand{\dotdashline}{\protect\tikz[baseline]{\protect\draw[dash dot,line width=0.7pt](0.0mm,0.5ex)--(5mm,0.5ex)}}
\newcommand{\capcirc}{\protect\tikz[baseline]{\protect\draw[solid,line width=0.5pt](2.mm,0.5ex) circle (0.6ex)}}

\begin{document}

\shorttitle{Resolvent-based study of 
supersonic turbulent boundary layers}
\shortauthor{H.~J.~Bae, S.~T.~M.~Dawson and B.~J.~McKeon}

\title{Resolvent-based study of compressibility effects
on supersonic turbulent boundary layers}
\author{H. Jane Bae\aff{1} \corresp{\email{hjbae@caltech.edu}}, Scott
T. M. Dawson\aff{1,2} \and Beverley J. McKeon\aff{1}}

\affiliation{\aff{1}Graduate Aerospace Laboratories, California
Institute of Technology, Pasadena, CA 91125, USA
\aff{2}Mechanical, Materials and Aerospace Engineering Department, Illinois
Institute of Technology, Chicago, IL 60616, USA}

\date{\today}
\maketitle

\begin{abstract} 
{\color{black}The resolvent formulation of \citet{mckeon2010} is
applied to supersonic turbulent boundary layers} to study the validity
of Morkovin's hypothesis, which postulates that high-speed turbulence
structures in zero pressure-gradient turbulent boundary layers remain
largely the same as its incompressible counterpart. Supersonic
zero-pressure-gradient turbulent boundary layers with adiabatic wall boundary conditions at
Mach numbers ranging from 2 to 4 are considered. Resolvent analysis
highlights two distinct regions of the supersonic turbulent boundary
layer in the wave parameter space: the relatively supersonic region
and the relatively subsonic region.  In the relatively supersonic
region, where the flow is supersonic relative to the freestream,
resolvent modes display structures consistent with Mach wave radiation
that are absent in the incompressible regime.  In the relatively
subsonic region, we show that the low-rank approximation of the
resolvent operator is an effective approximation of the full system
and that the response modes predicted by the model exhibit universal
and geometrically self-similar behaviour via a transformation given by
the semi-local scaling. Moreover, with the semi-local scaling, we show
that the resolvent modes follow the same scaling law as their
incompressible counterparts in this region, which has implications for
modelling and the prediction of turbulent high-speed wall-bounded
flows. We also show that the thermodynamic variables exhibit similar
mode shapes to the streamwise velocity modes, supporting the strong
Reynolds analogy.  Finally, we demonstrate that the principal
resolvent modes can be used to capture the energy distribution between
momentum and thermodynamic fluctuations.  
\end{abstract} 

\maketitle






\section{Introduction}\label{sec:intro}

The prediction and modelling of turbulent high-speed wall-bounded
flows remain an active field of study for their tremendous
technological importance in the aerospace industry with respect to
high-speed vehicles. The turbulent boundary layers determine the
aerodynamic drag and heat transfer, which are quantities of interest
for accurate performance assessment. For supersonic and hypersonic
turbulent boundary layers, a major role has been historically played
by experiments, with direct numerical simulations (DNS) becoming more
and more common in the last decade. 

Experimental investigations of supersonic and hypersonic turbulent
boundary layers have been conducted historically with hot-wire
anemometry
\citep{kistler1959,laderman1974,owen1975,spina1987,konrad1998}
\citep[see also][for a review]{roy2006}, which suffers from
uncertainties associated with mixed-mode sensitivity
\citep{kovasznay1953} and was later shown to suffer from poor
frequency response and spatial resolution \citep{williams2018}. In
addition to hot-wire anemometry, direct measurements of spatially
varying velocity fields of high-speed turbulent boundary layers have
been attempted using particle image velocimetry
\citep{ekoto2008,tichenor2013,peltier2016,williams2018}, which range
up to Mach number of 7.5 for a flat-plate turbulent boundary layer.

Complementary to experiments, DNS of high-speed turbulent boundary
layers have been conducted to overcome the experimental difficulties
and gain access to the full three-dimensional structure of the
turbulent flow field.  Several DNS have been conducted with an
emphasis on studying Morkovin's scaling in turbulent boundary layers
at moderate freestream Mach numbers
\citep{guarini2000,maeder2000,pirozzoli2004,martin2007,shahab2011,bernardini2011,pirozzoli2011,hadjadj2015,poggie2015,trettel2016,modesti2016}
for both adiabatic and isothermal wall boundary conditions. Hypersonic
studies for turbulent boundary layers at higher Mach numbers have also
recently been available
\citep{duan2010,duan2011a,duan2011b,lagha2011,zhang2014,zhang2018}
with data sets of up to Mach number of 20. 

In spite of the recent developments in numerical experiments,
simulations of supersonic turbulent boundary layers still remain a
daunting challenge. DNS in the incompressible regime
\citep{simens2009}, as well as earlier experiments \citep{erm1991},
show that a fully developed state of the boundary layer requires the
use of an extremely long computational domain, which makes accurate
numerical simulations computationally demanding.  Simulations in the
supersonic regime are further slowed down by the inherently larger
computational effort and the possible occurrence of `eddy shocklets'.
This creates a critical demand for model-based approaches that
describe and predict the behaviour of turbulent flows at
technologically relevant high Reynolds numbers in the supersonic
regime. In particular, assessing Morkovin's hypothesis
\citep{morkovin1962} has been the focus of many high-speed turbulent
boundary layer studies.  \citet{morkovin1962} concluded from the
analysis of supersonic boundary layer data available at the time that
for moderate Mach numbers ``the essential dynamics of these shear
flows will follow the incompressible pattern.'' The hypothesis was
used and reformulated by \citet{bradshaw1974} to indicate that
high-speed boundary layers can be computed using the same model as
that of low speeds as long as the density fluctuations are weak.
Another consequence of Morkovin's hypothesis is the analogy between
the temperature and velocity fields that leads to velocity-temperature
relations such as the classical Walz formula \citep{walz1969} and the
strong Reynolds analogy and its variants
\citep{morkovin1962,gaviglio1987,huang1995,zhang2014}. It also
motivates the so-called `compressibility transformations' that
transform the mean velocity and Reynolds stress profiles from a
compressible boundary layer to equivalent incompressible profiles by
accounting for mean property variations across the thickness of the
boundary layer
\citep{vandriest1951,brun2008,zhang2012,trettel2016,yang2018}.
{\color{black}Moreover, previous studies of coherent structures
in compressible boundary layers show that the structure of the flow
associated with the near-wall region shares several features with the
incompressible boundary layer as well as the so-called internal shear
layers \citep{pirozzoli2008,ringuette2008,lagha2011}, further supporting
the hypothesis.}

Resolvent analysis provides a tool to understand and predict flows
using a low-rank approximation of the linear sub-system with the
nonlinear interactions treated as a forcing term. Previous studies on
incompressible flows show that resolvent analysis can capture a range
of phenomena already observed in wall turbulence from near-wall
streaks and quasi-streamwise vortices \citep{mckeon2010}, hairpin
vortices \citep{sharma2013} and the corresponding pressure signature
\citep{luhar2014b}, very large scale motions \citep{mckeon2010}, to
scaling of the statistics \citep{moarref2013}.  {\color{black}A similar
approach, developed by \citet{jovanovic2005}, was
used to study the roles of various disturbances in laminar
boundary layer transition and optimal transient
growth in laminar and turbulent boundary layers
\citep{monokrousos2011,sipp2013,alizard2015}.} As such, resolvent 
analysis yields an
efficient basis for the flow, which can be used to provide a low order
representation of the key dynamical processes of turbulence.  Several
studies have investigated more sophisticated means of shaping the
nonlinear forcing from data and analytical considerations
\citep{moarref2014,zare2017,illingworth2018,towne2019,morra2019}. For
turbulent channel flows (and consequently for turbulent
boundary layers), the resolvent velocity modes have a universal
scaling with Reynolds number, and in the overlap region of the mean
velocity, admit geometric self-similarity
\citep{moarref2013,moarref2014}. This observation has strong
implications for modelling: resolvent modes throughout the entire
region can be described in terms of modes assessed at one wall-normal
plane. For example, under the correct scaling of the wall-parallel
wavenumbers, the self-similar hierarchies of response modes give rise
to self-similar families of vortical structures.  However, there has
been less work applying this operator based decomposition to
fully-developed compressible flows, with the exception of the recent
studies on compressible jet flows
\citep{jeun2016,towne2018,schmidt2018}, subsonic aerofoils
\citep{yeh2019}, turbulent Couette flows \citep{dawson2019}, and
hypersonic boundary layer transitions \citep{cook2018,dwivedi2018}.

The main goal of this work is to apply the resolvent analysis
framework for compressible flows to supersonic turbulent boundary
layers. The resolvent framework allows the decomposition of the
governing equations in the wavenumber-frequency space, permitting an
in-depth comparison of the underlying mechanisms in the flow. We
utilise this tool to compare the mechanisms driving the incompressible
and compressible boundary layer which will allow us to assess
Morkovin's hypothesis on a mode-by-mode basis.

The paper is organised as follows. We first introduce the compressible
Navier-Stokes equations and the resolvent operator in
\S\ref{sec:form}, where we discuss the relevant resolvent norm,
boundary conditions, and computational methods. In \S\ref{sec:modes},
we discuss the characteristics of the resolvent modes for the
supersonic turbulent boundary layer and define the relatively
supersonic and subsonic region. We show that the response modes in the
relatively supersonic region display Mach waves. We also highlight the
low-rank behaviour of the resolvent operator and discuss the necessary
conditions for the universality of the resolvent modes in the
relatively subsonic region. In \S\ref{sec:scaling}, we then provide
the Reynolds and Mach number scaling for the principal resolvent modes
and amplification factor for the inner, logarithmic and outer region
of the boundary layer and demonstrate that the leading response mode
is enough to predict the energy distribution between momentum and
thermodynamic fluctuations.  Finally, we conclude the paper in \S
\ref{sec:conclusions}.

\section{Resolvent formulation of compressible zero-pressure-gradient turbulent boundary layer}
\label{sec:form}
\subsection{Compressible Navier-Stokes equations}
\label{sec:form:compNS_eq}
The nondimensional compressible Navier-Stokes equations for a perfect
gas are given by 
\begin{align}
\rho\left(\frac{\partial u_i}{\partial t}+u_j\frac{\partial
u_i}{\partial x_j}\right) =& -\frac{1}{\gamma M^2}\frac{\partial
p}{\partial x_i} + \frac{1}{\Rey} \frac{\partial}{\partial
x_j}\left[\mu\left(\frac{\partial u_i}{\partial x_j} + \frac{\partial
u_j}{\partial x_i}\right) + \lambda\frac{\partial u_k} {\partial
x_k}\delta_{ij} \right],\label{eq:mom_eq}
\end{align}
\begin{align}
\frac{\partial\rho}{\partial t} + u_j\frac{\partial \rho}{\partial
x_j} =& -\rho\frac{\partial u_i}{\partial x_i},\label{eq:den_eq}\\
\rho\left(\frac{\partial T}{\partial t} + u_j\frac{\partial
T}{\partial x_j}\right) =& -(\gamma-1)p\frac{\partial u_i}{\partial
x_i} + \frac{\gamma}{\Pran\Rey} \frac{\partial}{\partial x_j}\left(k
\frac{\partial T}{\partial x_j}\right) \nonumber\\
&+\gamma(\gamma-1)\frac{M^2}{\Rey}{\mu}\left[\frac{\partial
u_i}{\partial x_j} \frac{\partial u_i}{\partial x_j}+\frac{\partial
u_i}{\partial x_j} \frac{\partial u_j}{\partial x_i} + \lambda
\left(\frac{\partial u_k}{\partial x_k} \right)^2\right],
\label{eq:temp_eq}
\end{align}
where $\rho$, $p$, $u_i$, $T$ are, respectively, density, pressure,
velocity components, and temperature. Variables $\mu$ and $\lambda$
are the coefficients of first and second viscosity, respectively, $k$
is the coefficient of thermal conductivity, $\gamma=c_p/c_v$ is the
ratio of specific heats, and $\delta_{ij}$ is the Kronecker delta. We
formulate the equations in nondimensionalised form using the Mach,
Reynolds, and Prandtl numbers, given respectively by
\begin{equation}
M = \frac{\breve{u}}{\sqrt{\gamma\mathcal{R}\breve{T}}},\quad \Rey =
\frac{\breve{\rho}\breve{u}\breve{l}}{\breve{\mu}},\quad \Pran = 
\frac{\breve{\mu}c_p}{\breve{k}},
\label{eq:nondim}
\end{equation}
where $\breve{(\cdot)}$ denotes reference (dimensional) quantities,
$l$ is a length scale, and $\mathcal{R}$ is the universal gas
constant. The system is closed with the equation of state, $p = \rho
T$.

Here, we assume constant specific heat coefficients and constant
Prandtl number, $\Pran = 0.72$, and we set $\gamma = 1.4$ (diatomic
gas).  Furthermore, we assume that viscosity varies with temperature
according to the Sutherland formula 
\begin{equation}
\mu(T) = \frac{T^{3/2}(1+C)}{T+C},
\label{eq:sutherland}
\end{equation}
with $C=S/\breve{T}$, where $S = 110.4$K, and that the second
coefficient of viscosity $\lambda$ follows the Stokes' assumption
$\lambda = -2/3\mu$.

\subsection{Resolvent operator}
\label{sec:form:resolvent}

Assuming a fully developed, locally parallel flow with the directions
$x_1$, $x_2$ and $x_3$ signifying the streamwise, wall-normal, and
spanwise directions, respectively, the state variable
$\boldsymbol{q}=[q_1,q_2,q_3,q_4,q_5]^\intercal=[u_1,u_2,u_3,\rho,T]^\intercal$
is decomposed using the Fourier transform in homogeneous directions
and time,
\begin{equation}
\boldsymbol{q}(x_1,x_2,x_3,t) = \iiint_{-\infty}^{\infty}
\hat{\boldsymbol{q}}(x_2;\kappa_1,\kappa_3,\omega)e^{\mi(\kappa_1 x_1+
\kappa_3x_3 - \omega t)}\mathrm{d}\kappa_1\mathrm{d}\kappa_3\mathrm{d}\omega,
\end{equation}
where $\hat{(\cdot)}$ denotes variables in the transformed domain,
$\mi = \sqrt{-1}$, and the triplet $(\kappa_1,\kappa_3,\omega)$
identifies the streamwise and spanwise wavenumbers and the temporal
frequency, respectively. Here, the superscript $\intercal$ denotes
transpose and $\dagger$ will denote conjugate transpose. 

The mean turbulent state, $\bar{\boldsymbol{q}}(x_2) =
[\bar{u}_1(x_2),0,0,\bar{\rho}(x_2),\bar{T}(x_2)]^\intercal$,
corresponds to $(\kappa_1,\kappa_3,\omega)=(0,0,0)$.
 Furthermore, with the
parallel-flow assumption, which is reasonable as the base flow
variations of the zero-pressure-gradient turbulent boundary layer 
depend on the viscous time scale compared to the much
faster convective time scale for fluctuations, the mean momentum
equation \eqref{eq:mom_eq} gives a constant $\bar{p}(x_2)$.  In the
remainder of the paper, we scale the pressure such that $\bar{p}=1$
for simplicity.  Note that the spatially-developing turbulent boundary
layer can still be studied using resolvent analysis, and the
associated increase in computational effort is not prohibitive.
However, the interpretation of the underlying physical mechanisms is
significantly more straightforward for the quasi-parallel,
one-dimensional mean. 

Following a similar approach to \citet{mckeon2010}, the governing
equations \eqref{eq:mom_eq}--\eqref{eq:temp_eq} can be rewritten in
the Fourier domain for each $(\kappa_1,\kappa_3,\omega)\neq(0,0,0)$ as
\begin{align}
&-\mi\omega\hat{u}_i+\bar{u}_1\partial_1\hat{u}_i+\hat{u}_2\partial_2\bar{u}_i
= -\frac{1}{\gamma M^2}
\left(\partial_i\hat{T}+\bar{T}^2\partial_i\hat{\rho}
+\hat{\rho}\bar{T}\partial_i\bar{T} +
\bar{T}\hat{T}\partial_i\bar{\rho}\right)\nonumber \\
&\hspace{5em}+
\frac{\bar{T}}{\Rey}\left[\bar{\mu}\partial_j\left(\partial_j\hat{u}_i
+\partial_i\hat{u}_j\right)
+\bar{\lambda}\partial_i\left(\partial_j\hat{u}_j\right) +
\frac{\partial\bar{\mu}}{\partial
T}\partial_j\hat{T}\left(\partial_j\bar{u}_i
+\partial_i\bar{u}_j\right)\right] + \hat{f}_i \\
&-\mi\omega\hat{\rho}+\bar{u}_1\partial_1\hat{\rho} +
\hat{u}_2\partial_2\bar{\rho} = -\bar{\rho}\partial_i\hat{u}_i +
\hat{f}_4 \\
&-\mi\omega\hat{T}+\bar{u}_1\partial_1\hat{T}+\hat{u}_2\partial_2\bar{T}
= -(\gamma-1)\bar{T}\partial_i\hat{u}_i \nonumber \\
&\hspace{5em}+ \frac{\gamma\bar{T}}{\Pran\Rey}
\left[\bar{\mu}\partial_j\partial_j\hat{T}
+\frac{\partial^2\bar{\mu}}{\partial T^2}
(\partial_2\bar{T})^2\hat{T}+2\frac{\partial\bar{\mu}}{\partial
T}\partial_2
\bar{T}\partial_2\hat{T}+\frac{\partial\bar{\mu}}{\partial T}
\partial_2^2\bar{T}\hat{T}\right] \nonumber \\
&\hspace{5em}+ \gamma(\gamma-1)\frac{M^2\bar{T}}{\Rey}
\left[2\bar{\mu}\partial_2\bar{u}_1\partial_2\hat{u}_1+
2\bar{\mu}\partial_2\bar{u}\partial_1\hat{u}_2+\frac{\partial\bar{\mu}}{\partial T}
(\partial_2\bar{u}_1)^2\hat{T}\right]+\hat{f}_5,
\end{align}
where $\hat{\boldsymbol{f}}$ contains the nonlinear terms and
$(\partial_1,\partial_2,\partial_3) =
(\mi\kappa_1,\mathrm{d}/\mathrm{d}x_2,\mi\kappa_3)$ \citep[see][for
details on the formation of the linear operator]{mack1984}, 
{\color{black}and the mean state equations as
\begin{equation}
\frac{\mathrm{d}}{\mathrm{d}x_2}\left(\bar{\mu}\frac{\mathrm{d}\bar{u}_1}{\mathrm{d}x_2}\right) = 0,\quad \frac{\mathrm{d}}{\mathrm{d}x_2}\left(\frac{\bar{\mu}}{\Pran}\frac{\mathrm{d}\bar{T}}{\mathrm{d}x_2}\right) + (\gamma-1)M^2\bar{\mu}\frac{\mathrm{d}\bar{u}_1}{\mathrm{d}x_2} = 0,\quad \bar{\rho}\bar{T}=1.
\end{equation}
The full set of equations characterised by the mean and the
fluctuating equations are unclosed; however, by assuming that the mean
velocity profile is known, the shape of the mean profile acts as a
constraint on the full closure of the nonlinear terms for the
fluctuating equations.  The Fourier domain equations can then} be
equivalently expressed as
\begin{equation}
\hat{\boldsymbol{q}}(x_2;\kappa_1,\kappa_3,\omega) = \left[-\mi\omega
\mathsfbi{I} + \mathsfbi{L}(\kappa_1,\kappa_3,\omega)\right]^{-1}
\hat{\boldsymbol{f}}(x_2;\kappa_1,\kappa_3,\omega),
\label{eq:resolvent}
\end{equation}
where $\mathsfbi{I}$ is the identity matrix and $\mathsfbi{L}$ is the
linearised operator of the governing equations around the supersonic
turbulent mean profile. The operator $\mathsfbi{H}=\left[-\mi\omega
\mathsfbi{I} + \mathsfbi{L}(\kappa_1,\kappa_3,\omega)\right]^{-1}$ is
called the resolvent operator and {\color{black}exists if there are no
eigenvalues of $\mathsfbi{L}$ equal to $\mi\omega$.} 

\subsection{Choice of resolvent norm}
\label{sec:form:norm}
From \eqref{eq:resolvent}, we wish to find a decomposition of the
resolvent operator that enables us to identify high gain input and
output modes with respect to the linear operator. For the resolvent
analysis, the decomposition is given by the Schmidt decomposition
(called the singular value decomposition for the discrete case).
However, this decomposition must be accompanied by a choice of inner
product and the corresponding norm. In the case of the incompressible
resolvent operator, the natural and physically meaningful norm is the
kinetic energy norm, which is defined as
\begin{equation}
2K = (\boldsymbol{q},\boldsymbol{q})_K = \|\boldsymbol{q}\|_K^2 =
\int_0^\infty \bar{\rho}u_i^\dagger u_i\, \mathrm{d}x_2.
\end{equation}
Unfortunately, there is no obvious choice for the compressible case
and the standard incompressible kinetic energy norm becomes a seminorm
on this space; however, \citet{chu1965} introduced a norm that
eliminates pressure-related energy transfer terms (compression work), 
\begin{align}
\label{eq:norm}
2E = (\boldsymbol{q},\boldsymbol{q})_E = \|\boldsymbol{q}\|_E^2 &=
\int_0^\infty\left(\bar{\rho}u_i^\dagger u_i + \frac{\bar{T}}
{\gamma\bar{\rho}M^2}\rho^\dagger\rho +
\frac{\bar{\rho}}{\gamma(\gamma-1)\bar{T}M^2}T^\dagger
T\right)\mathrm{d}x_2
\end{align}
which has been used in numerous other studies of compressible flows
where the definition of an inner product is required
\citep[e.g.,][]{hanifi1996,malik2006,ozgen2008,malik2008,depando2014,bitter2014,dawson2019},
and this norm will be used for the remainder of the paper. Discussion
of other possible inner products and assumptions for compressible
flows are given in \citet{rowley2004}. {\color{black}A study
on the sensitivity of the resolvent modes with respect to the
compressible inner product and the standard incompressible kinetic
energy inner product was performed by \citet{dawson2019} for  planar
Couette flow, and the differences in mode shapes were attributed to
changes in mean profiles and the compressible fluctuation equations
rather than the choice of the inner product.}

We take the Schmidt decomposition of the resolvent, namely,
\begin{equation}
\mathsfbi{H} =
\sum_{j=1}^\infty\boldsymbol{\psi}_j(x_2;\kappa_1,\kappa_3,\omega)
\sigma_j(\kappa_1,\kappa_3,\omega)\boldsymbol{\phi}_j^\dagger(x_2;\kappa_1,\kappa_3,\omega),
\label{eq:svd}
\end{equation}
with an orthogonality condition
\begin{align}
(\boldsymbol{\psi}_i(x_2;\kappa_1,\kappa_3,\omega),
\boldsymbol{\psi}_j(x_2;\kappa_1,\kappa_3,\omega) )_E= \delta_{ij},\\
(\boldsymbol{\phi}_i(x_2;\kappa_1,\kappa_3,\omega),
\boldsymbol{\phi}_j(x_2;\kappa_1,\kappa_3,\omega))_E = \delta_{ij},\\
\end{align}
and $\sigma_j \ge \sigma_{j+1}\ge 0$.  The $\boldsymbol{\phi}_j$ and
$\boldsymbol{\psi}_j$ form the right and left Schmidt bases (singular
vectors) for the forcing and {\color{black}response} fields, and the
real $\sigma_j$ are the singular values. This decomposition is unique
up to a pre-multiplying unitary complex factor on both bases
corresponding to a phase shift \citep{young1988}.

This basis pair can then be used to decompose any arbitrary forcing
and the resulting state vector at a particular Fourier component such
that
\begin{align}
\hat{\boldsymbol{f}}(x_2;\kappa_1,\kappa_3,\omega) &= \sum_{j=1}^\infty
\boldsymbol{\phi}_j(x_2;\kappa_1,\kappa_3,\omega)a_j(\kappa_1,\kappa_3,\omega)\\
\hat{\boldsymbol{q}}(x_2;\kappa_1,\kappa_3,\omega) &= \sum_{j=1}^\infty
\sigma_j(\kappa_1,\kappa_3,\omega)
\boldsymbol{\psi}_j(x_2;\kappa_1,\kappa_3,\omega)a_j(\kappa_1,\kappa_3,\omega).
\end{align}
Clearly the forcing shape that gives the largest energy is given by
$a_j = \delta_{1j}$, i.e. when the forcing is aligned with the
principal singular vector. Moreover, we later show that the resolvent
operator is low-rank, i.e. $\sigma_1 \gg \sigma_2$, where the flow is
most energetic. Thus, in this paper, we focus on the principal Schmidt
{\color{black}vectors} (singular vectors), i.e. the principal forcing
mode $\boldsymbol{\phi}_1$ and the principal response mode
$\boldsymbol{\psi}_1=[(q_1)_1, (q_2)_1, (q_3)_1, (q_4)_1,
(q_5)_1]^\intercal=[(u_1)_1, (u_2)_1, (u_3)_1, (\rho)_1,
(T)_1]^\intercal$, and the principal singular value $\sigma_1$.

\subsection{Boundary conditions for the resolvent operator}
\label{sec:form:bc}
For the compressible boundary layer, the boundary conditions at the
wall are given by
\begin{equation}
u_i(x_2=0) = 0,\quad (T-\bar{T})(x_2=0) = 0. 
\end{equation}
The boundary conditions on the velocity fluctuations are the usual
no-slip conditions, and the boundary condition on the temperature
fluctuation is consistent for a gas flowing over a solid wall. 

The boundary conditions at the freestream are given by
\begin{equation}
(u_i-\bar{u}_i)(x_2\rightarrow\infty),\,
(\rho-\bar{\rho})(x_2\rightarrow\infty),\, (T-\bar{T})(x_2\rightarrow\infty) < \infty,
\end{equation}
which are less restrictive than requiring all fluctuations to be zero
at infinity. However, in supersonic flow, waves may propagate to
infinity and this boundary condition allows the waves with constant
amplitude to be included. {\color{black} In practice, this is achieved
by imposing a free boundary condition on the top boundary.}

\subsection{Computational methods}
\label{sec:form:numerics}
Following the Schmidt decomposition in \eqref{eq:svd}, there are
formally an infinite number of singular values. However, we solve the
discrete equations using a {\color{black}spectral collocation method
with the number of points in the wall-normal direction given by
$N_2$}, limiting the number of singular values to $5N_2$, the size of
the state vector $\boldsymbol{q}$. 

The discrete points in $x_2$ are given by a rational transformation of
the Chebyshev collocation points. The Chebyshev collocation points are
defined as $x_2' = \cos({\pi j}/({N_2-1}))$ for $j = 0,1,\dots,N_2-1$
in the domain $-1\le x_2'\le 1$.  The rational transformation  $x_2 =
a{(1+x_2')}/{(1-x_2')}$ maps $x_2'$ to the semi-infinite domain, where
$a/\delta=2$ is the wall-normal location containing $(N_2-1)/2$ points
\citep{grosch1977,christov1982} and $\delta$ is the boundary layer
thickness corresponding to the location where mean velocity reaches
$99\%$ of the freestream velocity. With this transformation, the value
of a function $\boldsymbol{\chi}(x_2)$ can be expressed as
\begin{equation}
\boldsymbol{\chi}(x_2) = \sum_{n=0}^{N_2-1} b_n T_n\left(\frac{x_2-a}{x_2+a}\right)
= \sum_{n=0}^{N_2-1} b_n T_n(x_2'),
\end{equation}
where $T_n$ is the $n$th order Chebyshev polynomial and $b_n$ is the
coefficient for the corresponding Chebyshev polynomial. 

The rational transformation of the Chebyshev collocation points allows
the use of spectral methods for the semi-infinite domain. The
stretching of the collocation points such that the majority of the
points lie within $x_2/\delta \le 2$ is appropriate for most of the
energy-containing modes that are located within the boundary layer.
However, in supersonic flows, waves may propagate to infinity and thus
become under-resolved due to the fact the grid resolution $\Delta x_2
\rightarrow \infty$ as $x_2\rightarrow\infty$.  Nonetheless, the
limitation stems from resolving a semi-infinite domain with a finite
number of discrete points, and thus any choice of decomposition in the
semi-infinite domain suffers from this limitation. The resolvent modes
considered for this paper are primarily located within the boundary
layer, and thus we are not affected by this limitation. 

{\color{black}
The turbulent mean profiles for the supersonic case are obtained from
\citet{bernardini2011} and \citet{pirozzoli2011}, where DNS of a
spatially evolving zero-pressure-gradient supersonic turbulent
boundary layer with the wall temperature set to its nominally
adiabatic value are computed. The results of the resolvent analysis for the
compressible boundary layer are compared against the results from the
resolvent analysis of the incompressible boundary layer with the mean
velocity profile obtained from \citet{jimenez2010}.} The cases are
chosen such that $\Rey_\tau$ is similar to the Reynolds number for the
incompressible case ($\Rey_\tau\approx 450$). While $\Rey_\tau$ is not
the ideal Reynolds number scaling for the outer region and a better
comparison would be the momentum thickness Reynolds number
$\Rey_{\delta_2}$, we make use of the friction Reynolds number
regardless. An additional case with $M_\infty = 2$ and $\Rey_\tau =
900$ was chosen to observe Reynolds number effects. The specific cases
used are listed in table \ref{tab:cases} along with their respective
resolution in numerical computations, where $N_2$ is the number of
collocation grid points in the $x_2$ direction, $N_1$ and $N_3$ are
the number of spatial frequencies for $\kappa_1$ and $\kappa_3$, and
$N_c$ is the number of grid points for the wave speed
$c=\omega/\kappa_1$.
\begin{table}
\begin{center}
\setlength{\tabcolsep}{5pt}
\begin{tabular}{c c c c c c c c c c c}
$M_\infty$ & $\Rey_\tau$ & $\Rey_{\delta_2}$ & $N_1$ & 
$N_2$ & $N_3$ & $N_c$ & $\log\left(\Delta\lambda_1/\delta\right)$ & 
$\log\left(\Delta\lambda_3/\delta\right)$ & $\Delta x_{2,\min}/\delta$\\[0.3cm]
0          & 445.5       & 1100.0        &       & 
      &      &       &             &               &  \\
2          & 447.7       & 1327.4        &       & 
      &      &       &             &               & \\
2          & 898.5       & 3027.2        & {\color{black}81}    & 
{\color{black}401}   & {\color{black}81}   & {\color{black}25}    & {\color{black}0.05}        & {\color{black}0.05 }         & {\color{black}$3.08\times10^{-5}$}\\
3          & 502.0       & 1815.7        &       &  
      &      &       &             &               & \\
4          & 504.6       & 2129.6        &       & 
      &      &       &             &               & \\
\end{tabular}
\end{center}
\caption{The freestream Mach number $M_\infty$, friction Reynolds
number $\Rey_\tau$, momentum thickness Reynolds number based on the
wall dynamic viscosity $\Rey_{\delta_2}$, and grid resolutions for the
different cases. $N_i$ is the number of grid points in the $x_i$
direction and $N_c$ is the number of grid points for the wave speed
$c=\omega/\kappa_1$.  $\Delta\lambda_i$ and $\Delta x_2$ are the grid
resolutions in the wall-parallel and wall-normal directions,
respectively.  \label{tab:cases}}
\end{table}

\section{Characteristics of resolvent modes for compressible turbulent boundary layer}
\label{sec:modes}

\begin{figure}
\centerline{
\subfloat[]{{\includegraphics[width=0.49\textwidth]{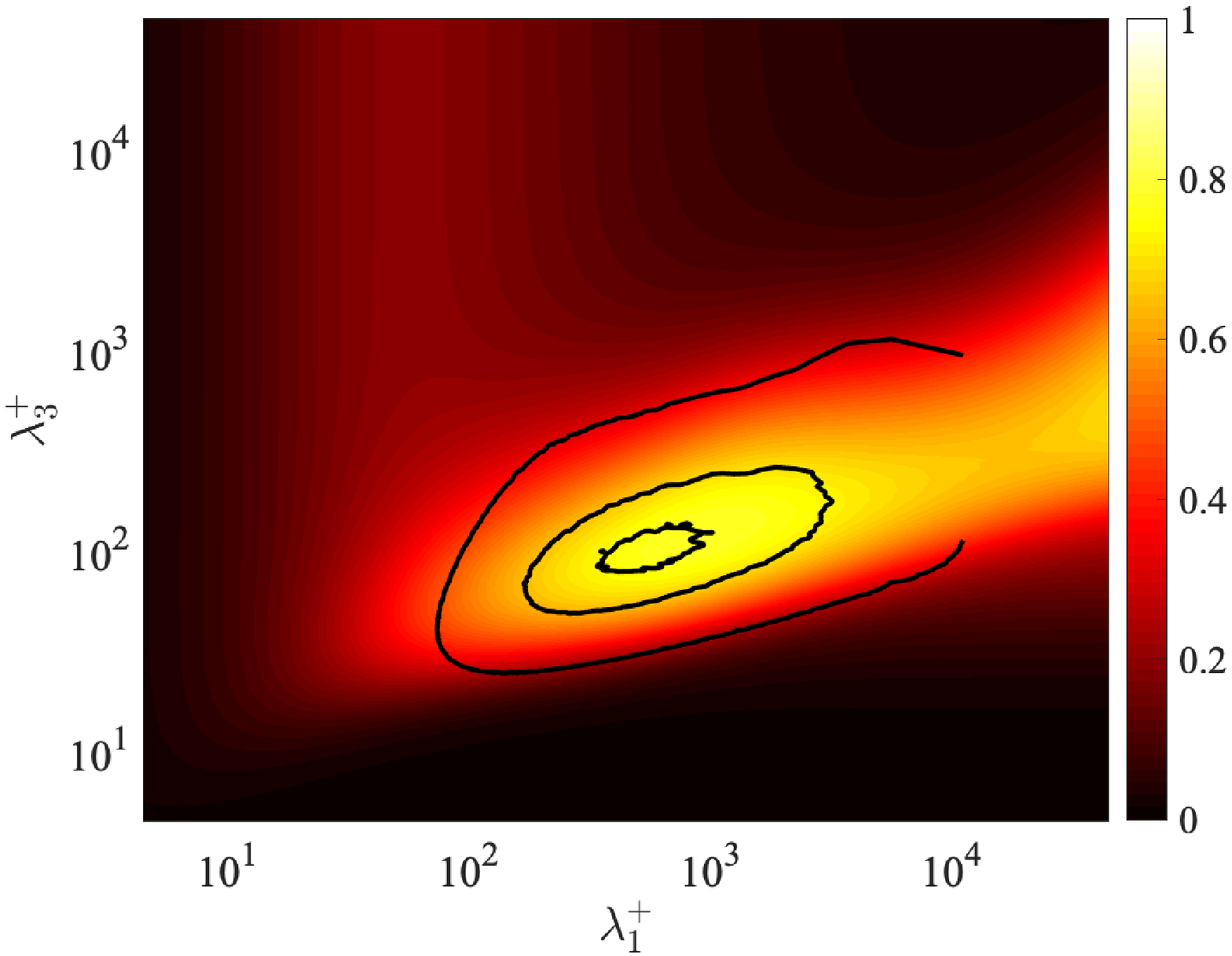}}}
\hspace{0.1cm}
\subfloat[]{{\includegraphics[width=0.49\textwidth]{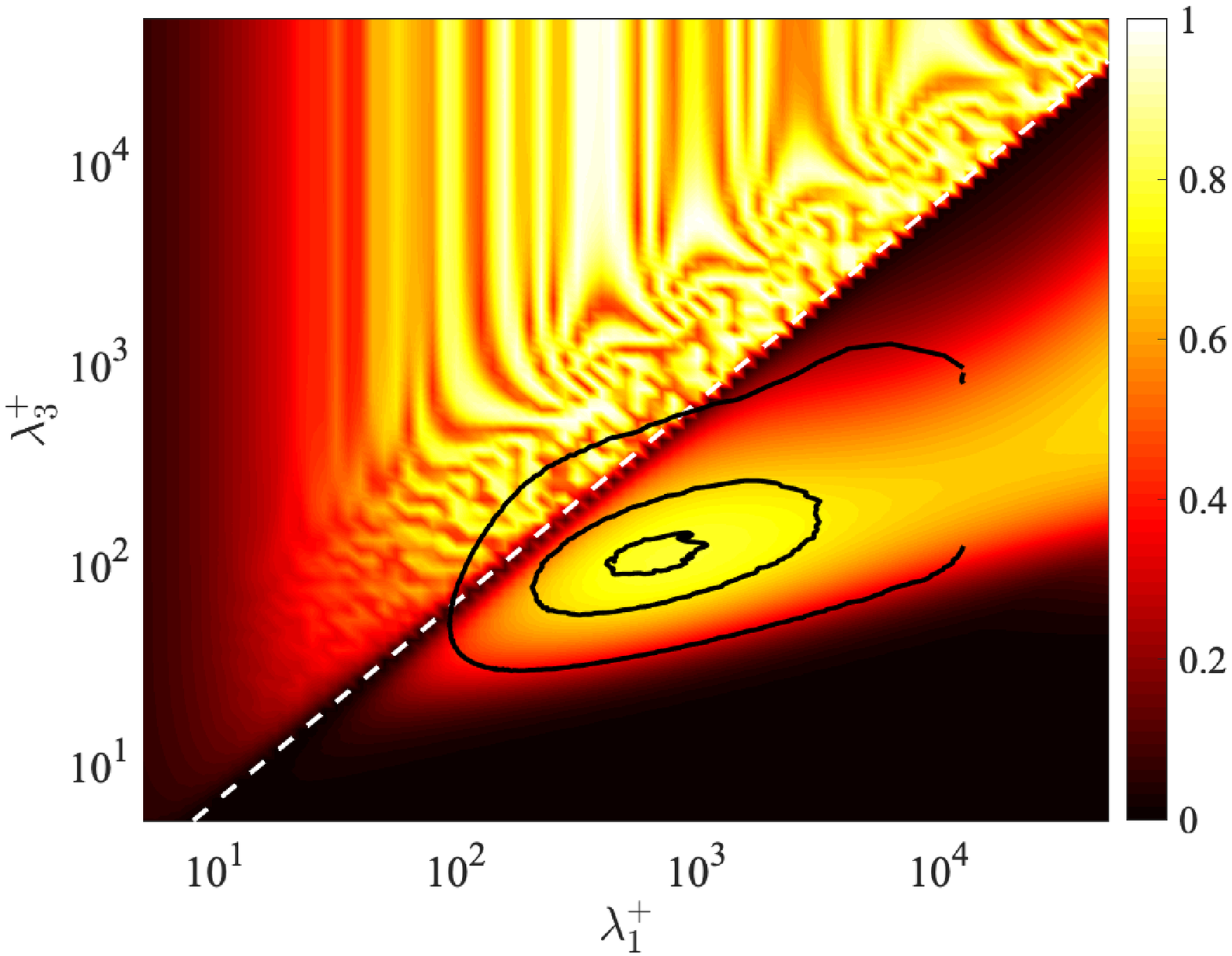}}}} 
\centerline{
\subfloat[]{{\includegraphics[width=0.49\textwidth]{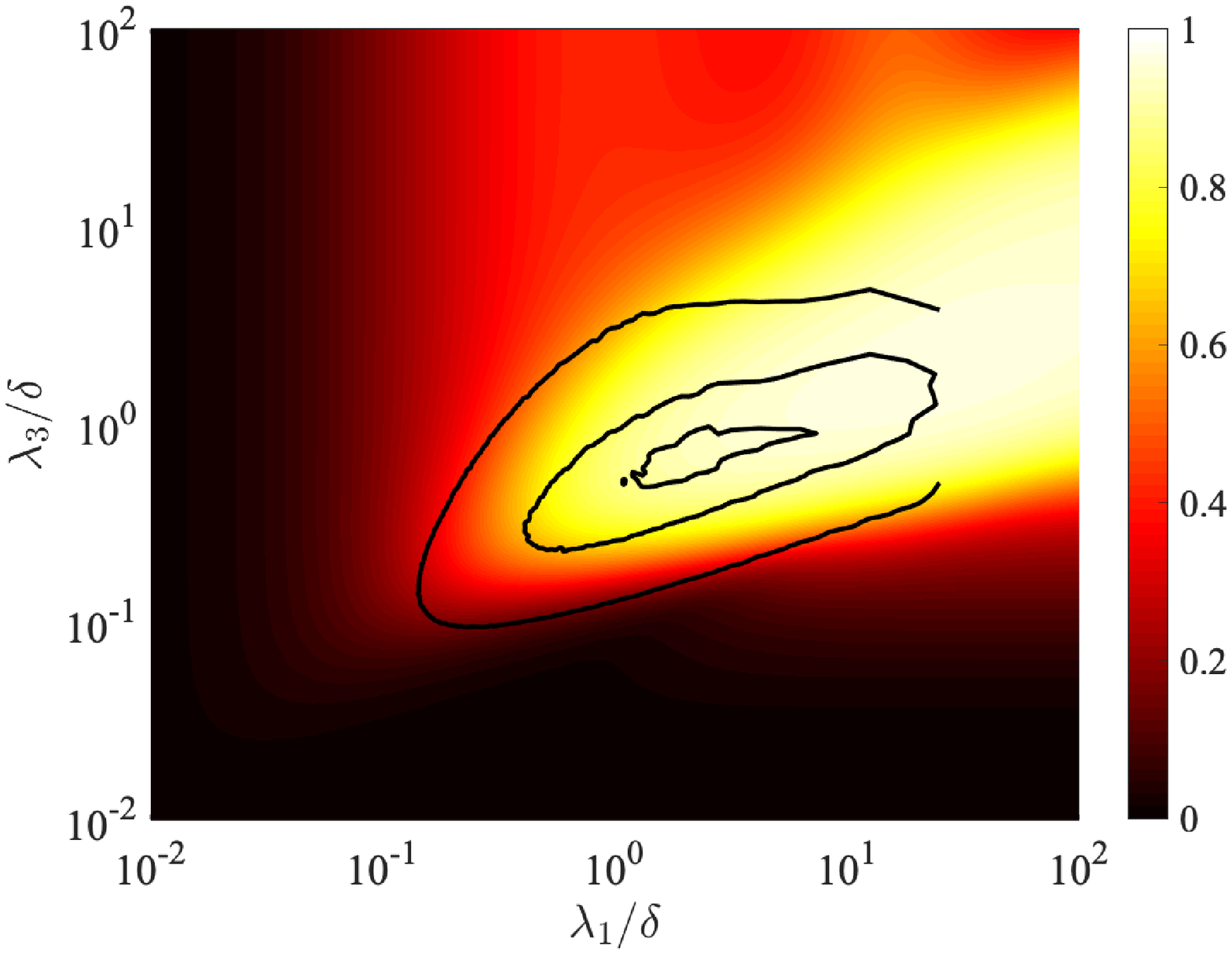}}}
\hspace{0.1cm}
\subfloat[]{{\includegraphics[width=0.49\textwidth]{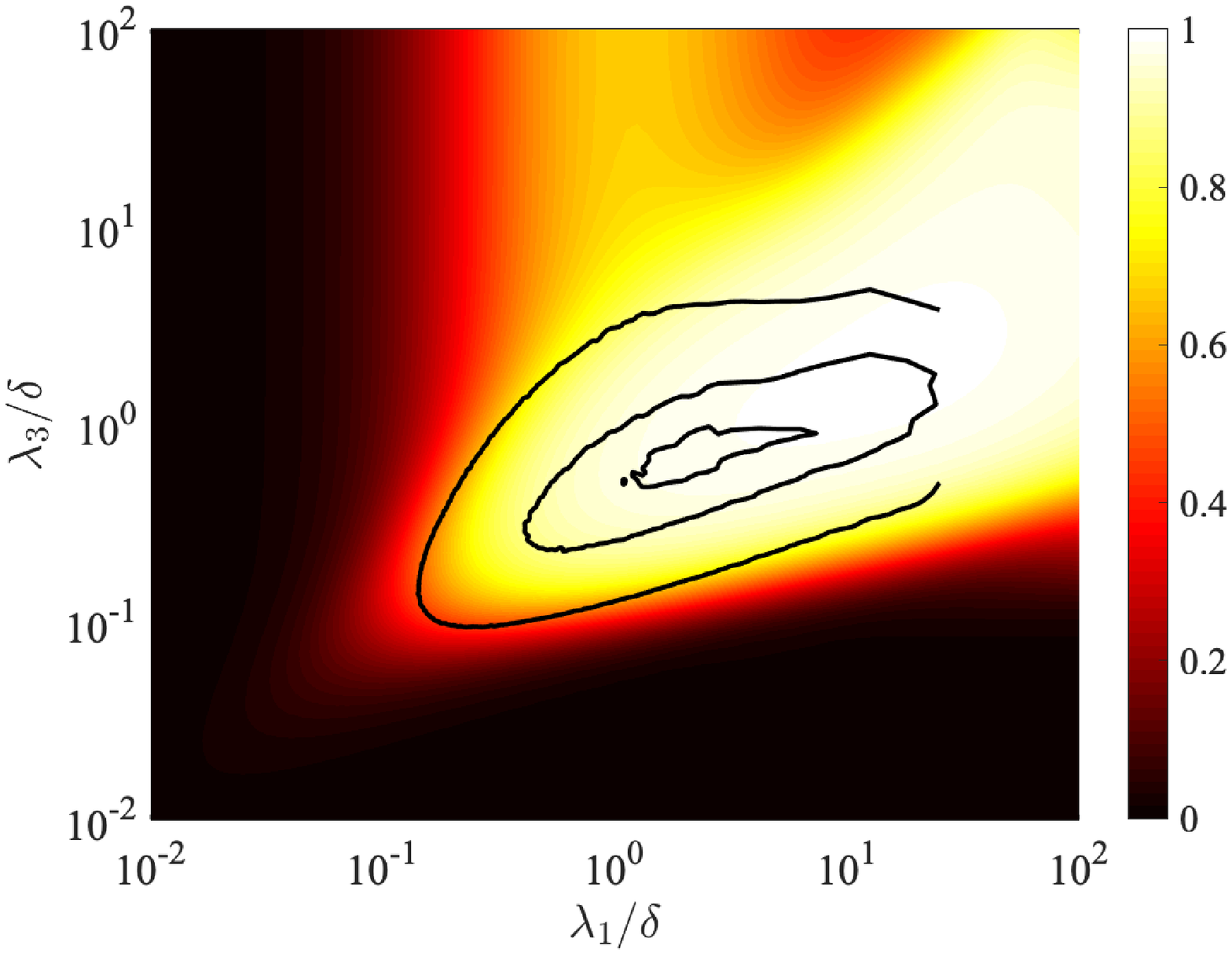}}}}

\caption{Energy contained in the principal response mode relative to
the total response, $\sigma_1^2/(\sum_j \sigma_j^2)$, for different
streamwise and spanwise wavelengths for the (a,c) incompressible and
(b,d) compressible ($M_\infty=4$) turbulent boundary layer at (a,b)
$x_2^+=15$ and (c,d) $x_2/\delta=0.2$. The contours are 10\%, 50\%,
and 90\% of the maximum energy of the premultiplied energy spectra for
channel flow at $\Rey_\tau\approx 550$ \citep{delalamo2004} at the
corresponding wall-normal locations. The white dashed line indicates
the relative Mach number of unity, $\overline{M}_\infty = 1$.
\label{fig:sv_energy}}
\end{figure}
%
\begin{figure}
\centerline{
\subfloat[]{{\includegraphics[width=0.45\textwidth]{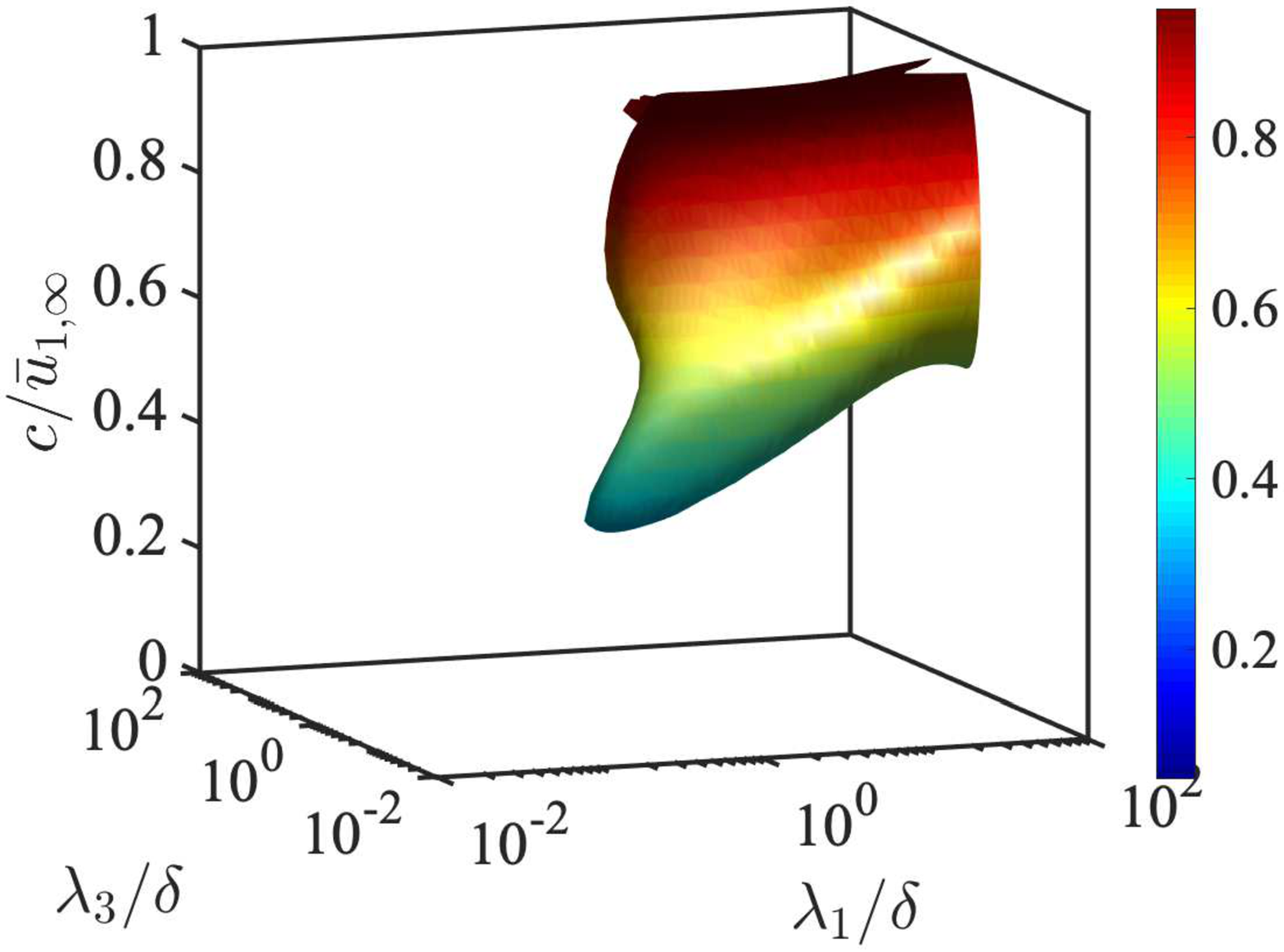}}}
\hspace{0.1cm}
\subfloat[]{{\includegraphics[width=0.45\textwidth]{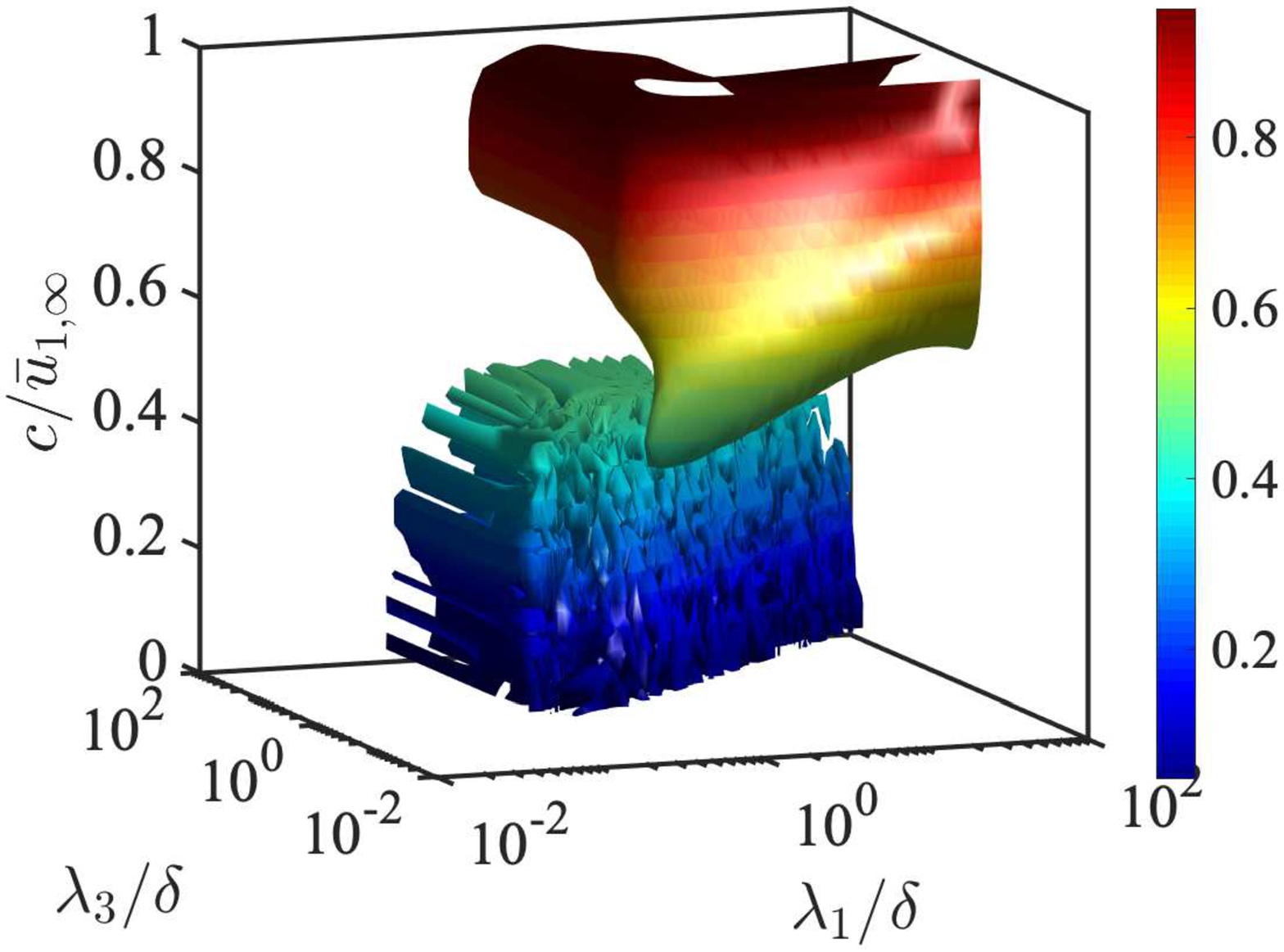}}}}
\centerline{
\subfloat[]{{\includegraphics[width=0.45\textwidth]{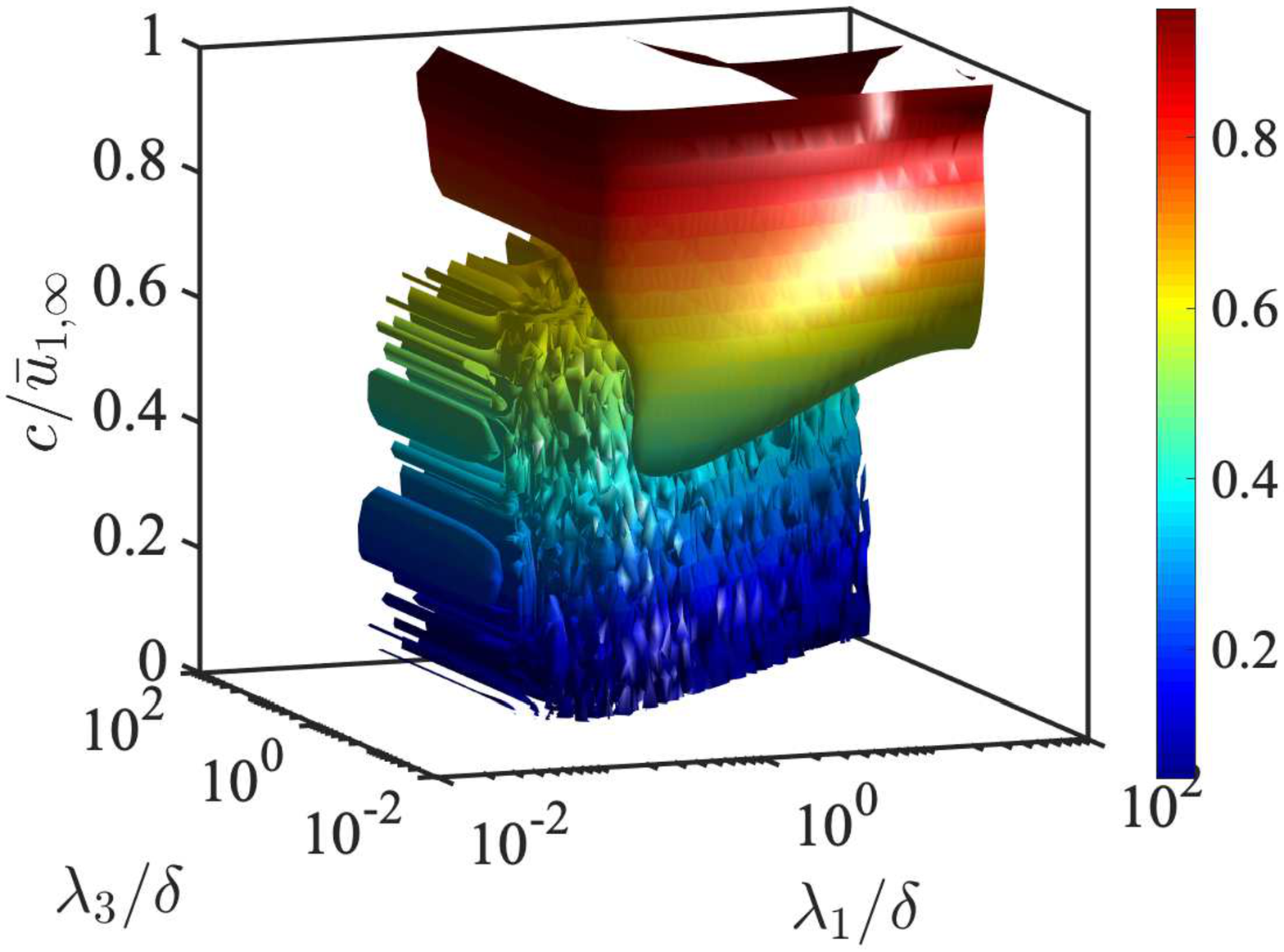}}}
\hspace{0.1cm}
\subfloat[]{{\includegraphics[width=0.45\textwidth]{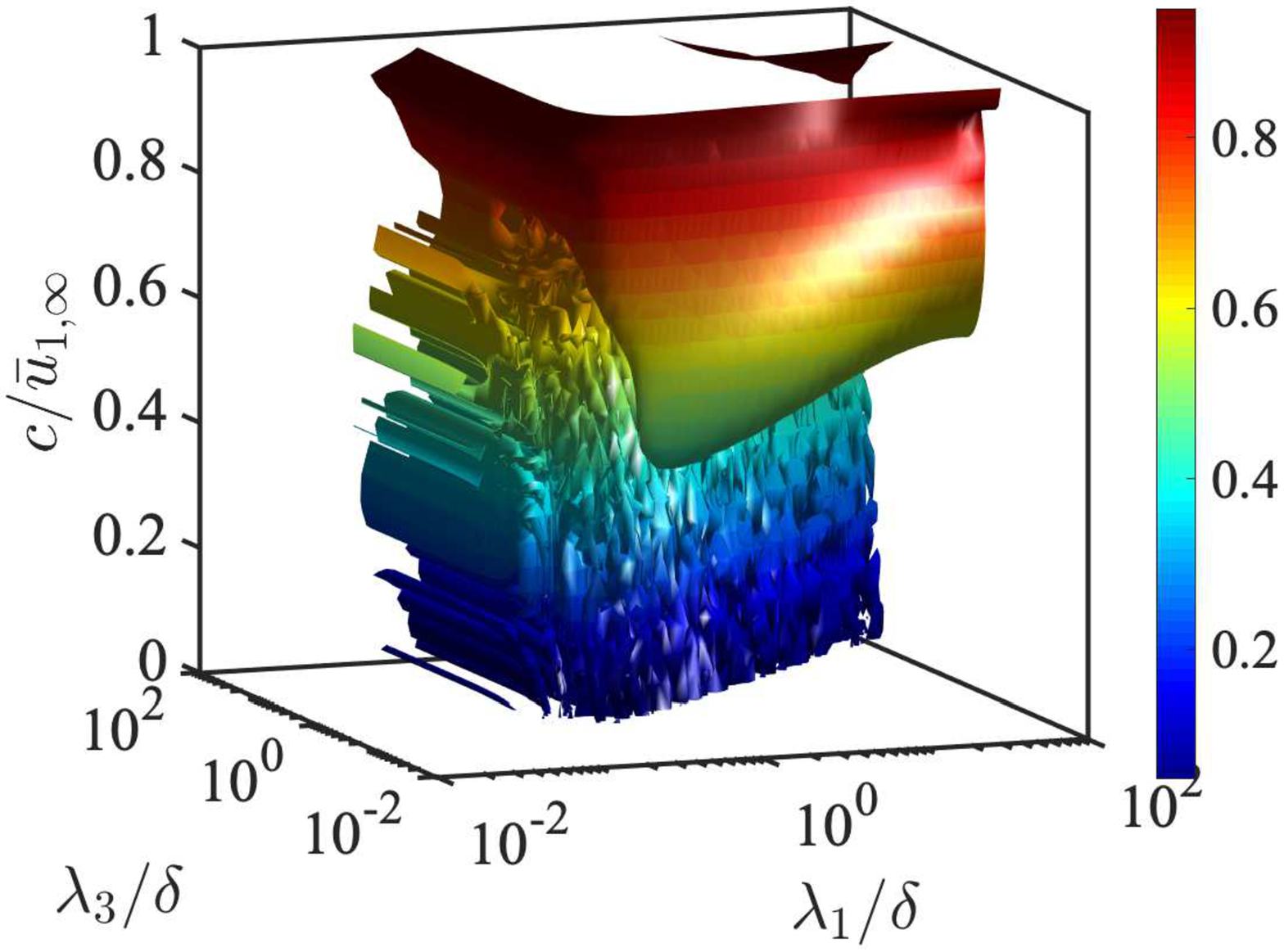}}}}

\caption{Energy contained in the principal response mode relative to
the total response, $\sigma_1^2/(\sum_j \sigma_j^2)$, for different
streamwise and spanwise wavelengths and wave speeds for
the (a) incompressible, (b) $M_\infty=2$, $\Rey_\tau=450$, (c)
$M_\infty=3$ and (d) $M_\infty=4$ cases. The contour surface is
$\sigma_1^2/(\sum_j \sigma_j^2) = 0.75$ coloured by wall-normal
distance from the wall.  \label{fig:sv_energy_3D}}
\end{figure}
We first examine the energy contained in the principal response mode.
The energy contribution of $\boldsymbol{\psi}_k$ to the total response
subject to broadband forcing in the wall-normal direction can be
quantified by {\color{black}$\sigma_k^2/(\sum_j\sigma_j^2)$}. Figure
\ref{fig:sv_energy} shows the principal energy contribution from the
principal response mode $\boldsymbol{\psi}_1$ for the incompressible
case and the $M_\infty=4$ case at two wall-normal locations $x_2^+=15$
and $x_2/\delta = 0.2$, where the superscript $+$ denotes wall units
defined in terms of $\bar{\rho}$ and $\mu$ at the wall and the
friction velocity $u_\tau$.

The results from the incompressible and compressible turbulent
boundary layers show similarities in the region where the principal
energy contribution of the incompressible boundary layer is
concentrated, i.e., where the low-rank approximation is valid for the
incompressible regime. This region coincides with the most energetic
wavenumbers from DNS of incompressible channel flows identified by the
premultiplied energy spectra given by the contour lines in the figure.
{\color{black} While the premultiplied energy spectra shown here are
from a DNS of channel flow, many studies in the past show similarities
in higher-order streamwise statistics for internal and external flows
\citep{mochizuki1996,metzger2001,delalamo2004,jimenez2008,monty2009},
especially in the near-wall region. Moreover, while comparisons 
between energy spectra of incompressible and compressible boundary 
layers are not available, first-order statistics of the two cases 
for adiabatic wall boundary conditions are known to collapse given 
a correct velocity  transformation \citep{pirozzoli2011}, making 
the comparison still meaningful.}
{\color{black} While data such as two-dimensional energy spectra at various
wall-normal heights for incompressible and compressible boundary
layers are not readily obtainable, access to such data could 
facilitate the analysis in the future.}

The most notable difference between the incompressible and
compressible results is the triangular region marked by freestream
relative Mach number \citep{mack1984}, $\overline{M}_\infty =
\overline{M}(x_2\rightarrow\infty)$, greater than unity (figure
\ref{fig:sv_energy}(b)). The relative Mach number, defined as 
\begin{equation} \overline{M}(x_2) =
\frac{(\kappa_1\bar{u}_1(x_2)-\omega)M_\infty}
{\left(\kappa_1^2+\kappa_3^2\right)^{1/2}\bar{T}(x_2)^{1/2}},
\end{equation}
can be understood as the local Mach number of the mean flow in the
direction of the wavenumber vector, $[\kappa_1,\kappa_3]^\intercal$,
relative to the wave speed at a given wall-normal location $x_2$.  A
three-dimensional depiction of the principal energy contribution as a
function of streamwise and spanwise wavenumbers and wave speeds is
given in figure \ref{fig:sv_energy_3D} for the incompressible case and
the compressible case at three different Mach numbers. It is clear
that the region with $\overline{M}_\infty
> 1$, i.e. the relatively supersonic region, increases with Mach
number and grows from the wall towards the freestream. In linear
stability theory, $\overline{M}_\infty$ has been used to classify
disturbances as subsonic, sonic, or supersonic \citep{mack1984,schmid2000}. Moreover, it has
been shown that if $\overline{M}_\infty>1$, a compressible boundary
layer is unstable to inviscid waves regardless of any other feature of
the velocity and temperature profiles \citep{mack1984}. Considering
that the family of modes with $\overline{M}_\infty>1$ does not have
any counterpart in incompressible boundary layers, it is expected that
the largest deviation between the compressible and incompressible
boundary layers occurs in this regime. In particular, the irregular
low-rank behaviour present in the relatively supersonic region in
figure \ref{fig:sv_energy}(b) and figure \ref{fig:sv_energy_3D}(b)--(d) is
due to the discrete acoustic eigenmodes of the system approaching the
wave speed $c$ (Appendix \ref{sec:append:modal}) and thus giving
resonant amplification of the resolvent operator \citep{dawson2019}.
{\color{black}This is a consequence of discretisation, and in the
continuous case, this resonant amplification effect will be present
everywhere in the relatively supersonic region.} {\color{black}
The extent of the relatively supersonic region, where the incompressible
and compressible boundary layers show a significant difference, increase with
increasing Mach number, and may be an indicator of why Morkovin’s
hypothesis fails for high Mach numbers.}

{\color{black} The second, although less significant, difference can be
seen for the two spectra at $x_2/\delta=0.2$ shown in figure
\ref{fig:sv_energy}(c) and (d)}, where the region of high energy
contribution covers a much wider range of wall-parallel wavelengths
for the $M_\infty=4$ case, with the `nose' of the spectrum located at a
smaller $(\lambda_1/\delta, \lambda_3/\delta)$.

The first observation regarding the relative Mach number of unity can
be explained by the formation of Mach waves in the relatively
supersonic region. The second can be comprehended in terms of the
correct scaling required for the compressible boundary layer. These main 
two observations will be discussed in the remainder of this section.

\subsection{Relatively supersonic region and Mach waves}
\label{sec:modes:acoustic}
As displayed by the energy contribution in figure \ref{fig:sv_energy},
the resolvent operator is shown to exhibit low-rank behaviour for the supersonic
turbulent boundary layer as well as for the incompressible case. \citet{moarref2013} showed that
significant understanding of the scaling of wall turbulence can be
obtained by using the simple rank-1 model. Here, we employ the same
rank-1 model by only keeping the most energetic forcing and response
directions corresponding to $\sigma_1$ and compute the premultiplied
one-dimensional energy density of the principal response of
$\mathsfbi{H}$ defined as
\begin{equation}
E_{qq}(x_2,\omega) = \sum_{i=1}^5\frac{\iint \kappa_1^2\kappa_3
\left[\sigma_1(\kappa_1,\kappa_3,\omega)|(q_i)_1|(\kappa_1,x_2,\kappa_3,\omega)\right]^2\,
\mathrm{d}\log\kappa_1 \mathrm{d}\log\kappa_3}
{\max\iint \kappa_1^2\kappa_3 \sigma_1(\kappa_1,\kappa_3,\omega)^2\,
\mathrm{d}\log\kappa_1 \mathrm{d}\log\kappa_3}.
\label{eq:Eqq}
\end{equation}
%
%
\begin{figure}
\centerline{
\subfloat[]{{\includegraphics[width=0.49\textwidth]{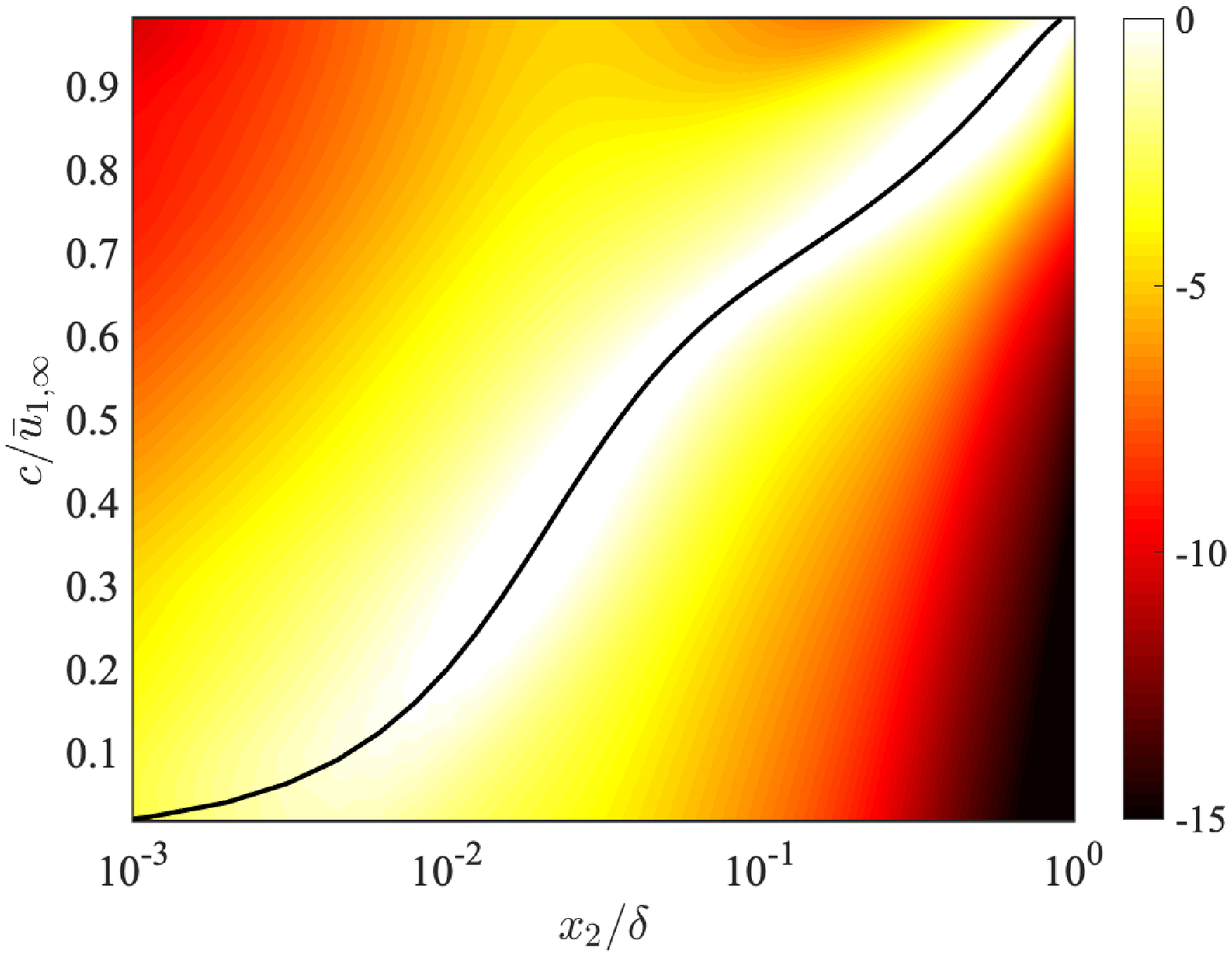}}}
\hspace{0.1cm}
\subfloat[]{{\includegraphics[width=0.49\textwidth]{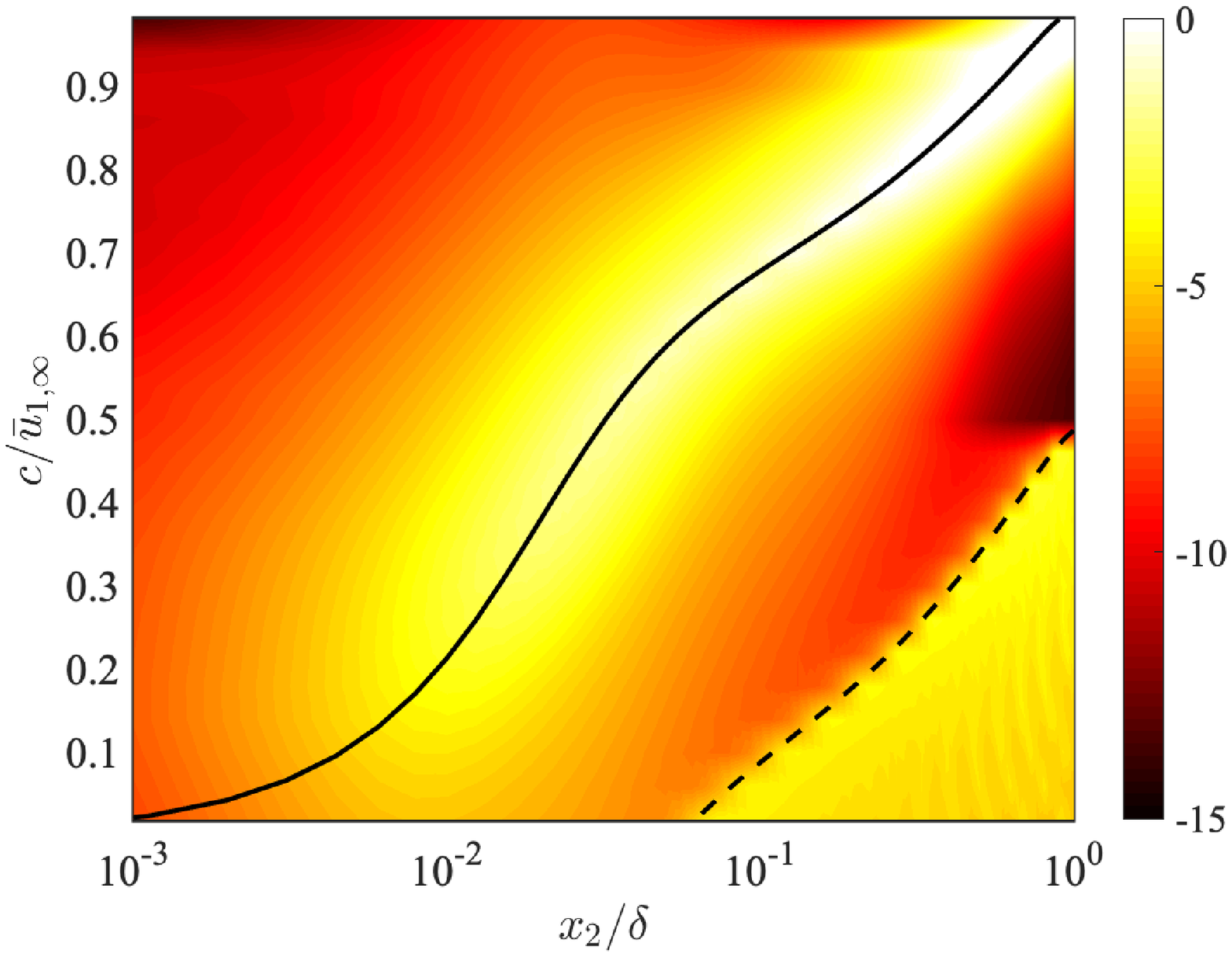}}} }
\centerline{
\subfloat[]{{\includegraphics[width=0.49\textwidth]{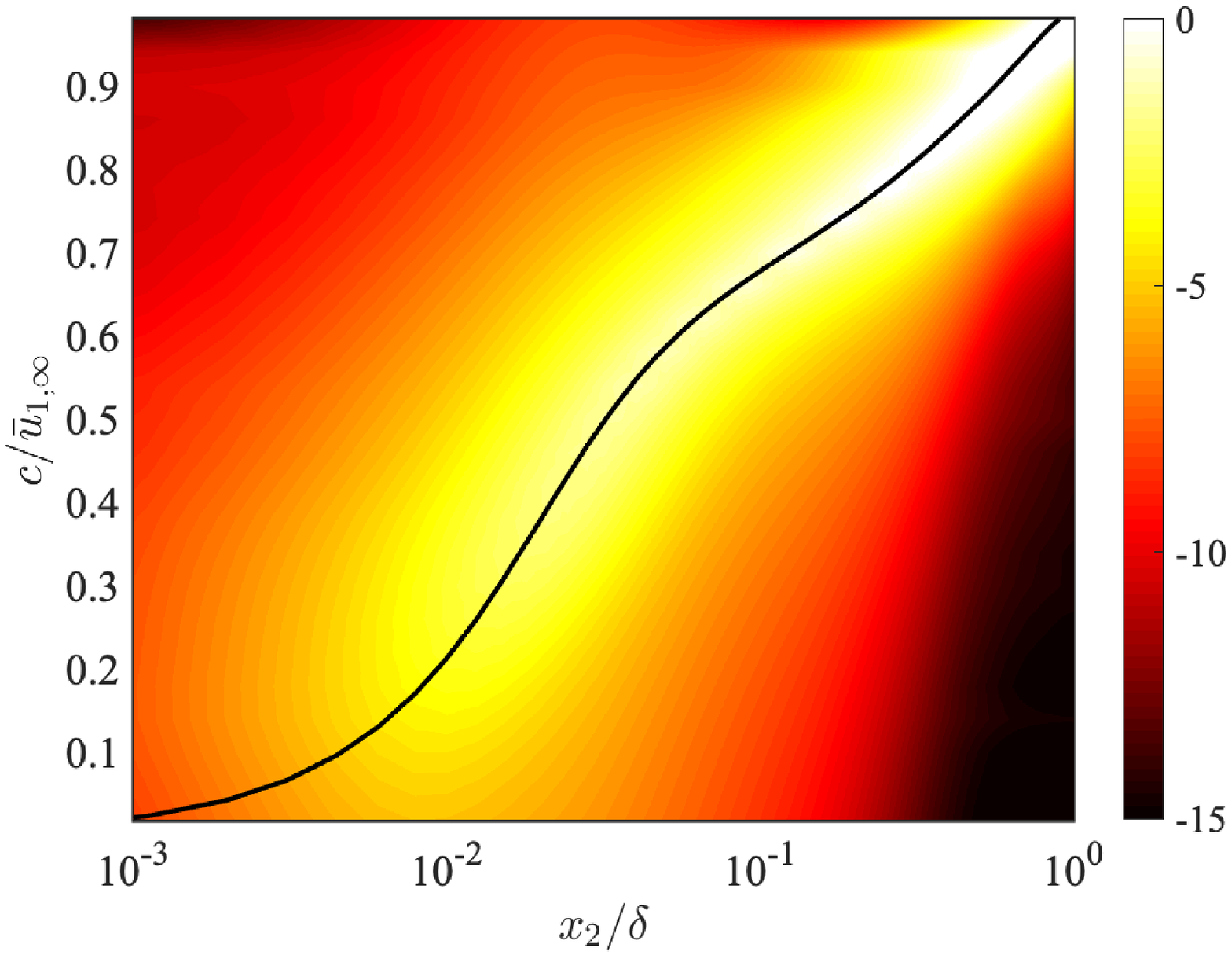}}}
\hspace{0.1cm}
\subfloat[]{{\includegraphics[width=0.49\textwidth]{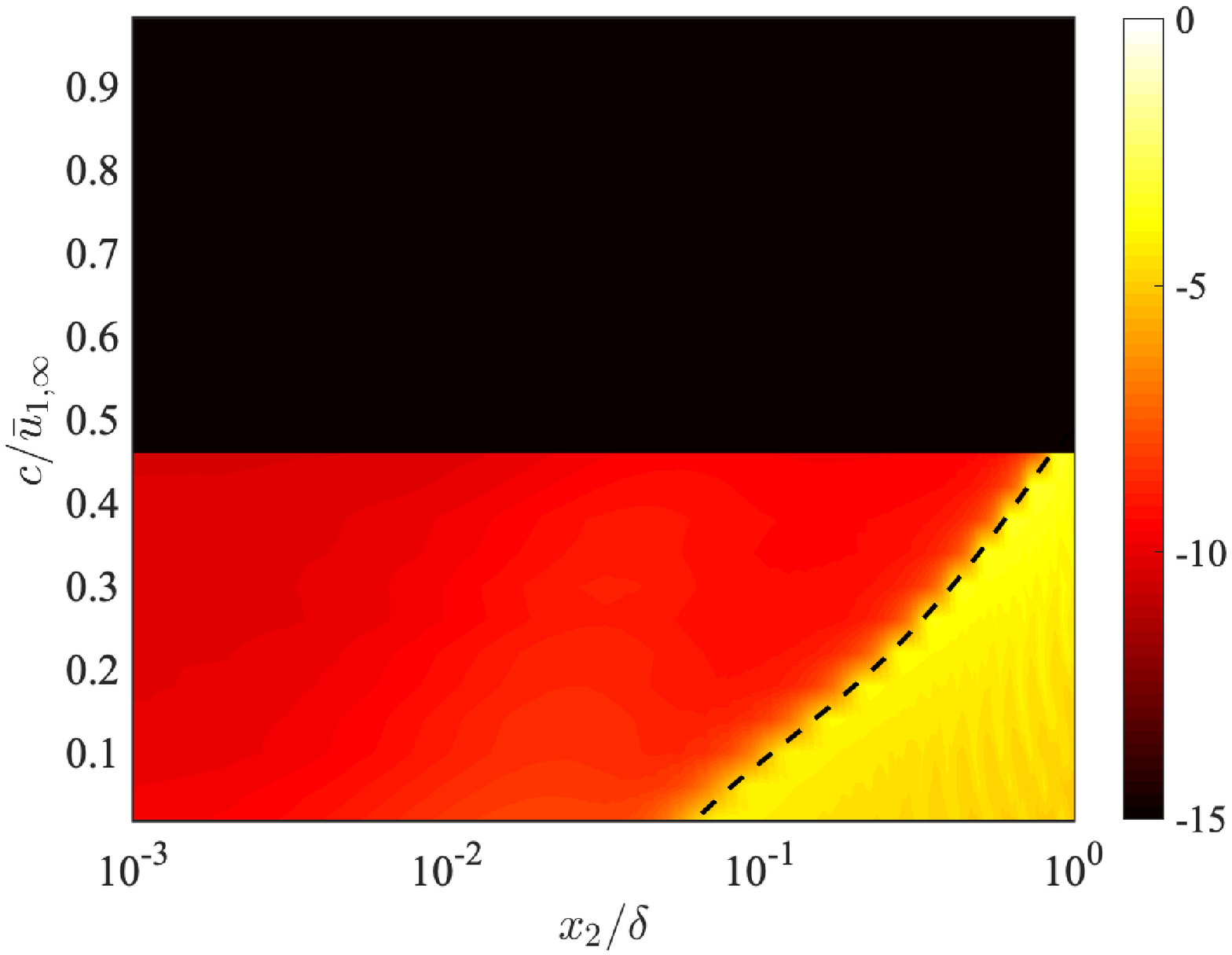}}} }
\caption{Premultiplied one-dimensional energy density
$E_{qq}$ for the (a) incompressible and (b) compressible
($M_\infty=2,\,\Rey_\tau=450$) cases, and the energy density for the
compressible case conditionally sampled to (c) $\overline{M}_\infty<1$ and
(d) $\overline{M}_\infty >1$. The turbulent mean velocity profile, $\bar{u}_1$
(\solidline) and the relative sonic line $\bar{c}$ (\dashline) are
shown for reference. Colours are in logarithmic scale.
\label{fig:energy_den}} \end{figure}
%

In figure \ref{fig:energy_den}(a) and (b), we plot the energy density
as a function of wave speed $c$ and $x_2$ for both the incompressible
and supersonic ($M_\infty=2$, $\Rey_\tau=450$) turbulent boundary
layers.  Unlike the incompressible case, where the energy density is
localised around the mean velocity profile (solid line), the
compressible case shows a second region which is displaced from the
mean velocity. The overlaid relative sonic line (dashed line), where
the velocity profile corresponds to relative streamwise Mach number of
unity at each wall-normal location, shows that the displacement of the
areas of high energy density is indicative of Mach waves. The relative
sonic line $\bar{c}$ is given by solving
\begin{equation}
\overline{M} = \frac{(\kappa_1\bar{u}_1 - \kappa_1\bar{c})M_\infty}
{\left(\kappa_1^2+\kappa_3^2\right)^{1/2}\bar{T}^{1/2}} = 1
\end{equation}
at each $x_2$ for $\kappa_3 = 0$, i.e. $\bar{c} = \bar{u}_1-
\bar{T}^{1/2}/M_\infty$, and indicates the minimum streamwise velocity
at each $x_2$ where a relatively supersonic region exists. From
conditionally sampling the energy intensity for
$\overline{M}_\infty<1$ or $\overline{M}_\infty>1$ by evaluating
\eqref{eq:Eqq} with the integrand of the numerator multiplied by an
indicator function for each case (as shown in figure
\ref{fig:energy_den}(c) and (d)), the two phenomena can clearly be
separated, and the second region of high energy density can be 
attributed entirely to the relatively supersonic region of
$\overline{M}_\infty>1$ and the existence of modes resembling Mach waves.

\begin{figure}
\vspace{0.2cm}
\centerline{
\subfloat[]{{\includegraphics[width=0.47\textwidth]{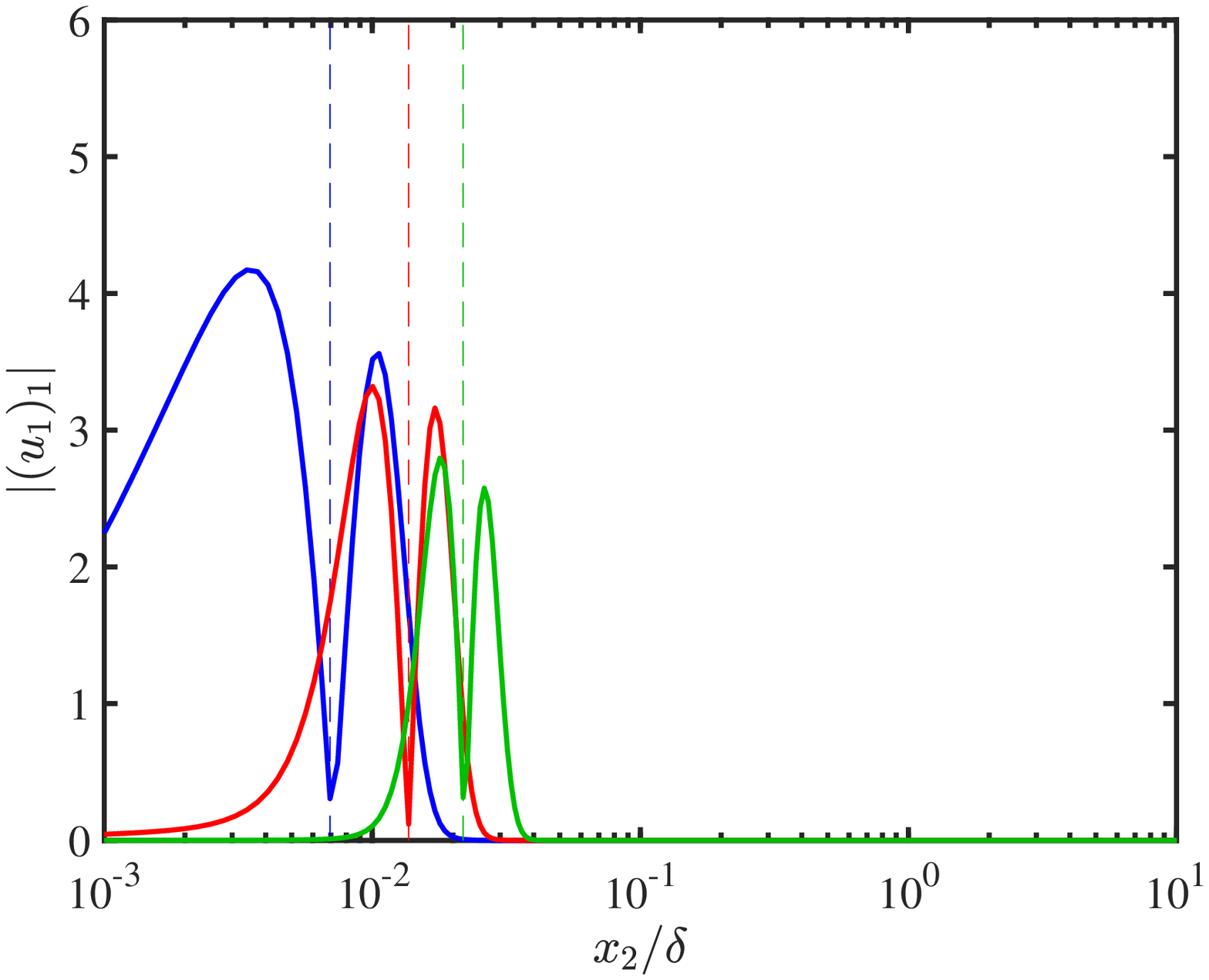}}}
\hspace{0.3cm}
\subfloat[]{{\includegraphics[width=0.47\textwidth]{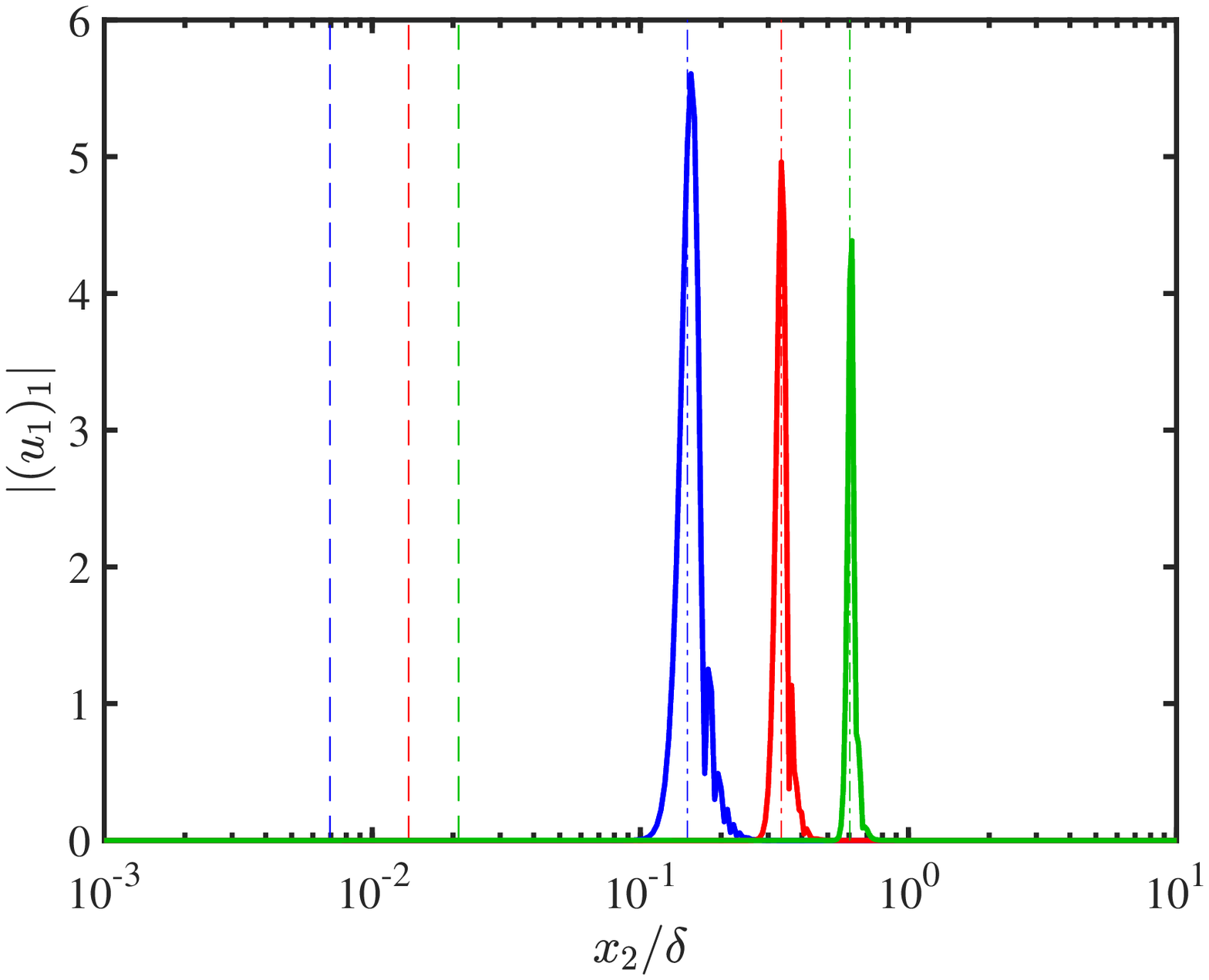}}} }
\caption{The response modes $(u_1)_1$ for the (a) incompressible and
(b) supersonic ($M_\infty=2,\,\Rey_\tau=450$) turbulent boundary layer
with $\lambda_1/\delta = 0.01$, $\lambda_3/\delta = 10$, and $c =
0.14$ ({\color{blue}blue}), $0.26$ ({\color{red}red}), $0.38$
({\color{green}green}). {\color{black}Reference lines are the wall-normal locations
of the critical layer,
$x_2=x_2^c$ (\dashline), and the sonic layer, $x_2=x_2^s$ (\dotdashline).}
\label{fig:acoustic_modes}}
\end{figure}

In figure \ref{fig:acoustic_modes}, we plot a few of the principal
streamwise velocity response modes $(u_1)_1$ for both the
incompressible and compressible ($M_\infty=2, \Rey_\tau=450)$ cases at
$(\lambda_1/\delta, \lambda_3/\delta) = (0.01,10)$, which lie within
the region $\overline{M}_\infty > 1$ for the compressible case for a
variety of wave speeds.  Note that the modes under consideration are
essentially two-dimensional, as we need $\lambda_1 \ll \lambda_3$ in
order for the relatively supersonic region to exist for a wide range
of wall-normal locations. This aspect ratio has not been studied in
the past in the context of unforced turbulent boundary layers and the
mode shapes that occur here are different from the nominal
three-dimensional case.  

Examining the principal response modes, we see the modes for the
incompressible case (figure \ref{fig:acoustic_modes}(a)) are centred
at the critical layer $x_2^c$, where $\bar{u}_1(x_2^c) = c$, but the
modes corresponding to $\overline{M}_\infty>1$ for the compressible
case (figure \ref{fig:acoustic_modes}(b)) are centred at the sonic
layer, $x_2^s$, where $\overline{M}(x_2^s) = 1$.  {\color{black}
Although not shown, the other velocity, density, and temperature
modes, and consequently the pressure modes, are also centred in the
sonic layer.} These response modes are consistent with Mach waves that
propagate towards the freestream and the concept of eddy shocklets
\citep{phillips1960,ffowcs1965}, where the instantaneous supersonic
events cause local shock-like structures in the boundary layer.
However, the analysis of the formation of eddy shocklets from
superimposed Mach waves requires additional knowledge of phase for
each wave parameter and is not considered here.
%

\subsection{Relatively subsonic region and universality of resolvent
modes}
\label{sec:modes:univ}

In order for the resolvent modes to exhibit universal behaviour for
different $\Rey_\tau$, {\color{black}the modes must have a narrow
footprint such that the support of the modes in the wall-normal
direction is localised.} This is because, otherwise, resolvent modes
are affected by the mean velocity at various wall-normal locations
that scale differently with $\Rey_\tau$ \citep{moarref2013}. Moreover,
the necessary condition for the existence of geometrically
self-similar resolvent modes is the presence of a logarithmic region
in the turbulent mean velocity profile.

\begin{figure}
\vspace{0.2cm}
\centerline{
\subfloat[]{{\includegraphics[width=0.47\textwidth]{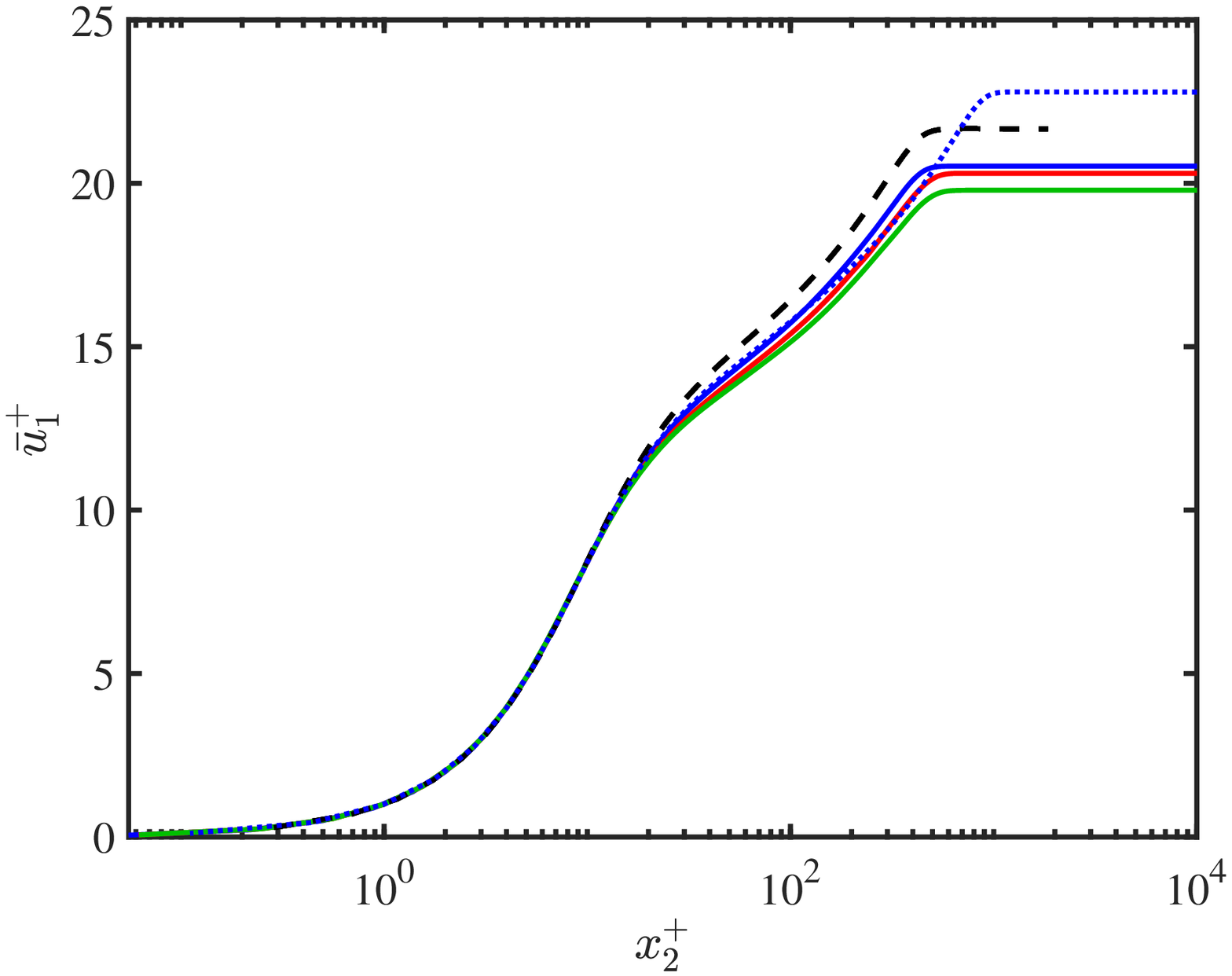}}}
\hspace{0.2cm}
\subfloat[]{{\includegraphics[width=0.47\textwidth]{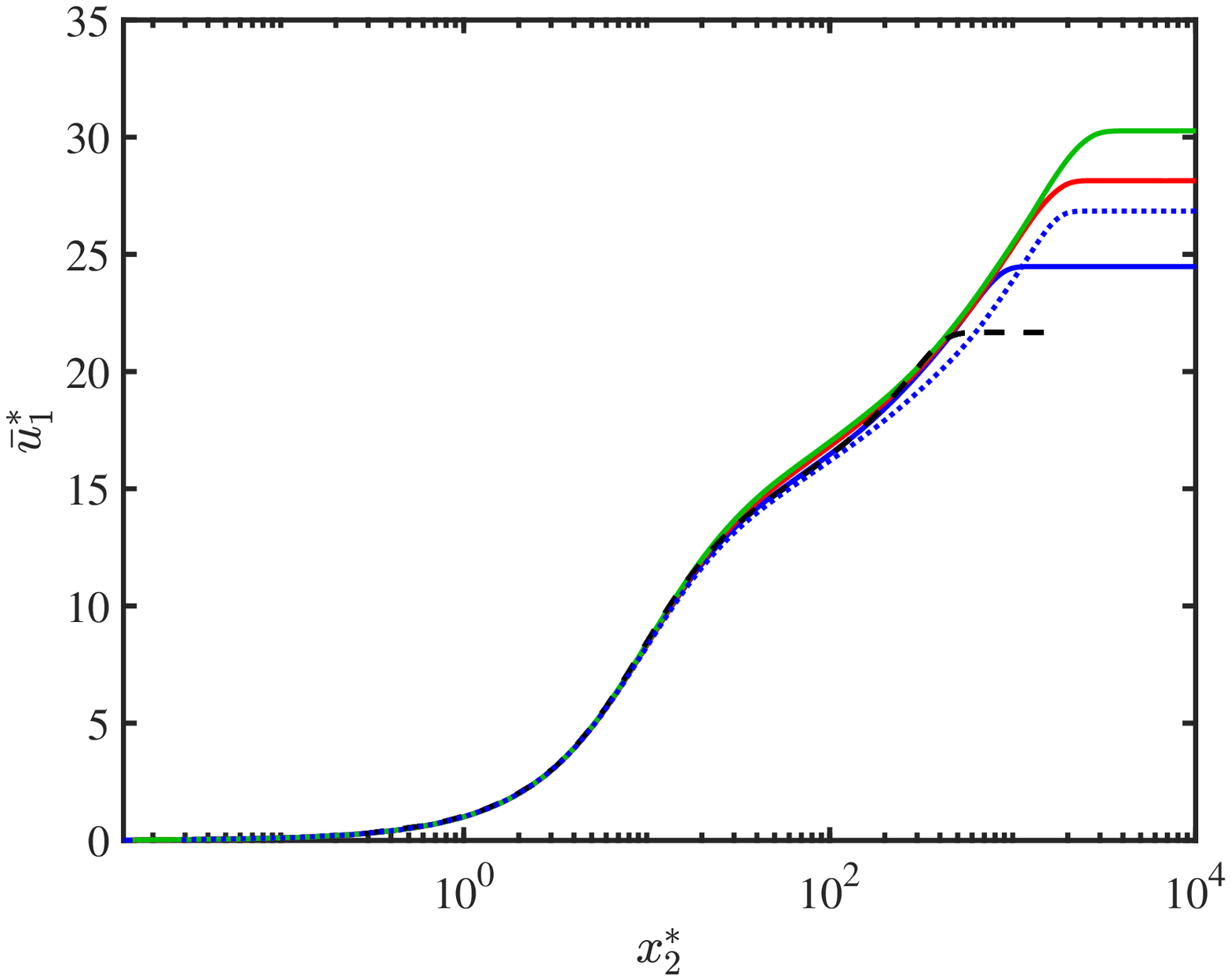}}} }
\centerline{
\subfloat[]{{\includegraphics[width=0.47\textwidth]{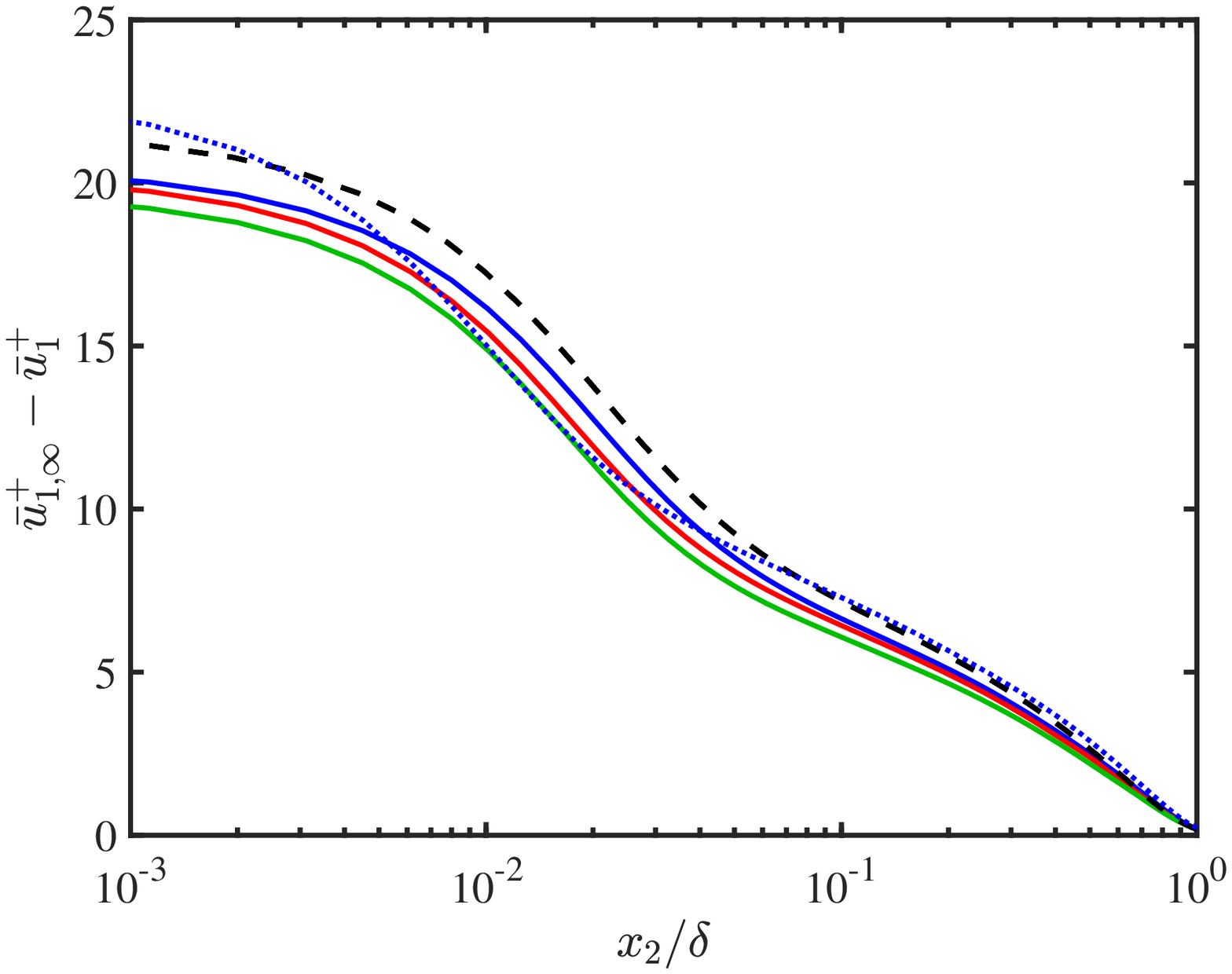}}}
\hspace{0.2cm}
\subfloat[]{{\includegraphics[width=0.47\textwidth]{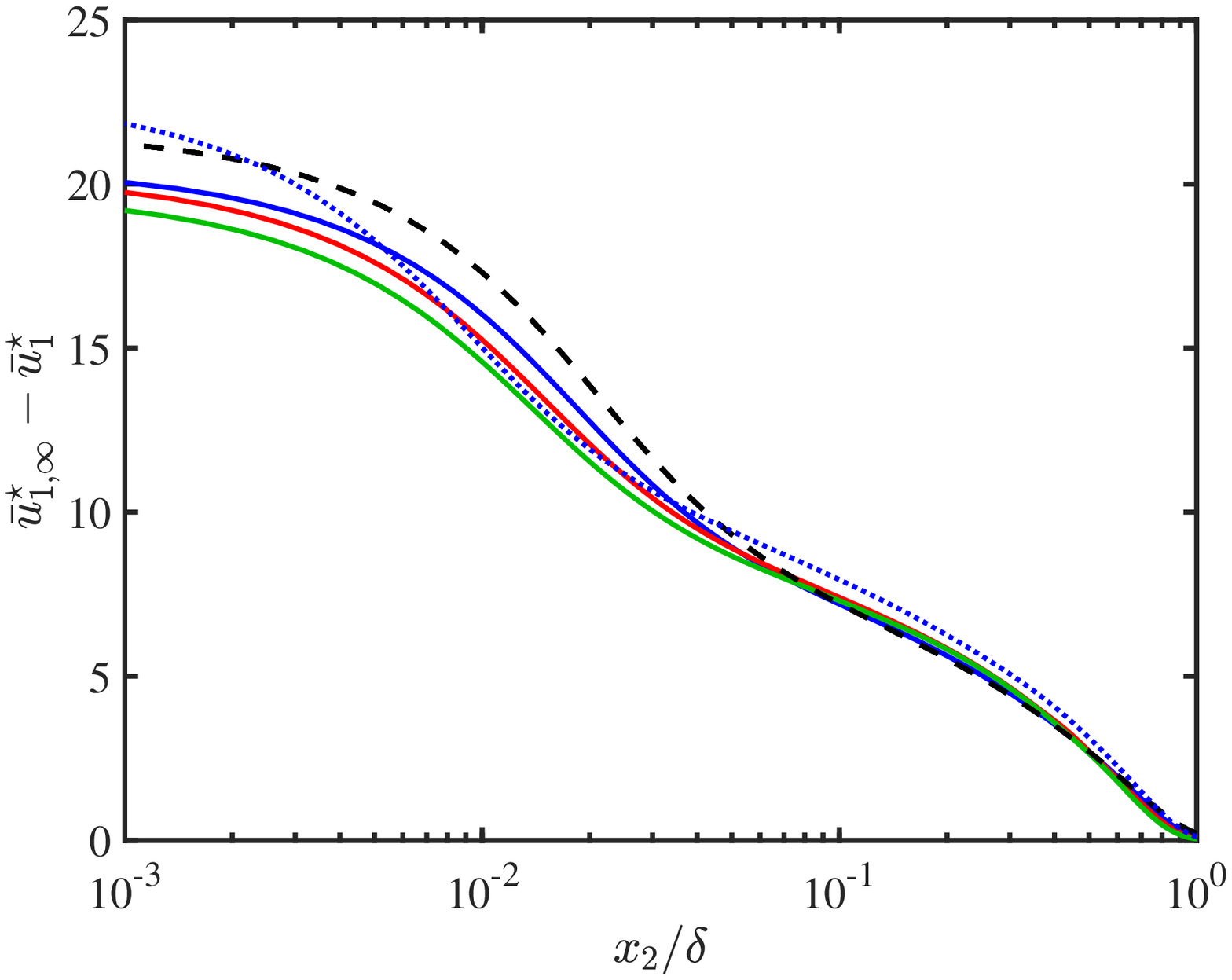}}} }
\caption{(a) Turbulent mean streamwise velocity profile
$\bar{u}_1^+(x_2^+)$, (b) the transformed velocity profile
$\bar{u}_1^*(x_2^*)$ given by \eqref{eq:TL_trans}, (c) the defect
velocity $\bar{u}_{1,\infty}^+-\bar{u}_1^+(x_2/\delta)$ with respect
to the freestream, and (d) the transformed defect velocity
$\bar{u}_{1,\infty}^\star-\bar{u}_1^\star(x_2/\delta)$ given by
\eqref{eq:outer_trans}. Lines indicate $M_\infty = 0$ (\dashline),
$M_\infty = 2,\,\Rey_\tau=450$ ({\color{blue}\solidline}), $M_\infty =
2,\,\Rey_\tau=900$ ({\color{blue}\dotline}), $M_ \infty = 3$
({\color{red}\solidline}), and $M_\infty = 4$
({\color{green}\solidline}).  \label{fig:mean_profile}}
\end{figure}
%
In order for the resolvent modes to be universal for the supersonic
boundary layer, not only do the modes have to be localised in $x_2$,
but the mean velocity profile must have a scaling law similar to that
of the incompressible case such that different regions of the mean
profiles collapse for various $M_\infty$ and $\Rey_\tau$.  In
compressible flows, viscous heating causes non-uniform mean density
and viscosity, which results in a mean velocity profile that no longer
satisfies the scaling of its incompressible counterpart. Many attempts
have been made to recover the scalings in this regime \citep[][among
others]{wilson1950,vandriest1951,coles1964,zhang2012,trettel2016},
with particular emphasis on the logarithmic region. Most of these
attempts have been made by seeking a transformation of $\bar{u}_1$ and
$x_2$ such that the compressible velocity profile maps onto an
equivalent incompressible profile. The most recent of these approaches
given by \citet{trettel2016} utilises a semi-local scaling in $x_2$
and an integrated stress-balance condition, which assumes that the sum
of viscous and Reynolds stresses in both the raw and transformed
states must be equal, for the scaling of $\bar{u}_1$ such that 
\begin{align}
x_2^* &= 
\frac{\bar{\rho}\left(\tau_w/\bar{\rho}\right)^{1/2}x_2}{\bar{\mu}},
\label{eq:semi_local}\\
\bar{u}_1^* &= 
\int_{0}^{\bar{u}_1^+}\left(\frac{\bar{\rho}}{\bar{\rho}_w}\right)^{1/2}
\left(1+\frac{1}{2\bar{\rho}}\frac{\mathrm{d}\bar{\rho}}{\mathrm{d}x_2}x_2 
-\frac{1}{\bar{\mu}}\frac{\mathrm{d}\bar{\mu}}{\mathrm{d}x_2}x_2\right)
\mathrm{d}\bar{u}_1^+.
\label{eq:TL_trans}
\end{align}
Here, the subscript $w$ indicates quantities evaluated at the wall,
and $\tau_w$ is the wall shear stress.  The results of this
transformation are illustrated in figure \ref{fig:mean_profile}(a) and
(b), and an improved collapse of the mean velocity profile in the
inner and logarithmic region for the various Mach numbers is achieved.
The semi-local scaling $x_2^*$ used here was introduced by
\citet{lobb1955}, revisited by \citet{huang1995} and
\citet{coleman1995} and generalised by \citet{trettel2016}. This
scaling gives rise to a {\color{black}semi-local} Reynolds number
\citep{cebeci2012,patel2015} at each wall-normal location 
\begin{equation}
\Rey_\tau^*(x_2) = \frac{\bar{\rho}(\tau_w/\bar{\rho})^{1/2}\delta}{\bar{\mu}}
\label{eq:semi_local_Re}
\end{equation} 
such that $x_2^* = (x_2/\delta)\Rey_\tau^*$.  While the transformation
proposed by \citet{trettel2016} works well for the inner and
logarithmic region, the collapse is not as good for the outer region.
We find that the best collapse for the outer region is achieved with
the transformation
\begin{equation}
\bar{u}_1^\star = \bar{u}_1^+\left(\frac{\bar{\rho}}{\bar{\rho}_w}\right)^{1/2},
\label{eq:outer_trans}
\end{equation}
which is equivalent to scaling the velocity with the semi-local
$u_\tau^* = \sqrt{\tau_w/\bar{\rho}}$ instead of $u_\tau =
\sqrt{\tau_w/\bar{\rho}_w}$, and the results are given in figure
\ref{fig:mean_profile}(c) and (d). A different transformation for the
outer layer is expected since the transformation given in
\eqref{eq:TL_trans} is based on the idea that momentum conservation
can be achieved by satisfying the stress balance condition, which only
holds in the inner layer of nearly parallel shear flow at reasonable
turbulence Mach numbers. Note that despite the better scaling in the
outer region, the collapse is not perfect, which is a known issue for
low Reynolds number boundary layer flows.  Still, a universal mean
velocity profile for the inner, logarithmic and outer layer can be
achieved by utilising the semi-local scaling. 

Given that the transformation of the turbulent supersonic mean
velocity profile in \eqref{eq:semi_local}--\eqref{eq:outer_trans}
produces a reasonable match to the incompressible profile in the
inner, outer and logarithmic regions, the relatively subsonic
($\overline{M}_\infty<1$) resolvent modes are expected to have a
universal behaviour in both Reynolds and Mach number, as the
one-dimensional energy density conditioned to $\overline{M}_\infty <
1$ in figure \ref{fig:energy_den}(c) shows a localisation with respect
to $x_2$.  Moreover, the existence of a logarithmic region in the
transformed mean streamwise velocity profile satisfies the necessary
condition for the geometrically self-similar modes to be present. Due
to the transformation in both $\bar{u}_1$ and $x_2$, the scaling of
the resolvent modes should be with respect to the semi-local
variables, $x_2^*$ and $\Rey_\tau^*$. The detailed scaling laws for
response modes of the inner, outer, and logarithmic region are given
in \S\ref{sec:scaling:prm}. 

\begin{figure}
\centerline{
\subfloat[]{{\includegraphics[width=0.49\textwidth]{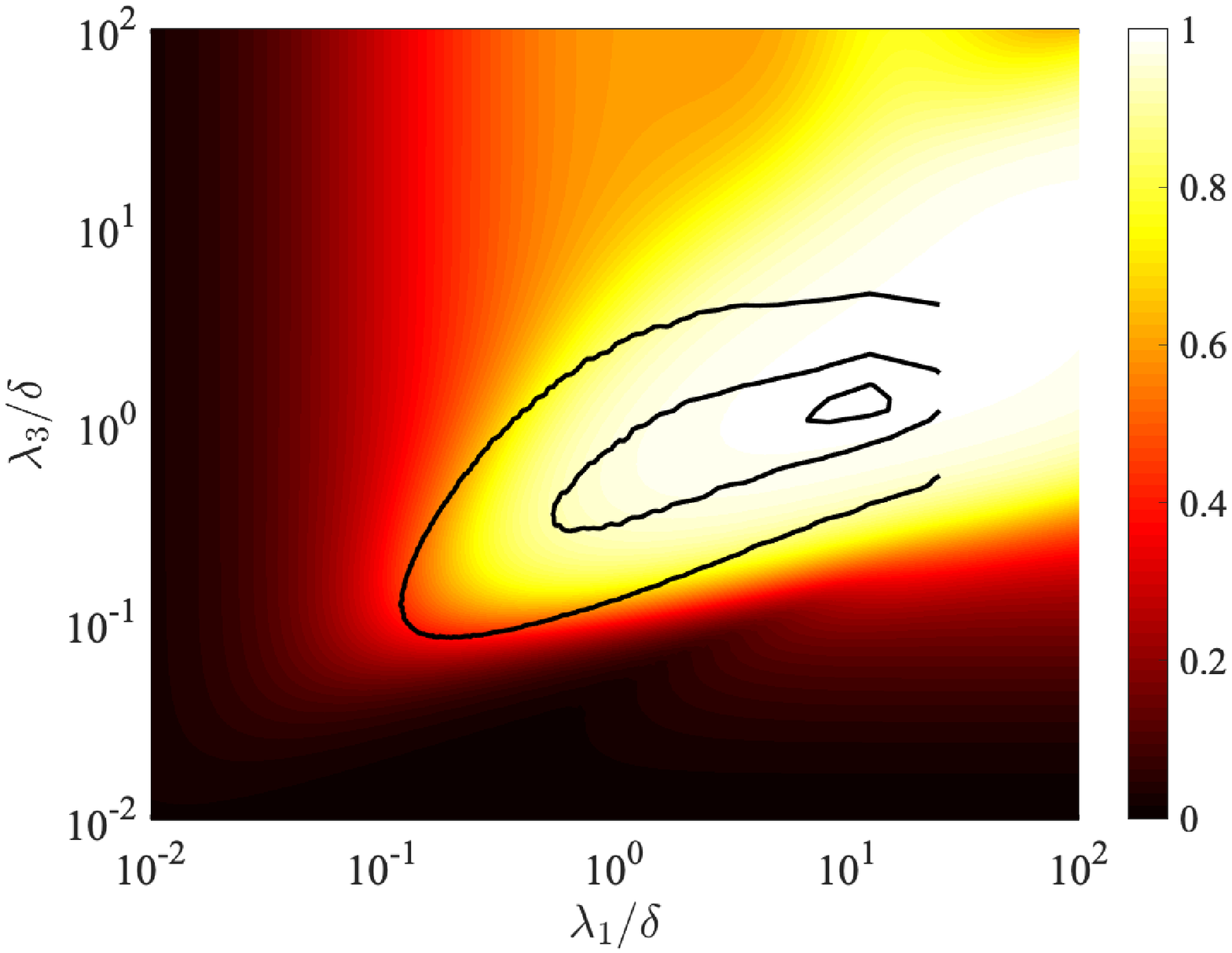}}}
\hspace{0.1cm}
\subfloat[]{{\includegraphics[width=0.49\textwidth]{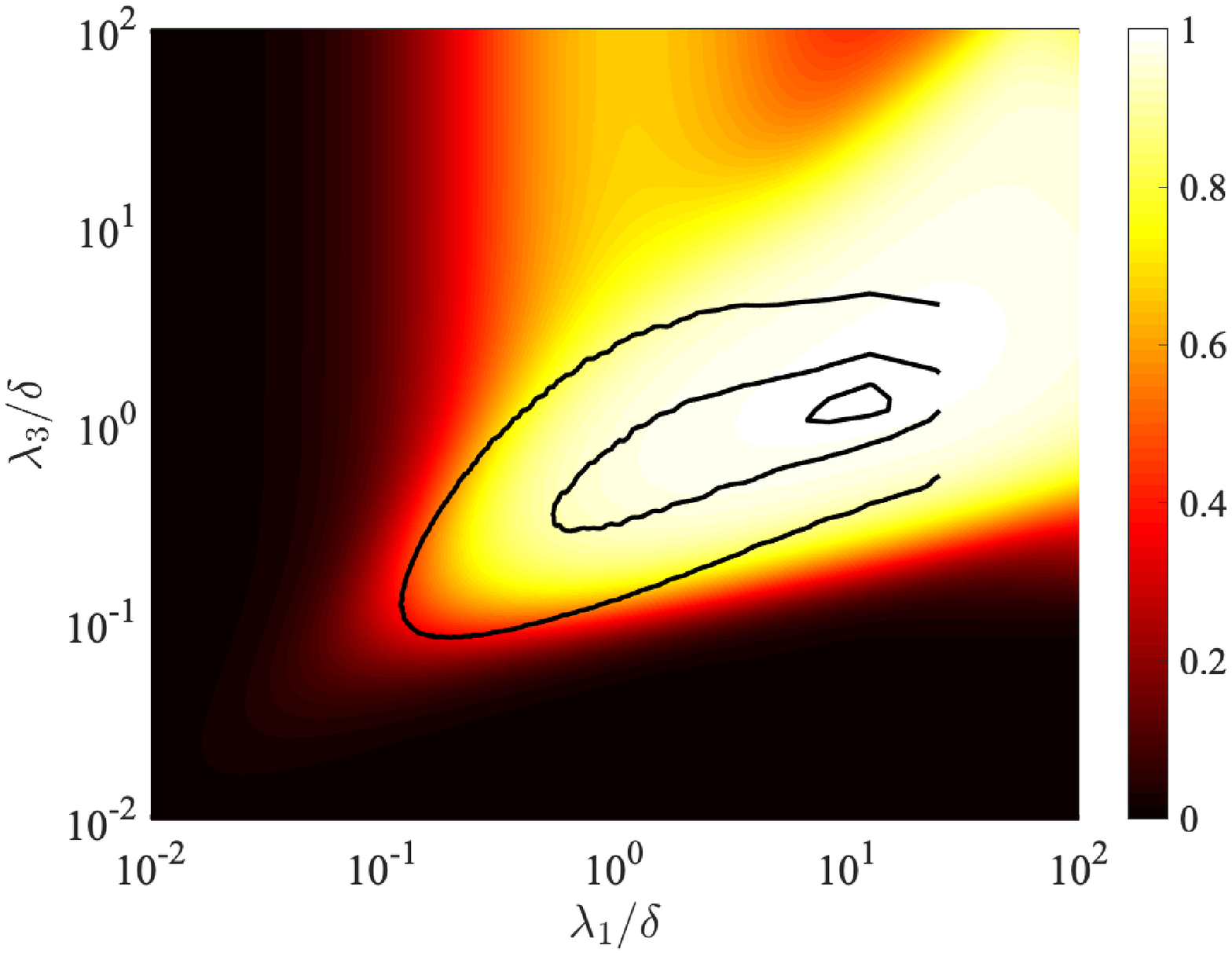}}} }

\caption{Energy contained in the principal response mode relative to
the total response for different streamwise and spanwise wavelengths
for the (a) incompressible ($\Rey_\tau = 1040$) and (b) compressible
($M_\infty=4$, $\Rey_\tau^* = 1020$) turbulent boundary layer at
$x_2/\delta=0.2$. The contours
are 10\%, 50\%, and 90\% of the maximum energy of the premultiplied
energy spectra for channel flow at $\Rey_\tau\approx 950$
\citep{delalamo2004} at the corresponding wall-normal location.
\label{fig:sv_energy_2}}
\end{figure}

The semi-local scaling also explains the discrepancy between the
principal energy contribution in the incompressible boundary layer
(figure \ref{fig:sv_energy}(c)) and the supersonic one (figure
\ref{fig:sv_energy}(d)). For the supersonic case of $M_\infty = 4$,
the semi-local Reynolds number at $x_2/\delta = 0.2$ is $\Rey_\tau^* =
1020$, which is significantly larger than the Reynolds number of the
incompressible case ($\Rey_\tau=450$). In figure
\ref{fig:sv_energy_2}, we instead compare against the results from the
incompressible turbulent boundary at $\Rey_\tau = 1040$ with the mean
velocity profile from \citet{schlatter2010} and the premultiplied
energy spectra from turbulent channel flow at $\Rey_\tau = 950$
\citep{delalamo2004}, which show a better qualitative comparison
between the two principal energy contribution spectra than the
comparison given in figure \ref{fig:sv_energy}. 

\section{Scaling of the principal singular value and resolvent modes}
\label{sec:scaling}

\subsection{Scaling of the principal response mode}
\label{sec:scaling:prm}

As discussed in \S\ref{sec:modes:univ}, the mean streamwise velocity
profile with the semi-local scaling collapses to the incompressible
boundary layer profile. This implies that the same scaling used in
\citet{moarref2013} for the incompressible channel flow can be
extended to the compressible boundary layer by using the length-scale
$x_2^*$ and Reynolds number $\Rey_\tau^*$.  

\begin{figure}
\vspace{0.2cm}
\centerline{
\includegraphics[width=0.6\textwidth]{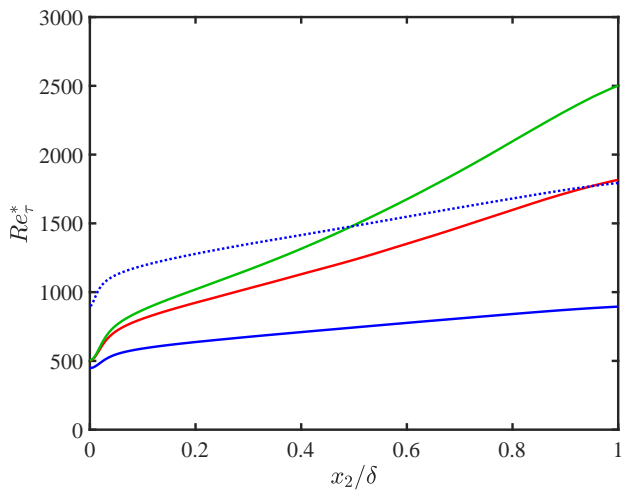}}
\caption{Semi-local friction Reynolds number $\Rey_\tau^*$ as a
function of $x_2$ for $M_\infty = 2,\,\Rey_\tau=450$
({\color{blue}\solidline}), $M_\infty = 2,\,\Rey_\tau=900$
({\color{blue}\dotline}), $M_ \infty = 3$ ({\color{red}\solidline}),
and $M_\infty = 4$ ({\color{green}\solidline}).  \label{fig:retaus}}
\end{figure}

Note that $\Rey_\tau^*$ depends on Mach number through the variation
in density and temperature (see figure \ref{fig:retaus}).  For an
adiabatic wall, the mean temperature profile is given by 
\begin{equation}
\frac{\bar{T}}{\bar{T}_\infty} = 1 + r \frac{\gamma -
1}{2}M_\infty^2\left[1-\left(\frac{\bar{u}_1}{\bar{u}_{1,\infty}}\right)^2
\right],
\end{equation}
where $r=\Pran^{1/3}$ is the recovery factor \citep{walz1969}.  Given
the definition of the semi-local Reynolds number
\eqref{eq:semi_local_Re} and the mean equation of state $\bar{p} =
\bar{\rho}\bar{T} = 1$, we have
\begin{equation}
\Rey_\tau^* = \Rey_\tau \frac{\bar{\mu}/\bar{\mu}_w}{\sqrt{\bar{\rho}/\bar{\rho}_w}}
= \Rey_\tau
\frac{\left(\bar{T}/\bar{T}_w\right)^{3/2}+C_w}{\left(\bar{T}/\bar{T}_w\right)^{3/2}\left(1+C_w\right)},
\end{equation}
where $C_w = S/T_w$. When $x_2=0$, we have that $\Rey_\tau^* =
\Rey_\tau$ as expected.  In the limit $x_2\rightarrow\infty$,
$\bar{T}/\bar{T}_w = 1/\left(1+r(\gamma-1)M_\infty^2/2\right)$ and we
have
\begin{equation}
\Rey_\tau^*(x_2\rightarrow\infty) = \Rey_\tau 
\frac{1+ C_w\left(1+r(\gamma-1)M_\infty^2/2\right)^{3/2}}{1+C_w}
\end{equation}
giving approximately a $M_\infty^3$ dependence for high Mach numbers.
\begin{table}
\begin{center}
\setlength{\tabcolsep}{10pt}
\begin{tabular}{l c c c c c}
Class                                                          & 
$\kappa_1$-scale                                               & 
$\kappa_3$-scale                                               & 
$x_2$-scale                                                    & 
$c$-scale                                                      &
$(u_1)_1,(\rho)_1,(T)_1$-scale                  
\\[0.4cm]
Inner                                                          &
$\dfrac{\kappa_1\delta}{\Rey_\tau^*}$                          & 
$\dfrac{\kappa_3\delta}{\Rey_\tau^*}$                          & 
$x_2^*$                                                        &
$c^*$                                                          &
$\widetilde{(q_i)}_1\dfrac{1}{\sqrt{\Rey_\tau^*}}$
\\[0.2cm]
Outer                                                          & 
$\kappa_1\delta\Rey_\tau^*$                                    & 
$\kappa_3\delta$                                               & 
$\dfrac{x_2}{\delta}$                                          &
$\dfrac{c}{\bar{u}_{1,\infty}^\star}$                          &
$\widetilde{(q_i)}_1$
\\[0.2cm]
Logarithmic                                                    & 
$\kappa_1x_2^cx_2^{c*}$                                        & 
$\kappa_3x_2^c$                                                & 
$\dfrac{x_2}{x_2^\text{c}}$                                    &
--                                                             &
$\widetilde{(q_i)}_1\sqrt{\frac{x_2^c}{\delta}}$

\end{tabular}
\end{center}
\caption{Expected length scales for the universal modes of the
resolvent operator for the turbulent boundary layer.
\label{tab:scaling}}
\end{table}

In order for a fair comparison between the incompressible and
compressible response modes, the compressible velocity response modes
must be normalised by the kinetic energy content in the response modes
due to the orthonormality constraint of the singular vectors and the
different norms used for the two cases. We define turbulent kinetic
energy and turbulent thermodynamic energy as 
\begin{equation}
E_K = (\boldsymbol{q},\boldsymbol{q})_K = \int_0^\infty \bar{\rho} u_i^\dagger u_i \mathrm{d}x_2, \quad
E_T = \int_0^\infty \frac{1}{\gamma M_\infty^2}
\left(\frac{\rho^\dagger\rho}{\bar{\rho}^2} + \frac{T^\dagger
T}{\bar{T}^2}\right)\mathrm{d}x_2,
\end{equation}
respectively. We normalise the velocity, density and temperature modes
such that 
\begin{equation}
\widetilde{(u_i)}_1 = \frac{\bar{\rho}^{1/2}(u_i)_1}{\sqrt{E_K}}, \quad
\widetilde{(\rho)}_1 = \frac{(\rho)_1/(\gamma M_\infty^2
\bar{\rho}^2)^{1/2}}{\sqrt{E_T}},\quad
\widetilde{(T)}_1 = \frac{(T)_1/(\gamma M_\infty^2
\bar{T}^2)^{1/2}}{\sqrt{E_T}}.
\end{equation}
The relationship between $E_K$ and $E_T$ is discussed later in
\S\ref{sec:scaling:ke_te}. The proposed scaling of the resolvent
for the different classes of the compressible boundary layer
is summarised in table \ref{tab:scaling}. Note that the combined
scaling of the normalised response modes, the singular values and the
ratio of $E_K$ to $E_T$ is of importance when examining the response
mode of the resolvent operator.

\subsubsection{Outer region}
\label{sec:scaling:prm:outer}

\begin{figure}
\vspace{0.2cm}
\centerline{
\subfloat[]{{\includegraphics[width=0.47\textwidth]{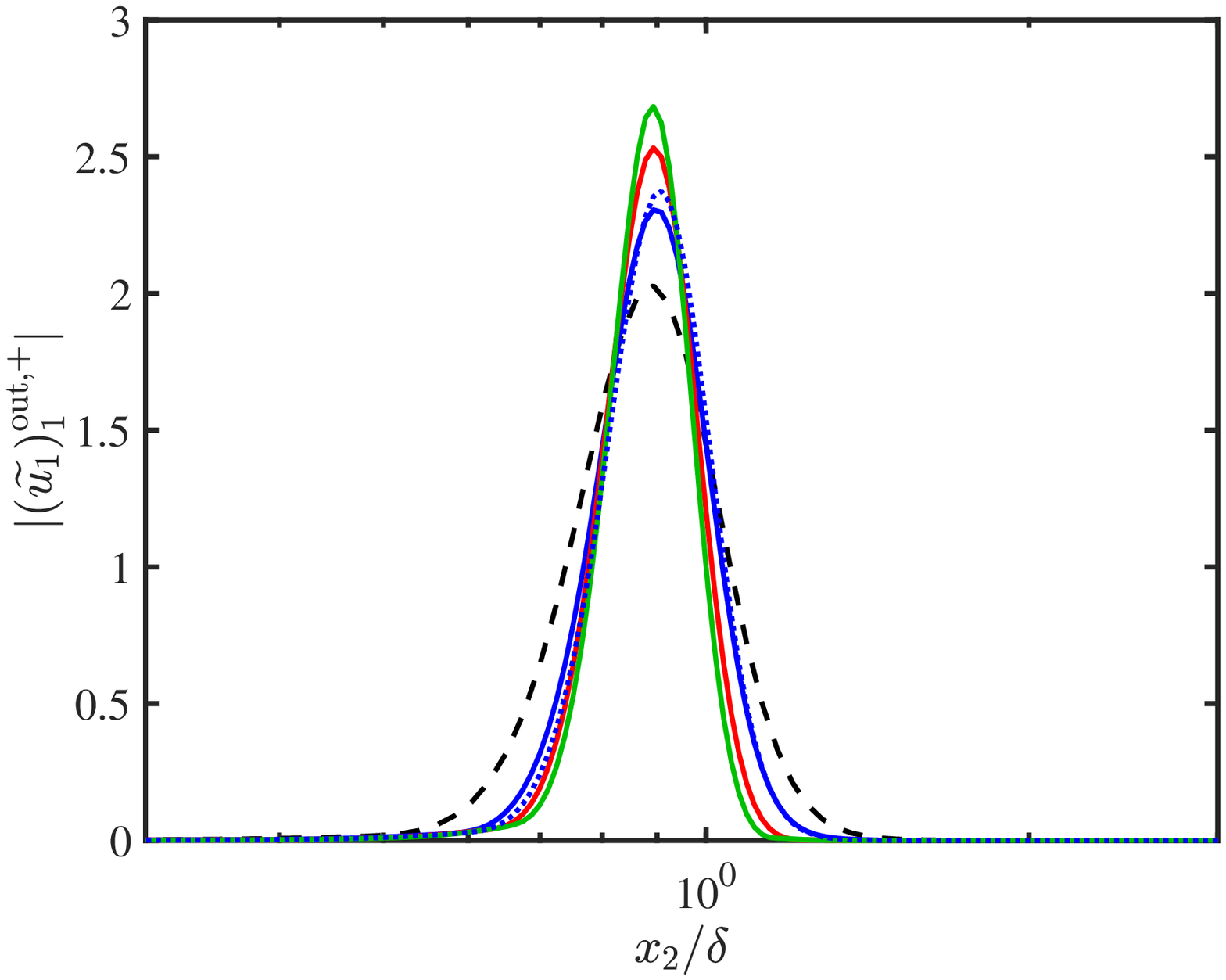}}}
\hspace{0.2cm}
\subfloat[]{{\includegraphics[width=0.47\textwidth]{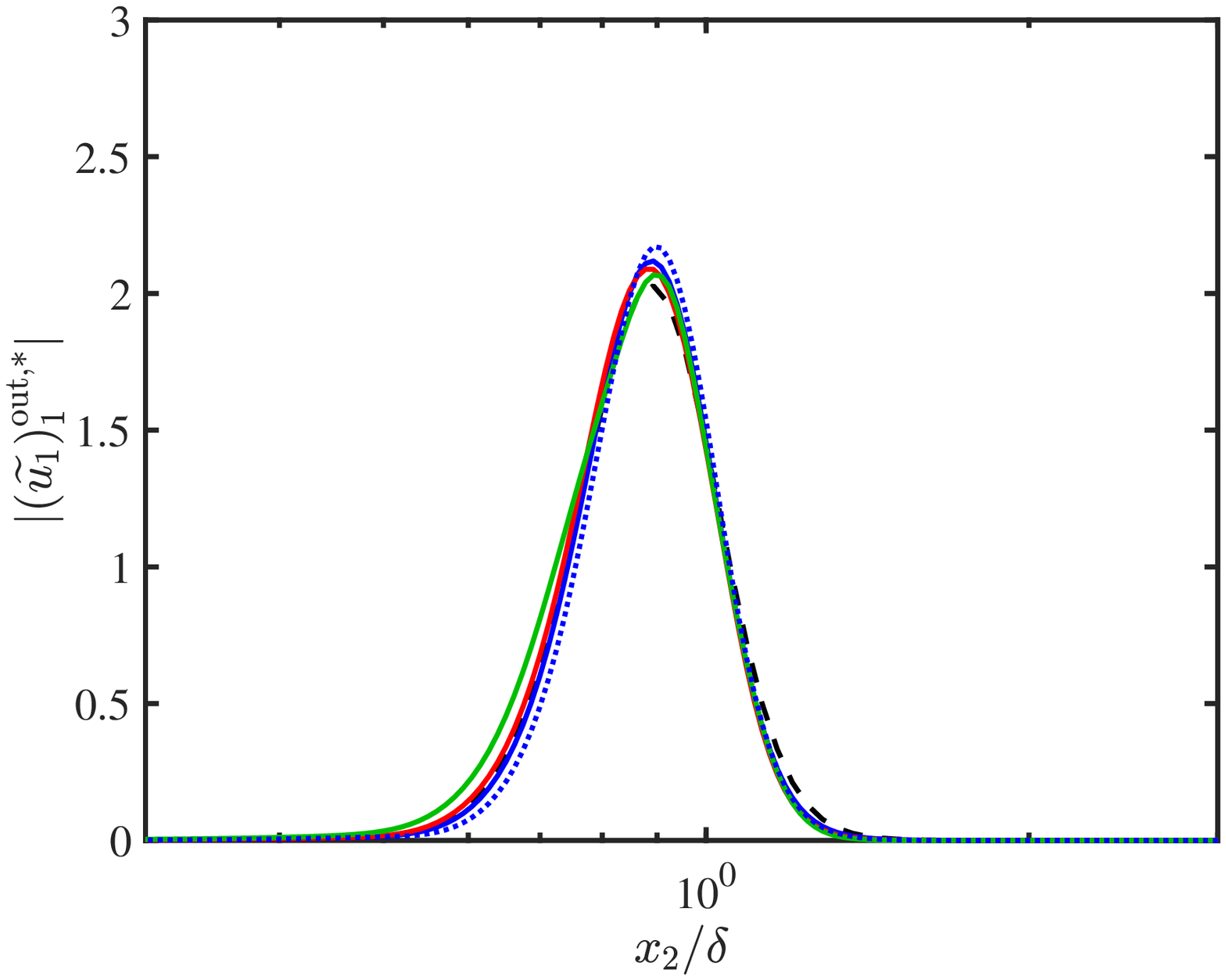}}} }
\centerline{
\subfloat[]{{\includegraphics[width=0.47\textwidth]{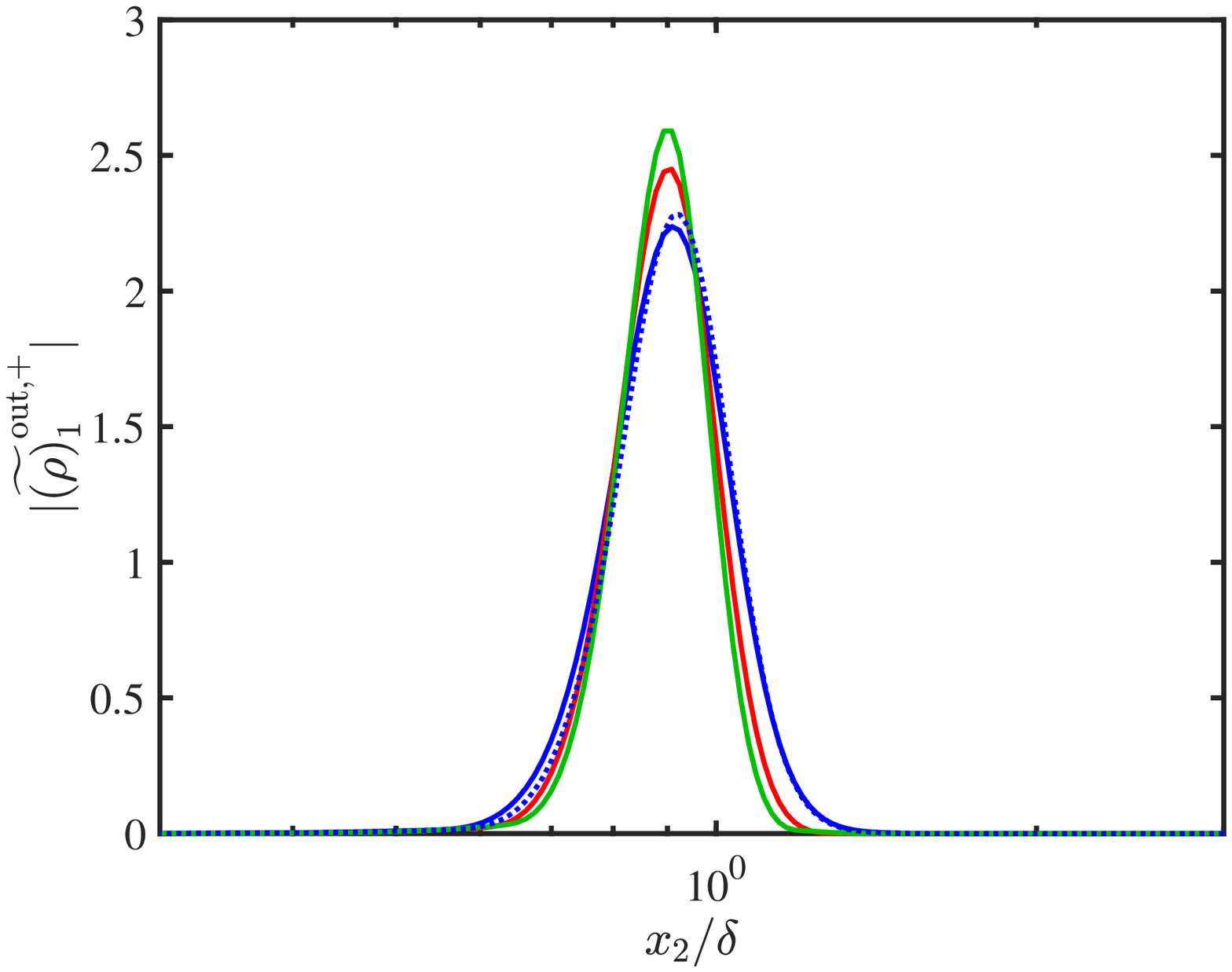}}}
\hspace{0.2cm}
\subfloat[]{{\includegraphics[width=0.47\textwidth]{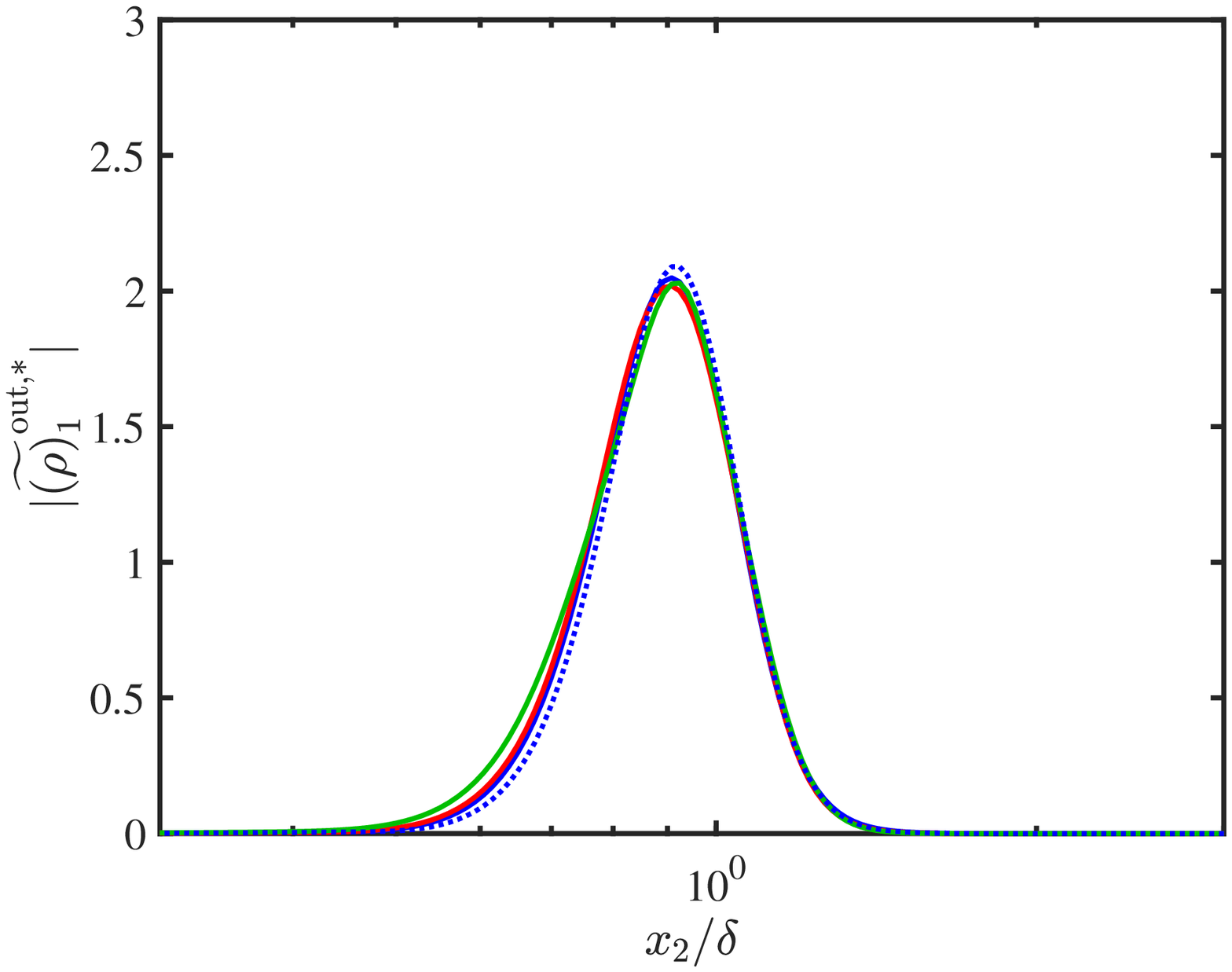}}} }
\centerline{
\subfloat[]{{\includegraphics[width=0.47\textwidth]{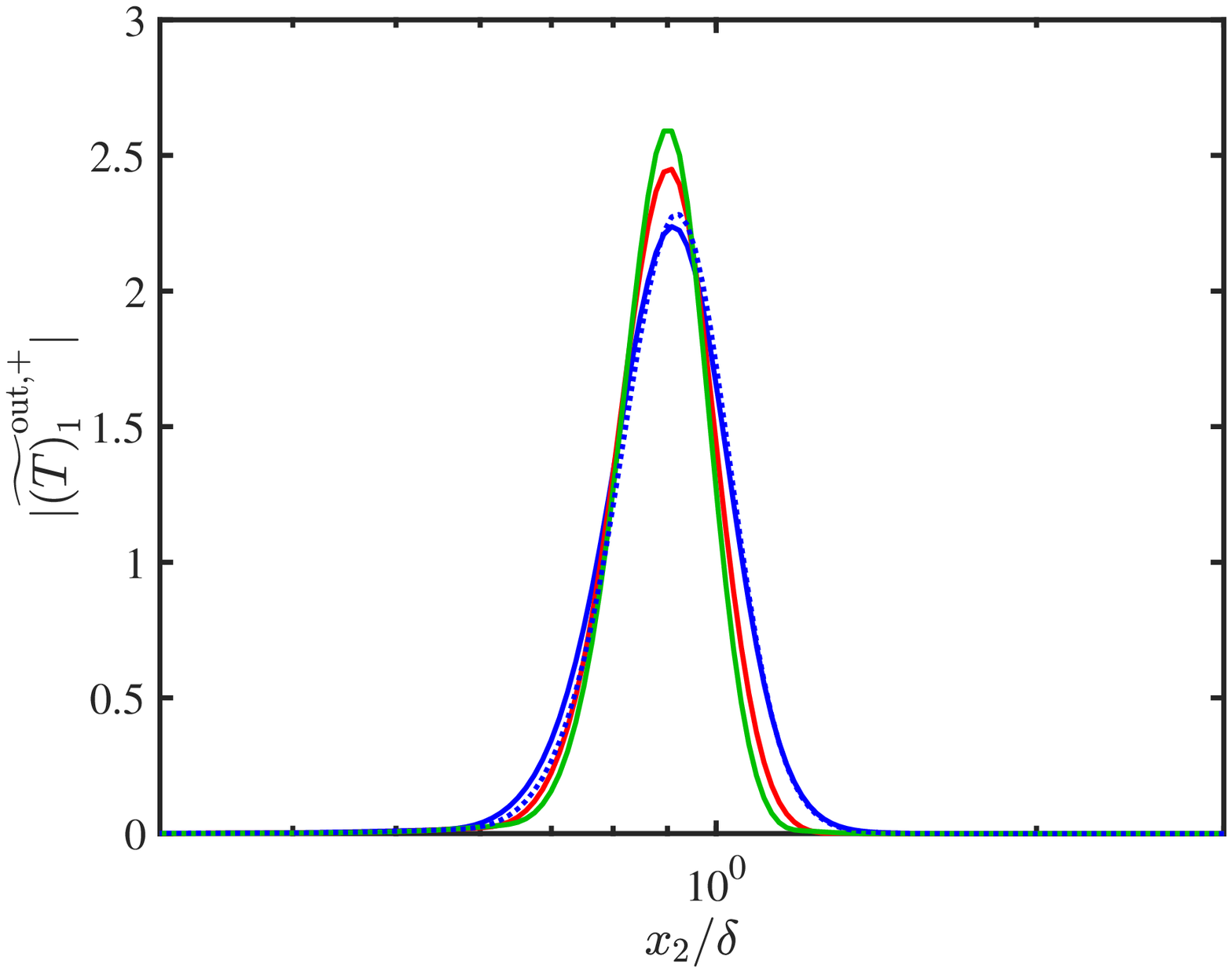}}}
\hspace{0.2cm}
\subfloat[]{{\includegraphics[width=0.47\textwidth]{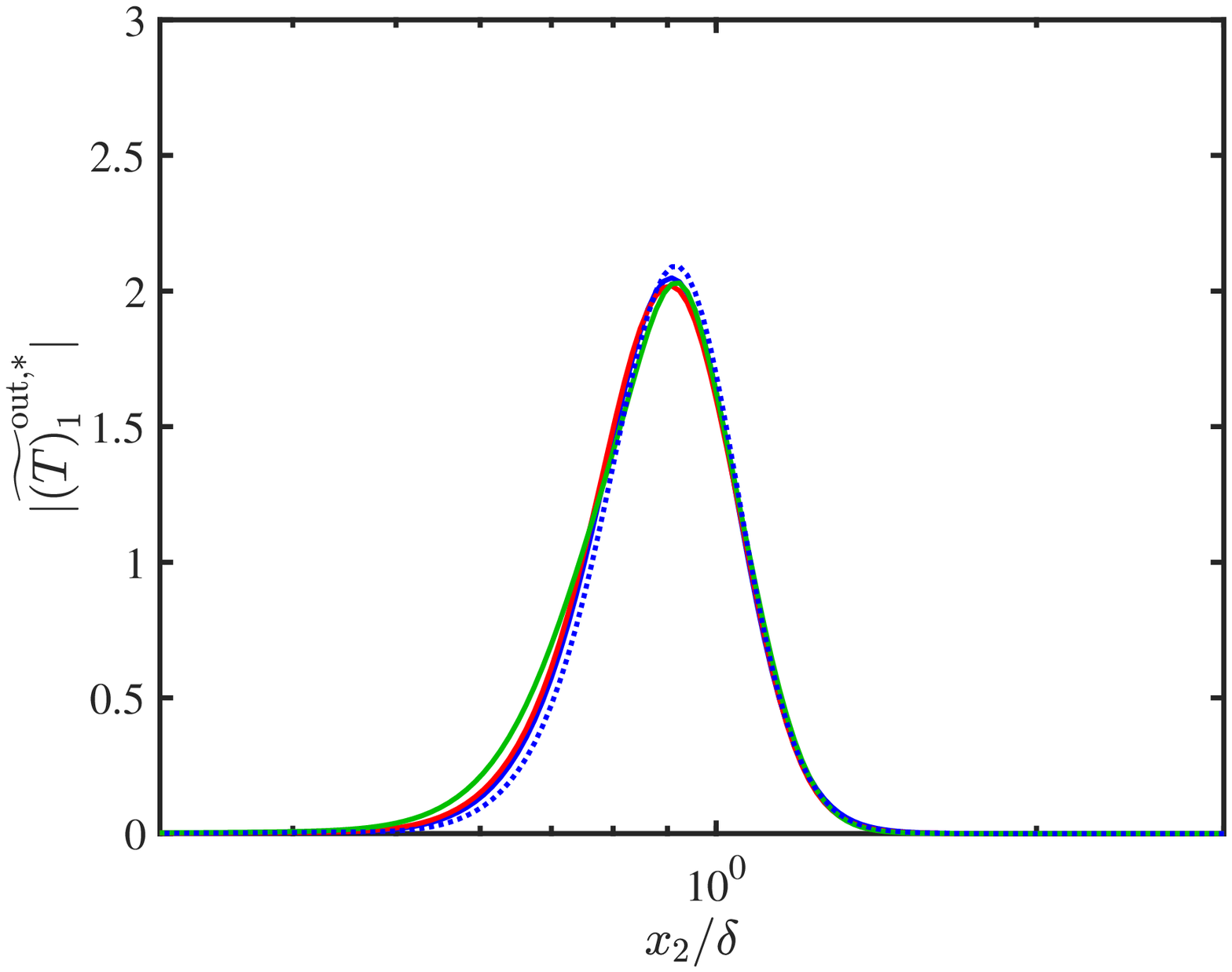}}} }
\caption{Normalised and scaled response modes in the outer region for
streamwise velocity (a) $\widetilde{(u_1)}_1^{\mathrm{out},+}$ and (b)
$\widetilde{(u_1)}_1^{\mathrm{out},*}$, density (c)
$\widetilde{(\rho)}_1^{\mathrm{out},+}$ and (d)
$\widetilde{(\rho)}_1^{\mathrm{out},*}$ and temperature (e)
$\widetilde{(T)}_1^{\mathrm{out},+}$ and (f)
$\widetilde{(T)}_1^{\mathrm{out},*}$. Lines indicate $M_\infty = 0$
(\dashline), $M_\infty = 2,\,\Rey_\tau=450$
({\color{blue}\solidline}), $M_\infty = 2,\,\Rey_\tau=900$
({\color{blue}\dotline}), $M_\infty = 3$ ({\color{red}\solidline}),
and $M_\infty = 4$ ({\color{green}\solidline}).
\label{fig:outer_scaling}} \end{figure}
Because the difference between $\Rey_\tau$ and $\Rey_\tau^*$ grows as
$x_2$ increases (see figure \ref{fig:retaus}), the difference between
the two scalings is expected to be most pronounced in the outer
region, defined as $(\bar{u}^\star_{1,\infty}-\bar{u}_1^\star)
\lesssim 6$ with $\kappa_3/\kappa_1\gtrsim \epsilon
\Rey_\tau^*/\Rey_{\tau,\min}^*$, $\epsilon^2 \approx 10$ \citep[see
Appendix \ref{sec:append:operator} and][for details]{moarref2013}.  In
this region, the $x_2$ dependent coefficients of $\mathsfbi{H}$ such
as $\bar{u}_1-c$ (and $\bar{T}$ and $\bar{\rho}$) are independent of
$\Rey_\tau$. And in the incompressible case, the balance between the
viscous dissipation term and the mean advection term in the resolvent
requires scaling of the spanwise coordinate in $\delta$ and the
streamwise coordinate with $\delta^+$ \citep{moarref2013}.  Thus, the
universal modes in the outer region for the incompressible case are
given by the wave parameters 
\begin{equation}
(q_i)_1^{\mathrm{out},+} = 
(q_i)_1\left(\kappa_1 = \kappa_1^\text{ref}\frac{\Rey_\tau^\text{ref}}{\Rey_\tau},
\kappa_3 = \kappa_3^\text{ref}, c = c^\text{ref}\right)
\end{equation}
in wall units. Using semi-local variables, we expect the wave
parameters for the compressible case to be
\begin{equation}
(q_i)_1^{\mathrm{out},*} = 
(q_i)_1\left(\kappa_1 = \kappa_1^\text{ref}\frac{\Rey_\tau^\text{ref}}{\Rey_\tau^*},
\kappa_3 = \kappa_3^\text{ref}, c = c^\text{ref}\right).
\end{equation}
Due to the scaling of the wall-normal distance in outer units as well
as the orthonormality condition for the response modes, i.e.
$((\boldsymbol{q})_1,(\boldsymbol{q})_1)_E = 1$, the mode height is
expected to scale in outer units. In figure \ref{fig:outer_scaling},
we plot the normalised and scaled principal streamwise velocity,
density and temperature response modes, with the reference parameters
$\Rey_\tau^{\text{ref}} = 445.5$, $\kappa_1^\text{ref}\delta = 1$,
$\kappa_3^\text{ref}\delta = 10$ and $c^\text{ref}/\bar{u}_{1,\infty}
= 0.98$. These reference parameters correspond to the very large scale
structures for the incompressible case. The collapse among different
Mach numbers is excellent for the semi-local scaling, and the
streamwise velocity modes for the supersonic cases are
indistinguishable from  the incompressible response modes.
{\color{black}By comparison, the modes do not collapse for various Mach
numbers when scaled in wall units, with the modes becoming narrower in
$x_2$ and the peak of the modes shifting closer to the wall with
increasing Mach number.} Although not shown, the wall-normal velocity
response modes exhibit a similar collapse when scaled with
$\Rey_\tau^*$ as opposed to $\Rey_\tau$, the proposed scaling in
\citet{sharma2017} for the incompressible case.  The collapse is not
as good for the spanwise velocity response modes (not shown), although
the mode shape is still very similar among different Mach numbers.
Furthermore, the mode shapes are comparable for the streamwise
velocity, density, and temperature. The resemblance of the density and
temperature response modes can be explained by the fact that from the
equation of state, the two are linked with the pressure fluctuations.
Moreover, the similarity between the streamwise velocity and thermal
modes reinforces the strong Reynolds analogy, which links the transport
of momentum and heat transfer, and it can be concluded that the velocity and
temperature profiles are correlated. 

\subsubsection{Inner region}
\label{sec:scaling:prm:inner}

\begin{figure}
\vspace{0.2cm}
\centerline{
\subfloat[]{{\includegraphics[width=0.47\textwidth]{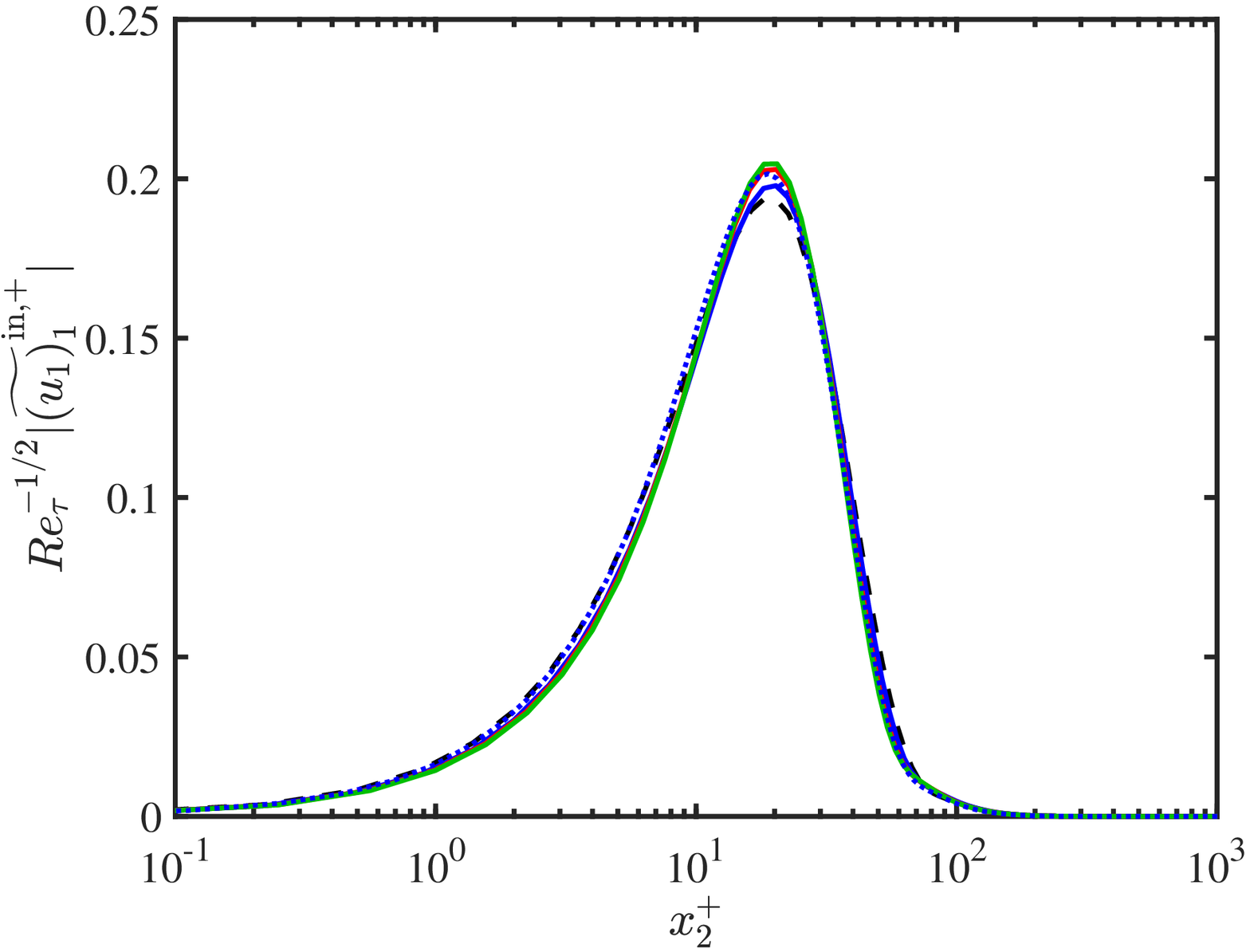}}}
\hspace{0.1cm}
\subfloat[]{{\includegraphics[width=0.47\textwidth]{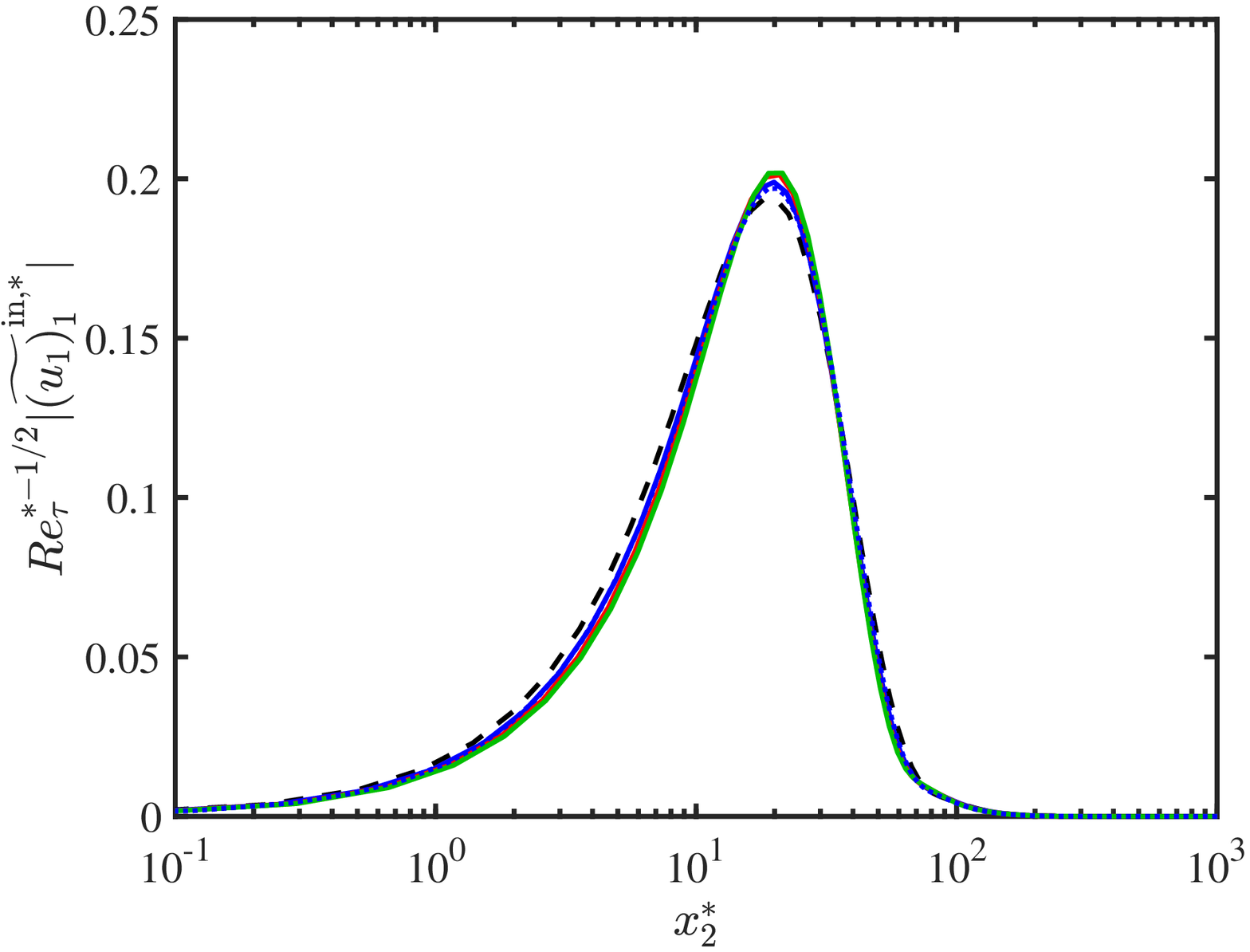}}} 
}
\centerline{
\subfloat[]{{\includegraphics[width=0.47\textwidth]{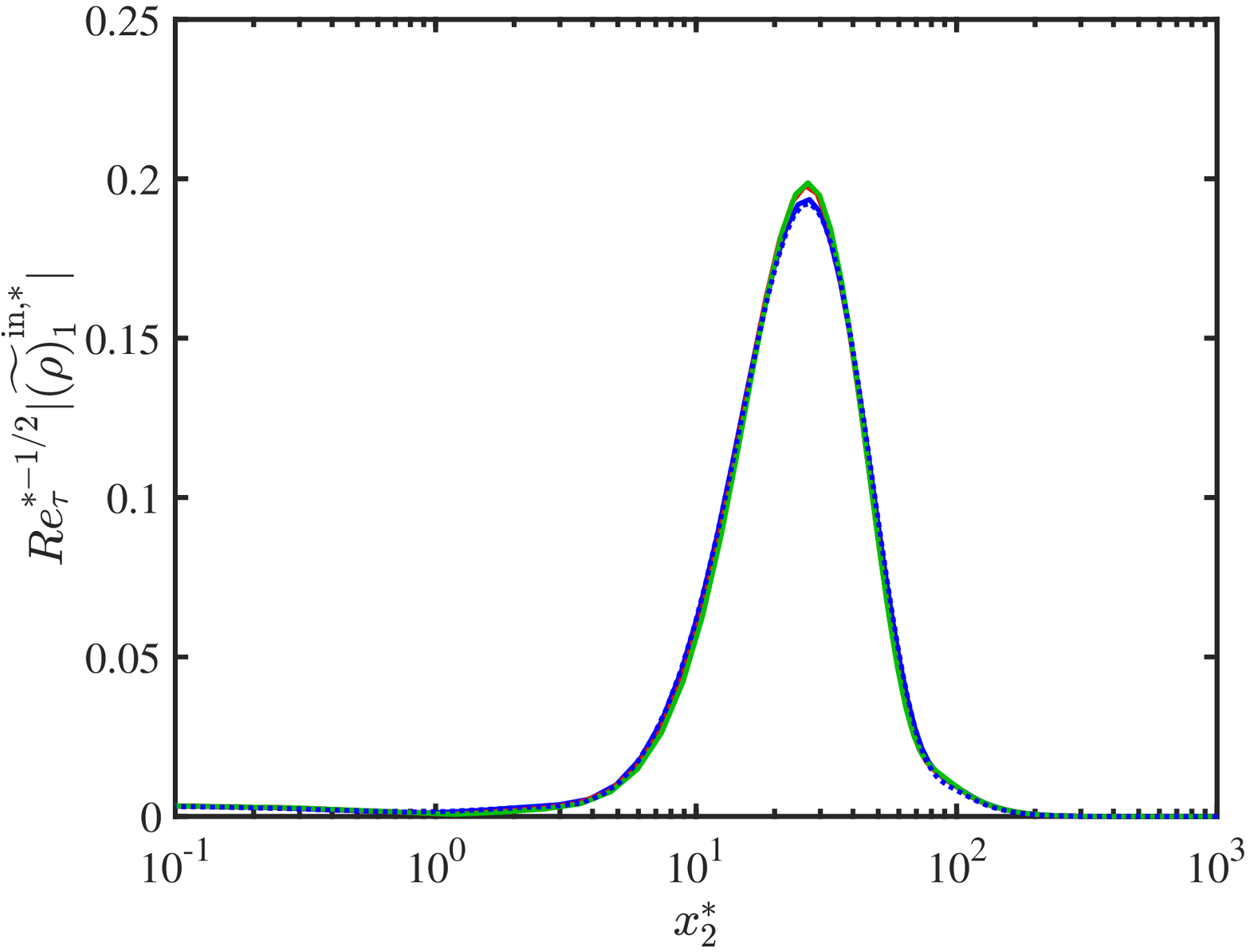}}}
\hspace{0.1cm}
\subfloat[]{{\includegraphics[width=0.47\textwidth]{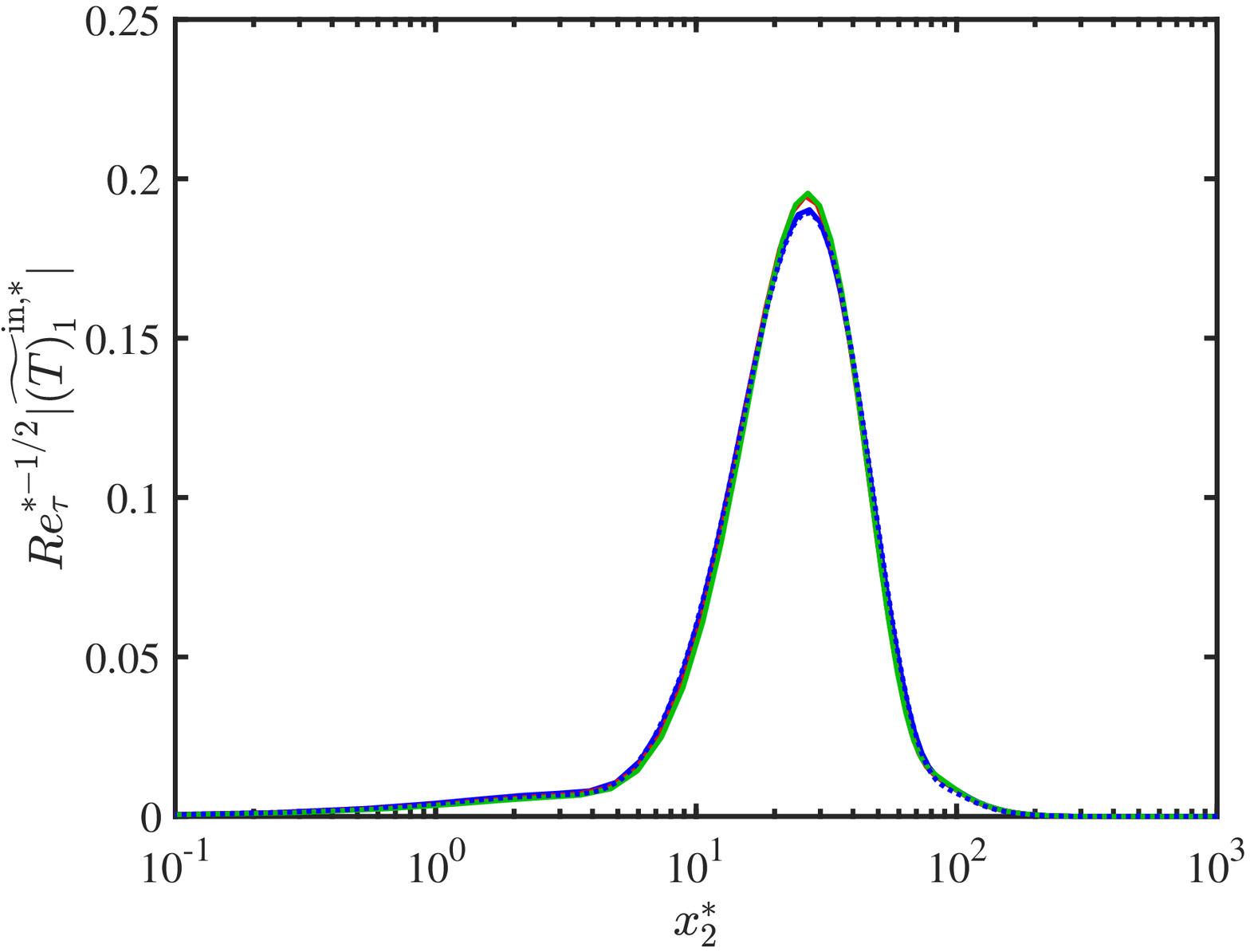}}} }
\caption{Normalised and scaled response modes in the inner region for
streamwise velocity (a) $\widetilde{(u_1)}_1^{\mathrm{in},+}$ and (b)
$\widetilde{(u_1)}_1^{\mathrm{in},*}$, (c) density,
$\widetilde{(\rho)}_1^{\mathrm{in},*}$ and (d) temperature
$\widetilde{(T)}_1^{\mathrm{in},*}$ scaled by either $\Rey_\tau^{-1/2}$
(a) or $\Rey_\tau^{*-1/2}$ (b,c,d) for reference parameters
$\Rey_\tau^\text{ref} = 445.5$, $\kappa_1^\text{ref}\delta = 1$,
$\kappa_3^\text{ref}\delta = 10$ and $x_2^{*/+,\text{ref}} = 10$. Lines
indicate $M_\infty = 0$ (\dashline), $M_\infty = 2,\,\Rey_\tau=450$
({\color{blue}\solidline}), $M_\infty = 2,\,\Rey_\tau=900$
({\color{blue}\dotline}), $M_ \infty = 3$ ({\color{red}\solidline}),
and $M_\infty = 4$ ({\color{green}\solidline}).
\label{fig:inner_scaling}}
\end{figure}
%
In the case of the inner region, the deviation of $\Rey_\tau$ and
$\Rey_\tau^*$ is not as significant, {\color{black}as can be seen in
figure \ref{fig:retaus}}, leading to a similar result for both the
wall-unit scaling and the semi-local scaling. For the universal modes
in the inner region, where the streamwise and spanwise coordinates are
given in inner units, we expect the universal wave parameters to be 
\begin{equation}
(q_i)_1^{\mathrm{in},*} = 
(q_i)_1\left(\kappa_1 = \kappa_1^\text{ref}\frac{\Rey_\tau^*}{\Rey_\tau^\text{ref}},
\kappa_3 = \kappa_3^\text{ref}\frac{\Rey_\tau^*}{\Rey_\tau^\text{ref}},
c^* = c^{\text{ref}*}\right)
\end{equation}
for $x_2^{\text{ref}*} < 30$ and $\kappa_3/\kappa_1\gtrsim \epsilon$,
and analogously defined for $(q_i)_1^{\mathrm{in},+}$.  In figure
\ref{fig:inner_scaling}, we plot the normalised and scaled principal
streamwise velocity, density and temperature response modes for
reference Reynolds number, $\Rey_\tau^\text{ref} = 445.5$ and wave
parameters $\kappa_1^\text{ref}\delta = 1$, $\kappa_3^\text{ref}\delta
= 10$ and $x_2^{\text{ref}*} = 10$ (and $x_2^{\text{ref}+} = 10$ for
comparison).  Here, the scaling of the wall-normal distance is in
semi-local (or wall) units. Thus, the orthonormality condition for the
velocity and thermodynamic modes requires that the response mode
height is expected to scale as $1/\sqrt{\Rey_\tau^*}$ (or
$1/\sqrt{\Rey_\tau}$). Again, the collapse of the response modes among
different Mach numbers is good for both cases, and the streamwise
velocity response modes also collapse with the incompressible case.
The wall-normal and spanwise velocity response modes also collapse for
the various Mach numbers to the incompressible case but are not shown
for brevity.  The mode shapes for density and temperature are almost
identical for the same reasons discussed above. The temperature and
density modes scaled in wall units are not shown, but they are almost
identical to the ones given in semi-local units.

\subsubsection{Logarithmic region}
\label{sec:scaling:prm:log}
%
\begin{figure}
\vspace{0.2cm}
\centerline{
\subfloat[]{{\includegraphics[width=0.47\textwidth]{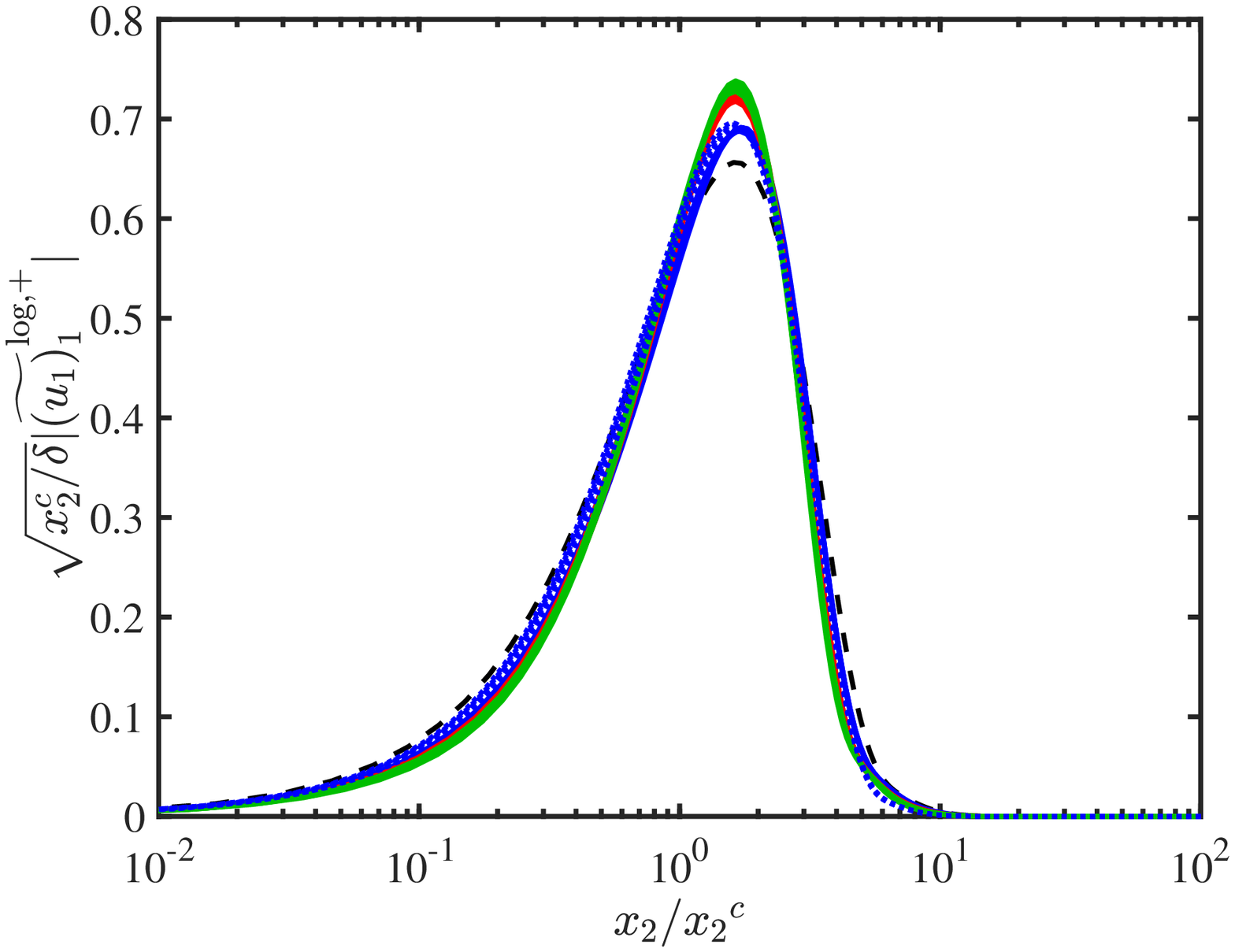}}}
\hspace{0.2cm}
\subfloat[]{{\includegraphics[width=0.47\textwidth]{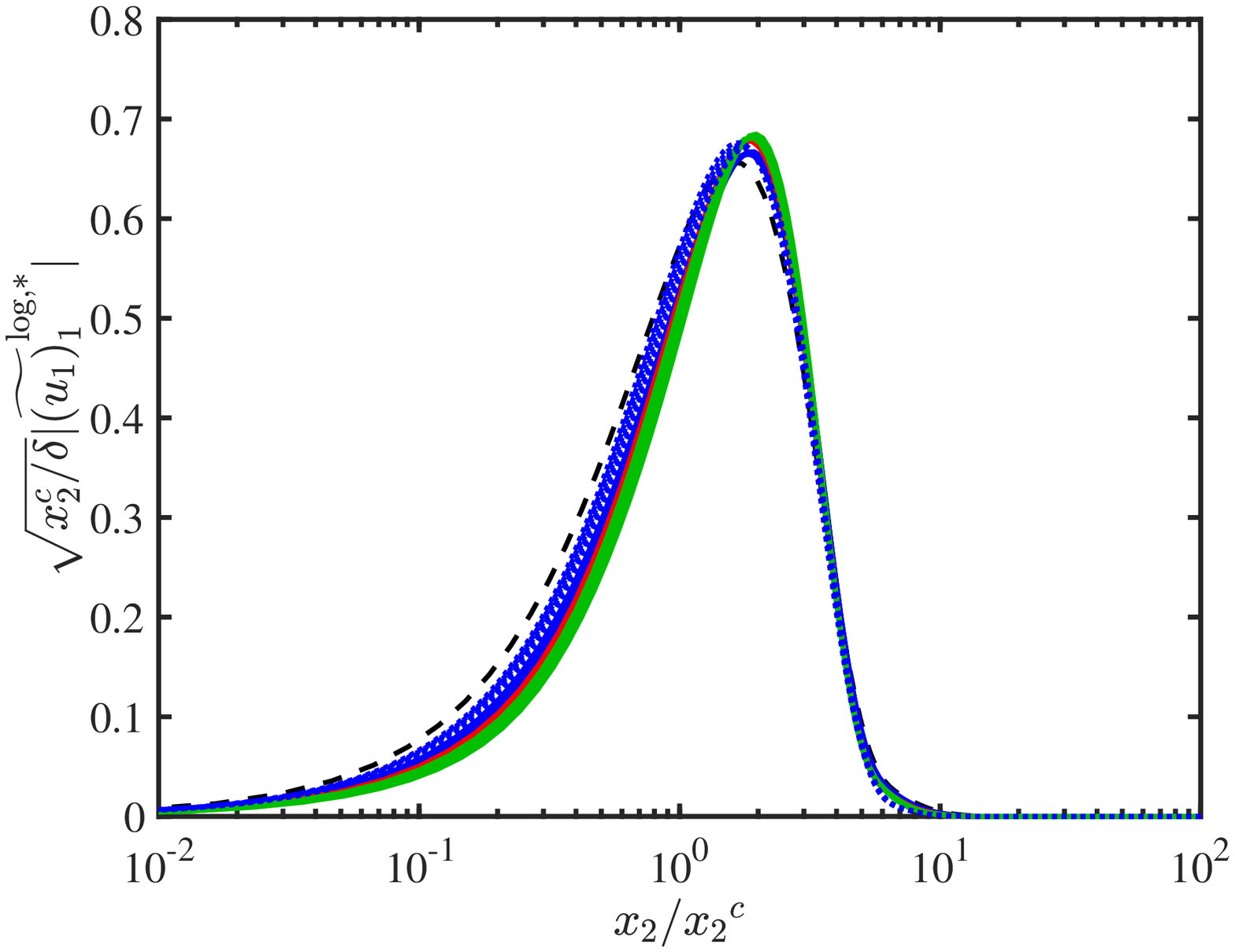}}} }
\centerline{
\subfloat[]{{\includegraphics[width=0.47\textwidth]{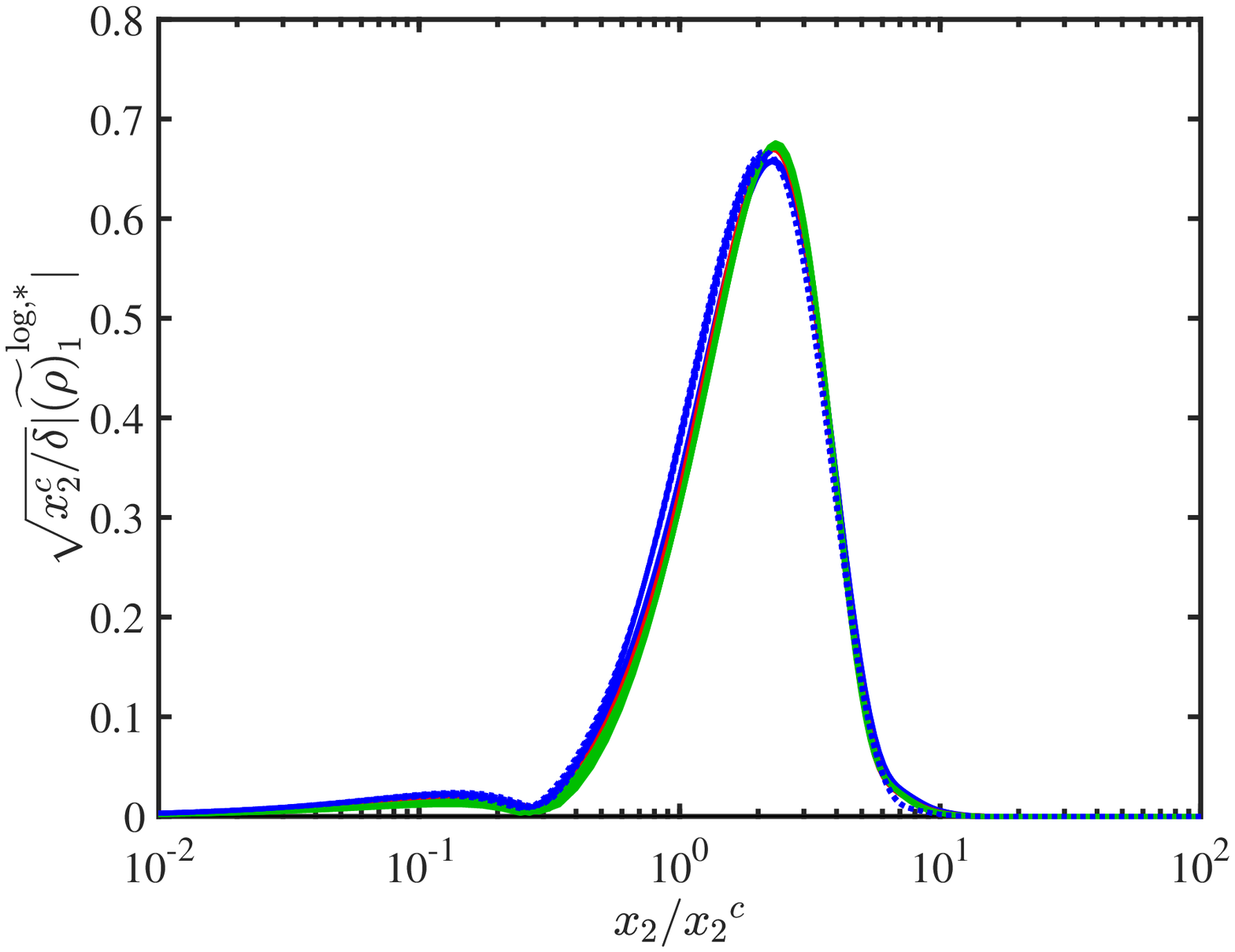}}}
\hspace{0.2cm}
\subfloat[]{{\includegraphics[width=0.47\textwidth]{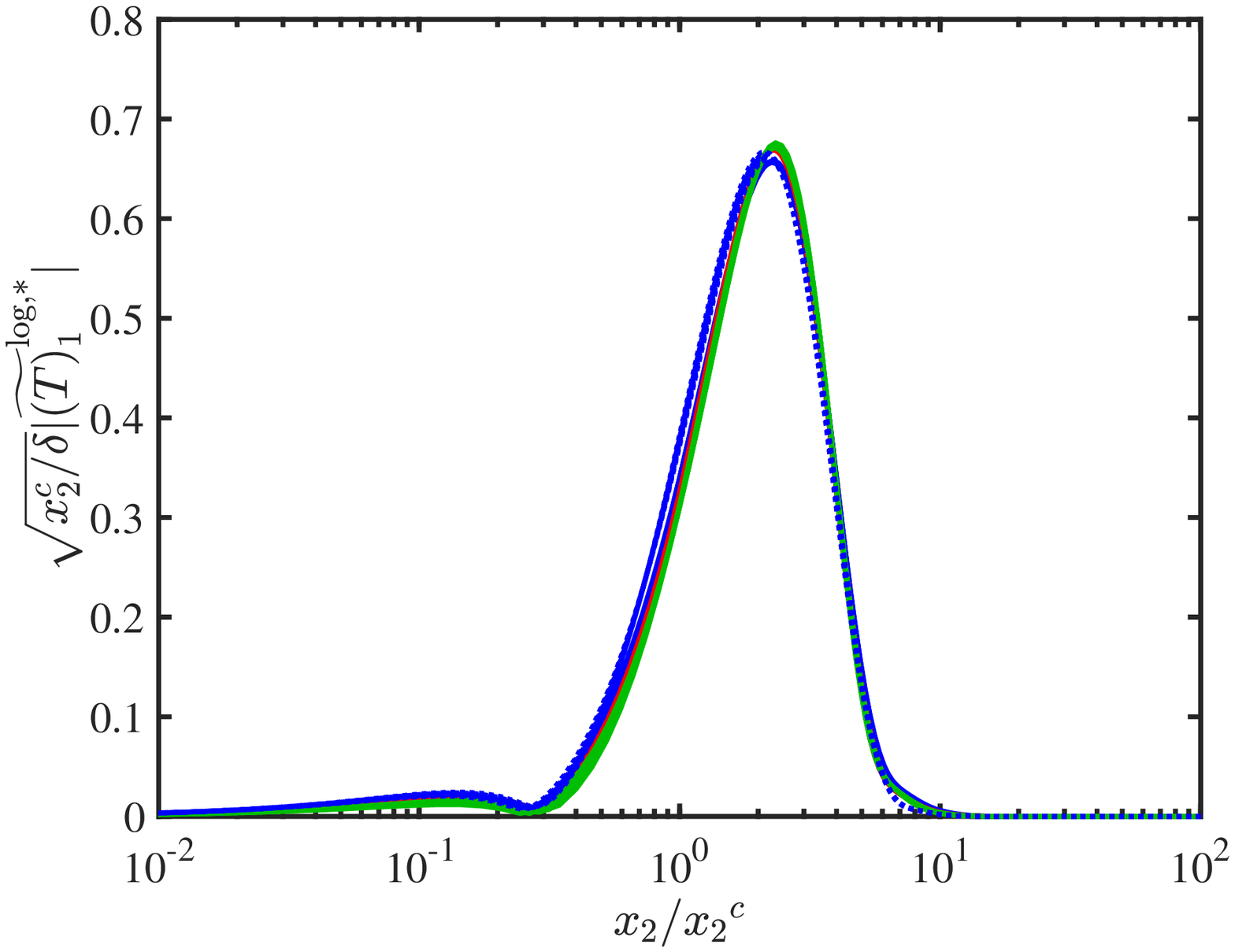}}} }
\caption{Normalised and scaled response modes in the logarithmic
region for streamwise velocity (a)
$\widetilde{(u_1)}_1^{\mathrm{log},+}$ and (b)
$\widetilde{(u_1)}_1^{\mathrm{log},*}$, (c) density
$\widetilde{(\rho)}_1^{\mathrm{log},*}$ and (d) temperature
$\widetilde{(T)}_1^{\mathrm{log},*}$ scaled by $\sqrt{x_2^c}$. Lines
indicate $M_\infty = 0$ (\dashline), $M_\infty = 2,\,\Rey_\tau=450$
({\color{blue}\solidline}), $M_\infty = 2,\,\Rey_\tau=900$
({\color{blue}\dotline}), $M_\infty = 3$ ({\color{red}\solidline}),
and $M_\infty = 4$ ({\color{green}\solidline}).
\label{fig:ss_scaling}}
\end{figure}
%
Finally, while the $\Rey_\tau$ for the cases under consideration is
too small for a clearly defined logarithmic region, we consider the
self-similar response modes. The wave parameters in this region are
given by
\begin{equation}
(q_i)_1^{\mathrm{log},*} = 
(q_i)_1\left(\kappa_1 = \kappa_1^\text{ref}
\frac{x_2^\text{ref}x_2^\text{ref*}}{x_2^cx_2^{c*}},
\kappa_3 = \kappa_3^\text{ref}\frac{x_2^\text{ref}}{x_2^c}, c\right)
\end{equation}
for wave speeds $c$ in the logarithmic region
($30/\Rey_\tau^*<x_2/\delta<0.15$) \citep{tennekes1972}, where
$x_2^\text{ref}$ denotes the critical layer for $c^\text{ref}$. The
wave parameters in wall units are given analogously for
$(q_i)_1^{\mathrm{log},+}$. In figure \ref{fig:ss_scaling}, we plot
the normalised and scaled principal streamwise velocity, density and
temperature response modes for reference Reynolds number,
$\Rey_\tau^\text{ref} = 445.5$, and reference wave parameters
$\kappa_1^\text{ref}\delta = 1$, $\kappa_3^\text{ref}\delta = 10$, and
$c^\text{ref}/\bar{u}_{1,\infty}=0.5$.  The orthonormality condition
for the velocity and thermodynamic quantities give the response mode
height scaling of $\sqrt{x_2^c/\delta}$.  Similar to the outer region,
the semi-local scaling gives a better collapse among the supersonic
response modes for various Mach numbers. While the agreement with the
incompressible case is not perfect, some improvement is made from the
use of the semi-local scale. As mentioned earlier, the $\Rey_\tau$ for
cases under consideration is too low for an actual logarithmic layer
and a better collapse is expected for higher Reynolds numbers.
Additionally, as in the case of the inner and outer region, the
temperature and density modes are identical and are similar to the
streamwise velocity mode shape as well.

\subsection{Scaling of the principal singular values}
\label{sec:scaling:psv}

\begin{figure}
\vspace{0.2cm}
\centerline{
\subfloat[]{{\includegraphics[width=0.47\textwidth]{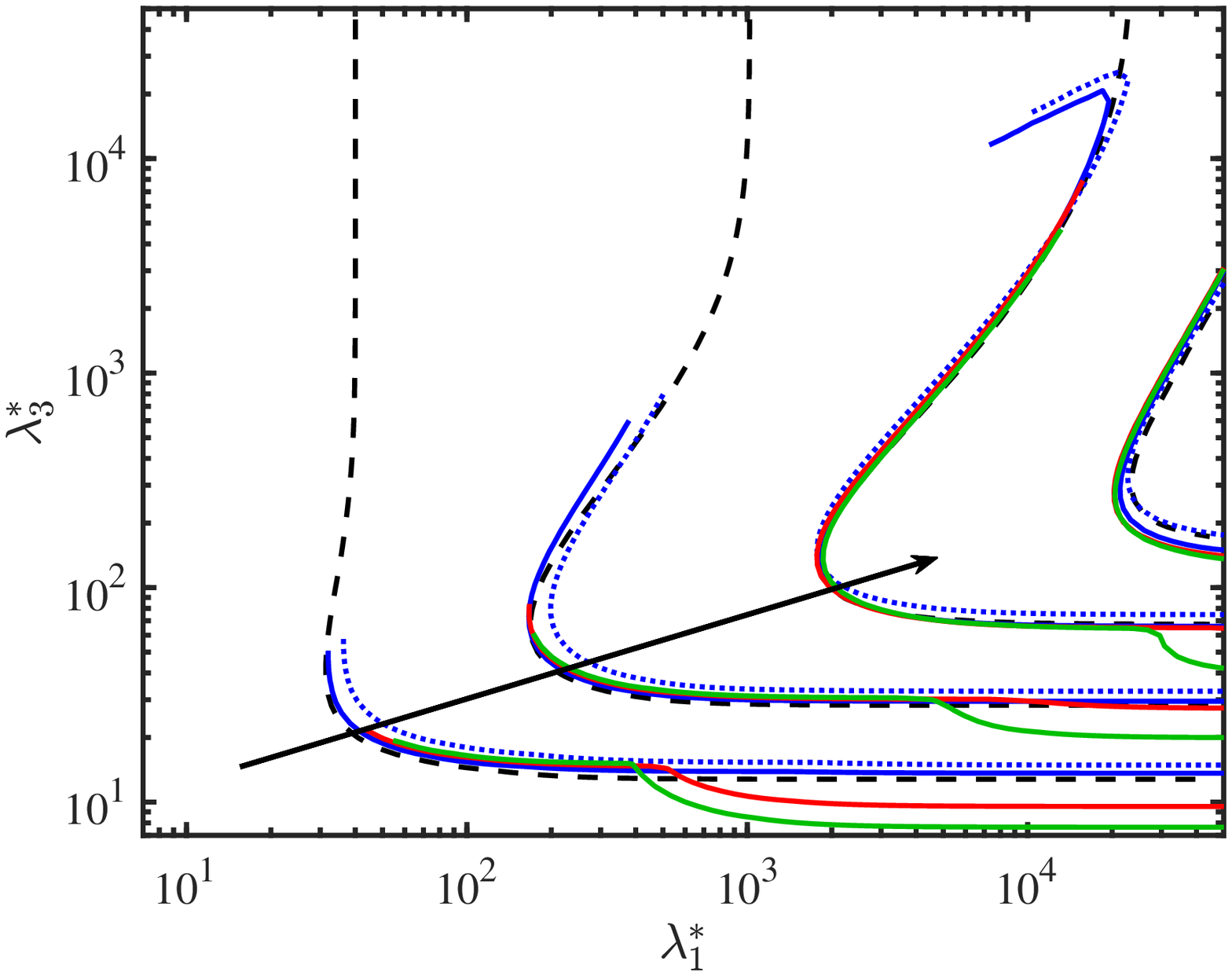}}}
\hspace{0.2cm}
\subfloat[]{{\includegraphics[width=0.47\textwidth]{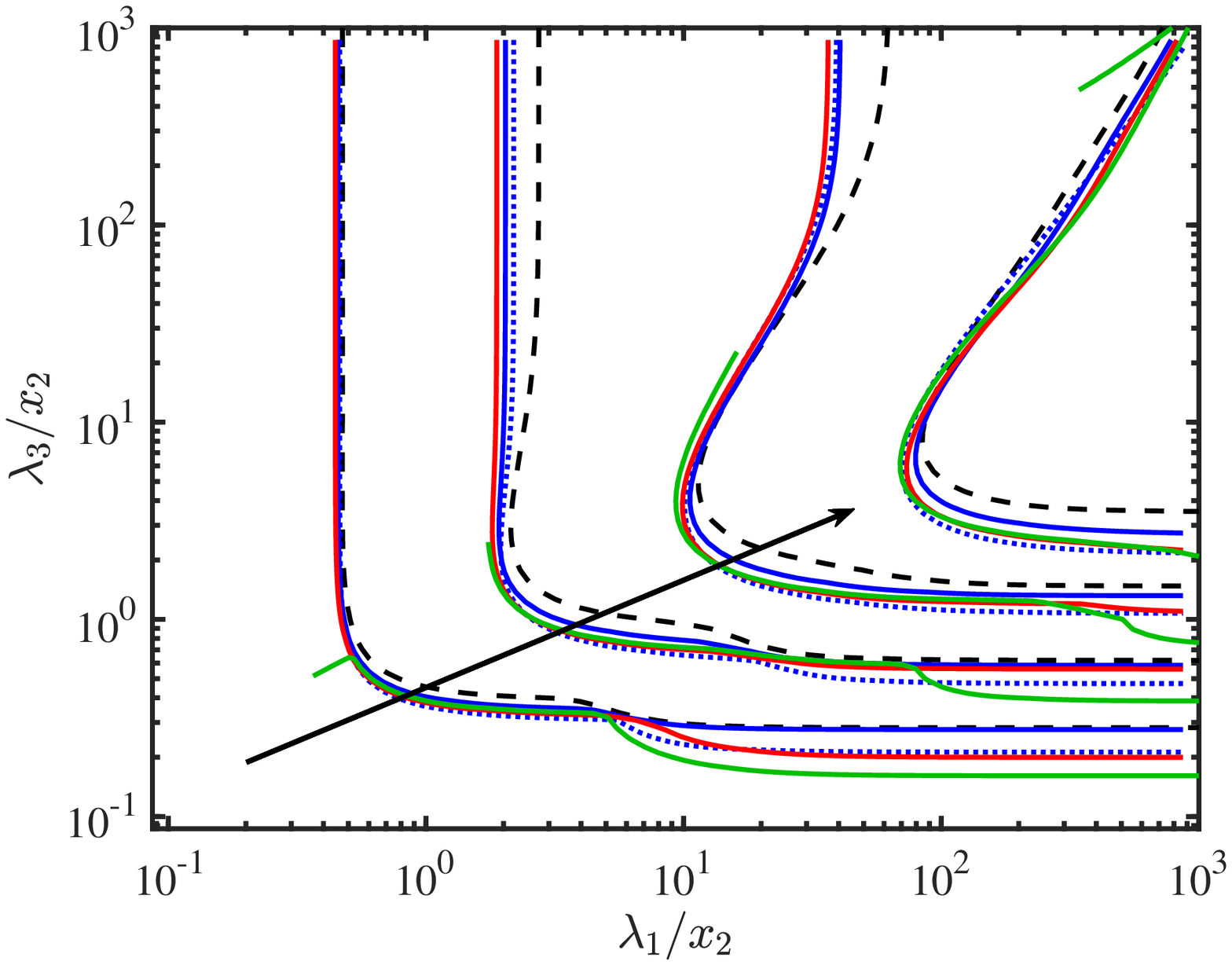}}} }
\centerline{
\subfloat[]{{\includegraphics[width=0.47\textwidth]{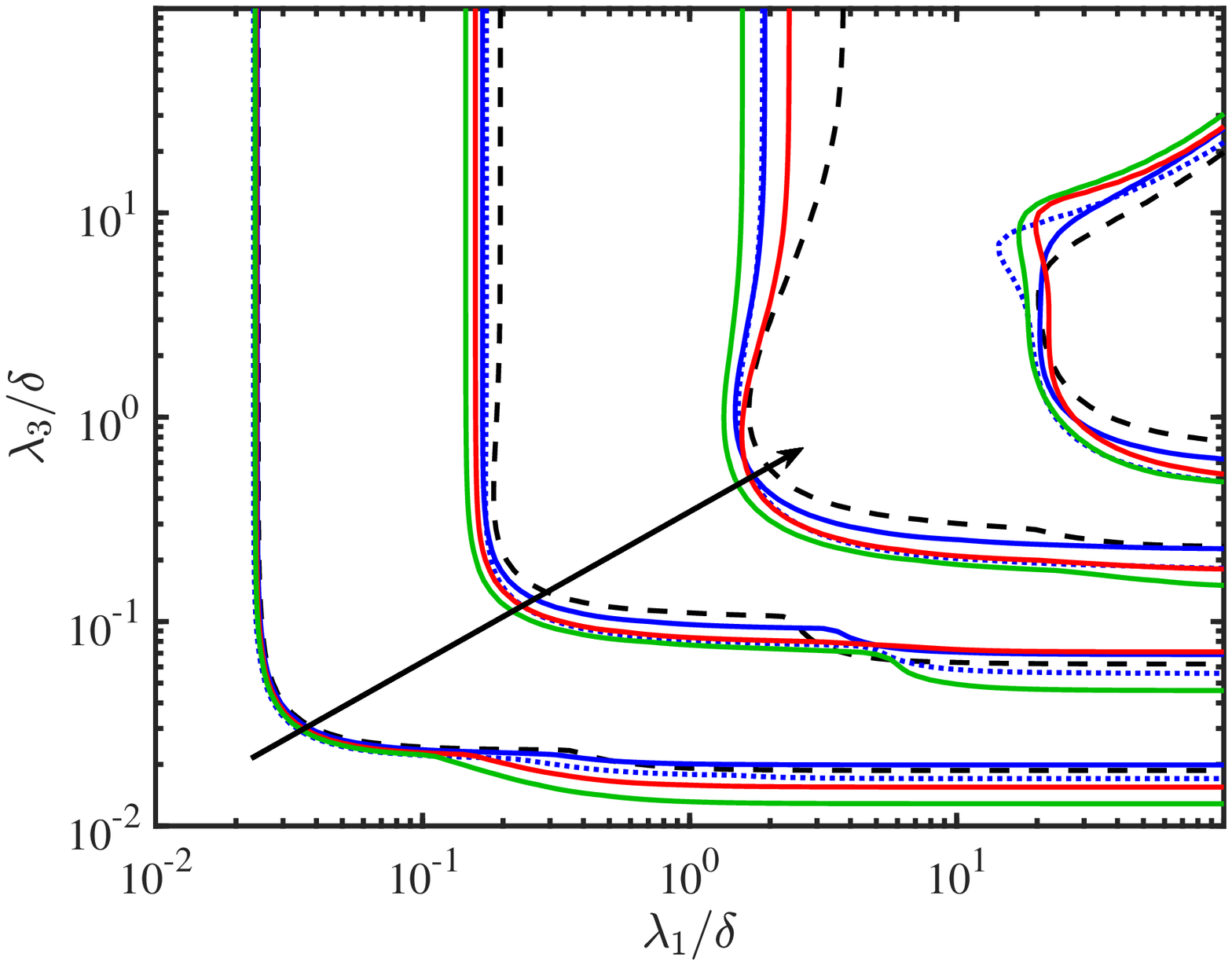}}}
}
%
\caption{Principal singular value in the relatively subsonic region
for (a) $c = \bar{u}_{1,\infty}(x_2^* = 10)$ for the inner layer, (b)
$c/\bar{u}_{1,\infty} = 0.7$ for the logarithmic layer, and (c)
$c/\bar{u}_{1,\infty} = 0.88$ for the outer layer.  Contour lines are
(a) $(10^2, 10^4,10^6,10^8)/\sqrt{\Rey_\tau^*\Rey_\tau^\star}$, (b)
$(10^{-3}, 10^{-1}, 10^1, 10^3)\times x_2^{*}x_2^{\star}x_2/\delta$,
and (c) $(10^{-7},10^{-4},10^{-1},10^2)\times
\Rey_\tau^\star\Rey_\tau^*$.  Lines indicate $M_\infty = 0$
(\dashline), $M_\infty = 2,\,\Rey_\tau=450$
({\color{blue}\solidline}), $M_\infty = 2,\,\Rey_\tau=900$
({\color{blue}\dotline}), $M_ \infty = 3$ ({\color{red}\solidline}),
and $M_\infty = 4$ ({\color{green}\solidline}). Arrows indicate
direction of increasing $\sigma_1$.
\label{fig:sv_scaling}} \end{figure}

In addition to the universality and self-similarity of the response
modes, the scaling of the principal singular values is also
investigated. The expected scaling of the singular values for the
incompressible case is given by $1/\Rey_\tau$ in the inner region,
$x_2^{+2}x_2/\delta$ in the logarithmic region and $\Rey_\tau^2$ in
the outer region \citep{moarref2013,sharma2017}. The scaling of the
principal singular values can be found by  performing a scaling
analysis on the resolvent operator $\mathsfbi{H}$ by assessing the
Reynolds number dependency of the terms in the linearised operator
$\mathsfbi{L}$ (see Appendix \ref{sec:append:operator} for details).
The elements of the resolvent operator matrix, and thus the  leading
singular value, can be shown to follow
$1/\sqrt{\Rey_\tau^*\Rey_\tau^\star}$ in the inner region, $x_2^\star
x_2^*x_2/\delta$ in the logarithmic region, and
$\Rey_\tau^*\Rey_\tau^\star$ in the outer region, where
$\Rey_\tau^\star = \bar{\rho}u_\tau\delta/\bar{\mu}$ and $x_2^\star =
\Rey_\tau^\star x_2/\delta$. This is due to the presence of the factor
$\bar{T}\bar{\mu}/\Rey$ (or equivalently $\bar{\mu}/(\Rey\bar{\rho})$)
in the governing equations combined with the semi-local scaling of 
$\kappa_1$, $\kappa_3$ and $x_2$. The proposed scaling for the
principal singular values is given in figure \ref{fig:sv_scaling}.
Note that this scaling will converge to the scaling given in
\citet{moarref2013} in the incompressible limit.

\subsection{Scaling of the kinetic and thermodynamic energy ratio}
\label{sec:scaling:ke_te}

Due to the orthonormality constraint of the resolvent modes, the
comparison between the compressible and incompressible resolvent modes
in the previous section was for normalised response modes
$\widetilde{(q_i)}_1$. However, it is equally important to assess the
distribution of energy among the kinetic and thermodynamic variables
for the supersonic cases. 

\begin{figure}
\vspace{0.25cm}
\centerline{
\subfloat[]{\includegraphics[width=0.47\textwidth]{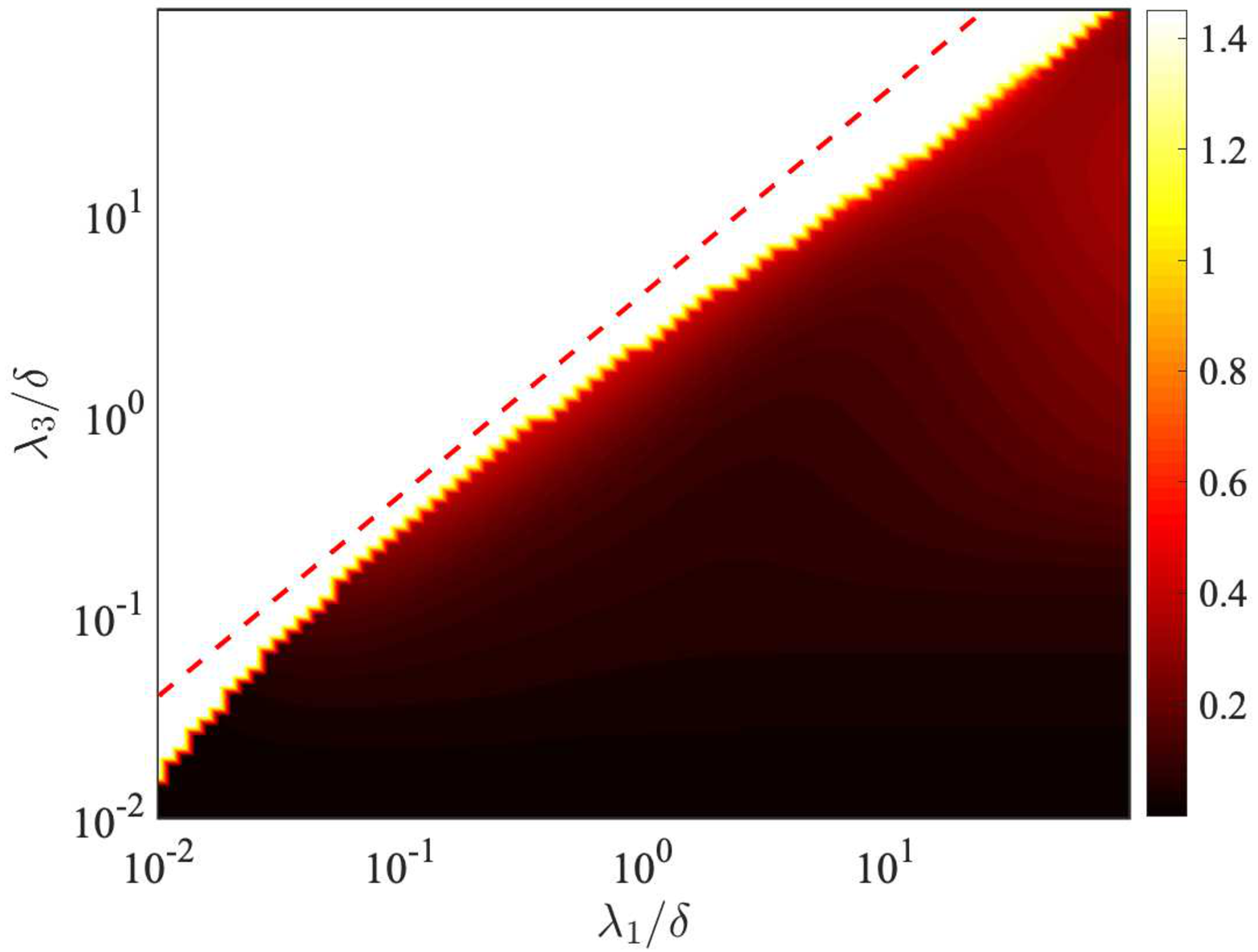}}
\hspace{0.2cm}
\subfloat[]{\includegraphics[width=0.47\textwidth]{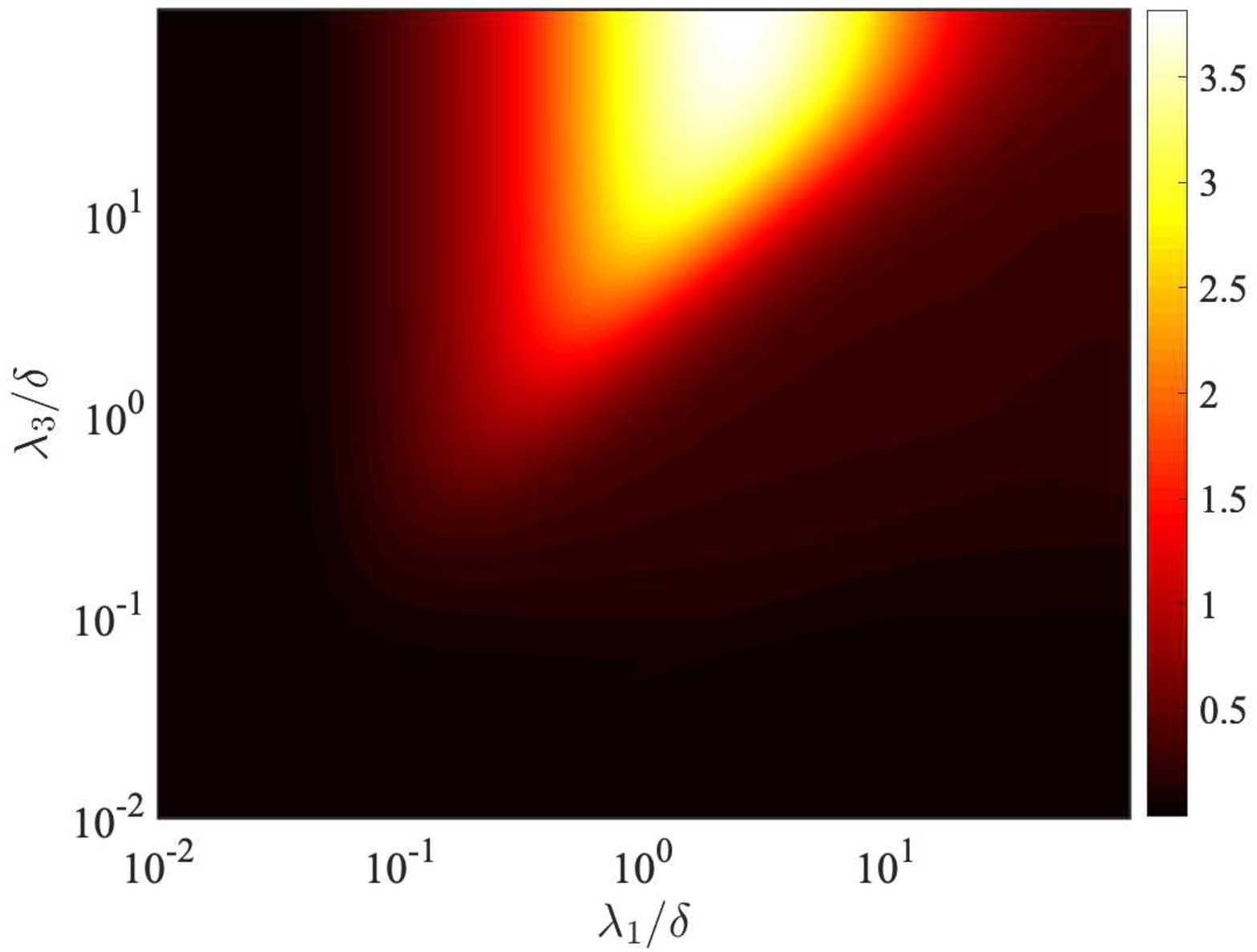}}
}
\centerline{
\subfloat[]{\includegraphics[width=0.47\textwidth]{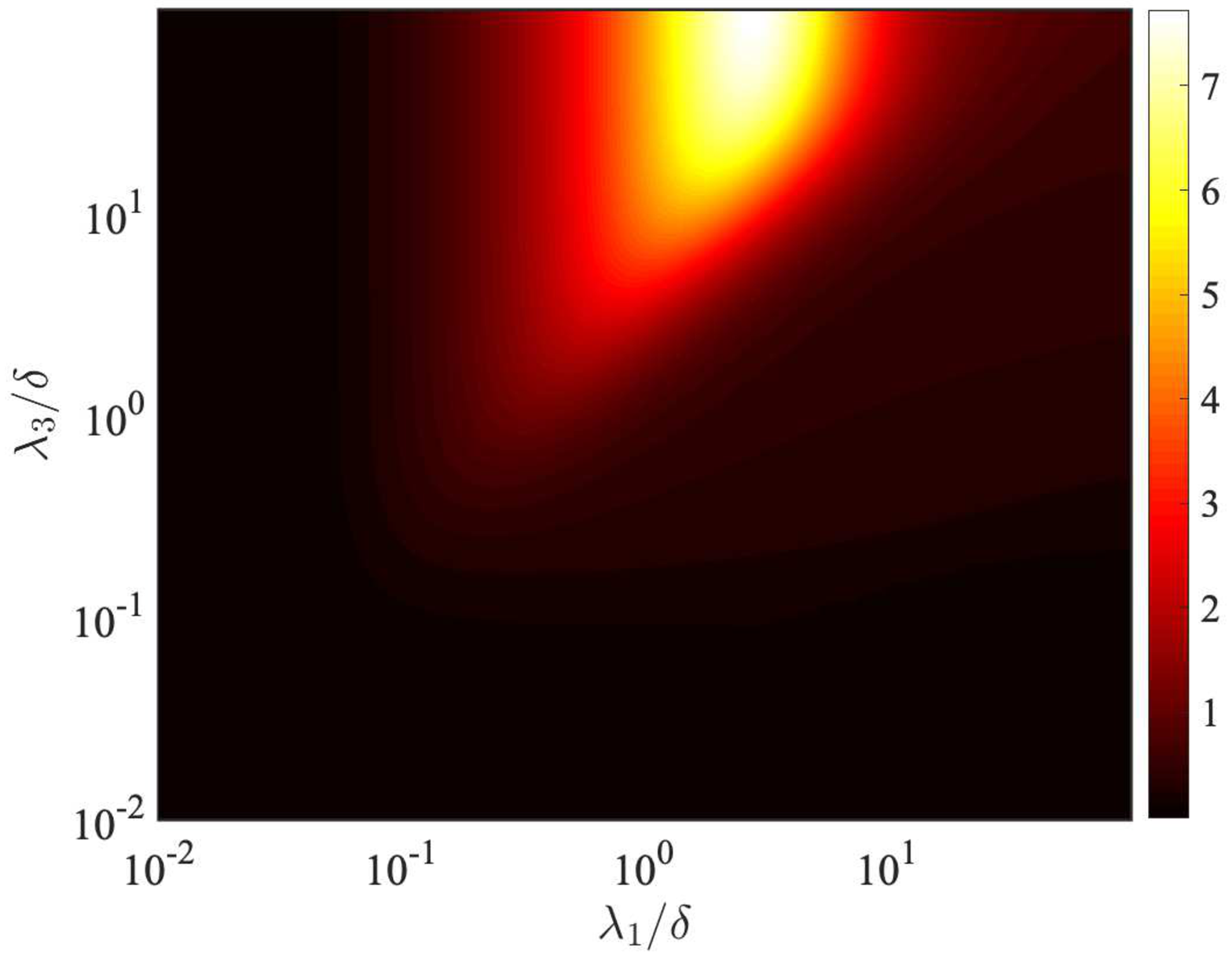}}
\hspace{0.2cm}
\subfloat[]{\includegraphics[width=0.47\textwidth]{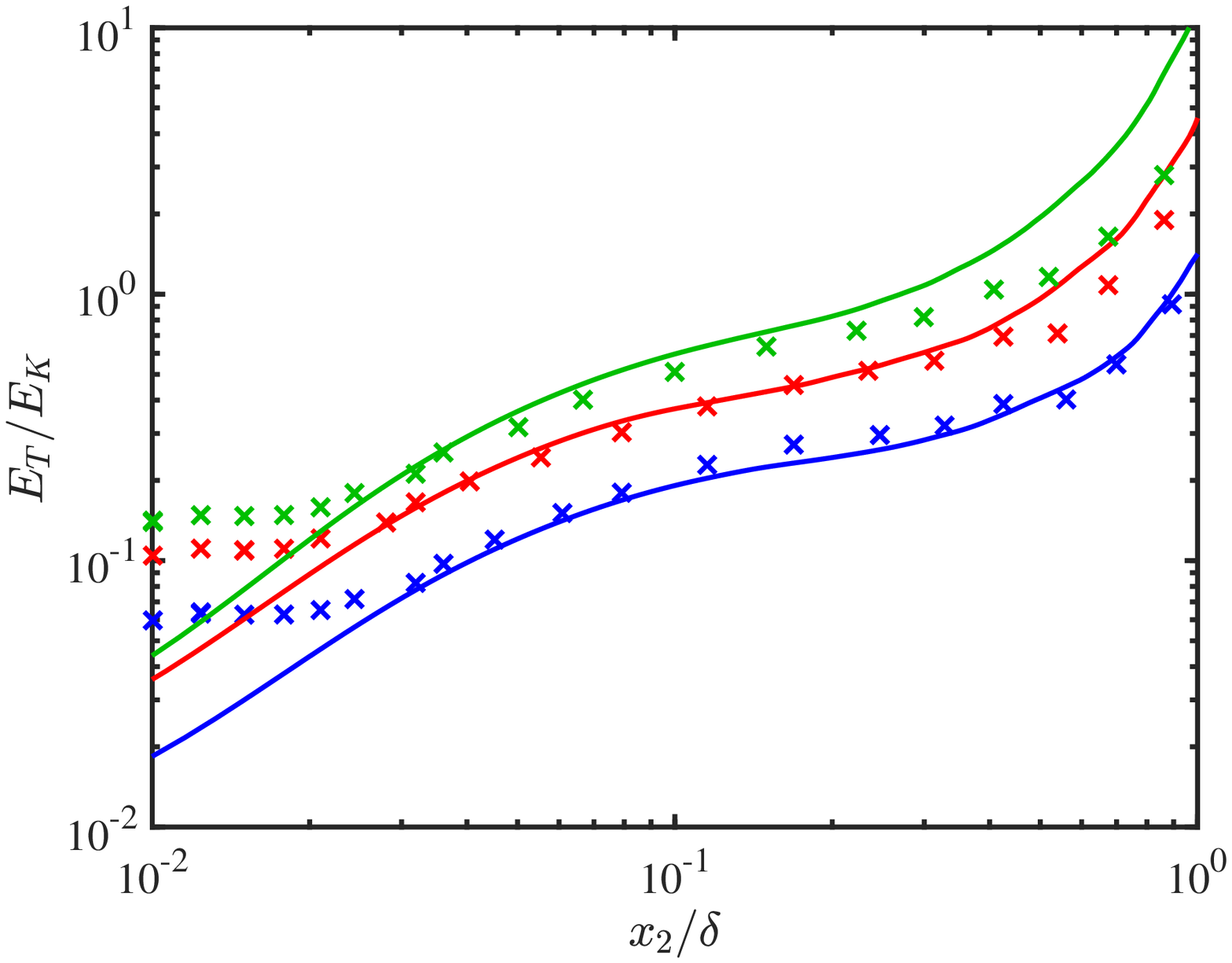}}
}
\caption{The spectra of the ratio of turbulent thermodynamic and
kinetic energy, $E_T/E_K$ computed from the principal resolvent modes
at (a) $x_2^s = 15$, (b) $x_2/\delta = 0.2$ and (c) $x_2/\delta = 0.5$
for the $M_\infty = 2$, $\Rey_\tau = 450$ case. The relative sonic
line $\bar{c}$ ({\color{red}\dashline}) is shown for reference. (d)
The values of $E_T/E_K$ at the most energetic wave parameters as
defined in \eqref{eq:ke_te_energetic} for the principal resolvent
modes ($\times$) and DNS (\solidline) for $M_\infty = 2$, $\Rey_\tau =
450$ ({\color{blue}blue}), $M_\infty=3$ ({\color{red}red}), and
$M_\infty=4$ ({\color{green}green}).  \label{fig:ke_te}}
\end{figure}
In figure \ref{fig:ke_te}(a), (b) and (c), we plot the spectra of the
ratio of thermodynamic energy and kinetic energy $E_T/E_K$ for the
resolvent modes for $M_\infty = 2$ and $\Rey_\tau=450$ for two
different wall-normal heights. For the incompressible case, this ratio
would be uniformly zero. Close to the wall, where a relatively
supersonic region is present, the thermodynamic energy clearly
dominates in the relatively supersonic region as expected. Further
away from the wall at $x_2/\delta = 0.2$ and $x_2/\delta = 0.5$, the
thermodynamic energy still dominates in a smaller region with
$\lambda_3 > \lambda_1$ with the strongest amplification centred
around $\kappa_1\delta \approx 0.5$. This may be explained by the
observation from \citet{pirozzoli2011} that in this region, the
thermal streaks are more elongated in the spanwise direction compared
to the velocity streaks. It is also consistent with the observation
that large-scale pressure-carrying eddies or wavepackets, which are
correlated with thermodynamic fluctuations, are more coherent in the
spanwise direction for both incompressible \citep{sillero2014} and
hypersonic \citep{duan2016} boundary layers. 

We also plot the ratio of turbulent kinetic energy to the sum of the
mean-square density and temperature fluctuations obtained from DNS
\citep{bernardini2011,pirozzoli2011} as a function of $x_2$ in figure
\ref{fig:ke_te}(d). In particular, we plot
\begin{equation}
\left(\frac{E_T}{E_K}\right)^\text{DNS} = 
\gamma M_\infty^2
\frac{\bar{\rho}u_{i,rms}u_{i,rms}/(\bar{u}_j\bar{u}_j)}
{\rho_{rms}^2/\bar{\rho}^2 + T_{rms}^2/\bar{T}^2},
\end{equation}
where $rms$ denotes the root-mean-squared fluctuations from DNS. For
all wall-normal locations, the ratio increases with Mach number.
Moreover, the ratio increases as a function of $x_2$ for a fixed Mach
number. In order to compare the results from the resolvent modes to
DNS, we define the energy ratio of the most energetic mode as 
\begin{equation}
\left(\frac{E_T}{E_K}\right)^\text{res} =
\frac{E_T}{E_K}\left(\underset{\kappa_1,\kappa_3}{\mathrm{arg}\hspace{0.1cm}\mathrm{max}}\, \Phi_{u_1u_1}(\kappa_1,\kappa_3) \right), 
\label{eq:ke_te_energetic}
\end{equation}
where $\Phi_{u_1u_1}$ is the premultiplied streamwise energy spectra
for the channel flow at $\Rey_\tau = 550$ obtained \emph{a priori}
from \citet{delalamo2004}.  The $(E_T/E_K)^\text{res}$ given by these
wave parameters is plotted in figure \ref{fig:ke_te}(d). The agreement
between the ratio of kinetic and thermodynamic energy given by the
most energetic principal response mode of the resolvent analysis and
the DNS is excellent in the logarithmic region. While DNS data are used
to identify the wave parameters of the most energetic mode, note that
the spectra are for incompressible channel flow rather than the
compressible turbulent boundary layer. The discrepancy in the outer
region, especially for the higher Mach numbers, may be due to the fact
that the difference in the spectra of the boundary layer and channel
flow,  which was used to choose the wave parameters for the most
energetic modes, becomes more pronounced in this region. This may
possibly be mitigated by  utilising energy spectra from boundary layer
simulations. In the inner region, the estimated energy ratio plateaus,
deviating from the DNS profile. This could be due to the increased
contributions from relatively supersonic region, which is more
prevalent in the near-wall region (see figures \ref{fig:sv_energy} and
\ref{fig:sv_energy_3D}).  Additionally, it is shown in
\citet{lehew2011} that the energetic contribution of structures with
convection velocities less than $10u_\tau$ is negligible in real
turbulent flows, which corresponds to the region where the mismatch is
pronounced. 

\section{Conclusions}\label{sec:conclusions}



We have applied resolvent analysis for the compressible
Navier-Stokes equations to a supersonic zero-pressure-gradient
turbulent boundary layer. From the low-rank approximation formulated
for individual wall-parallel wavenumbers and frequencies, we have
identified two distinct regions in the wave parameter space: the
relatively supersonic region and the relatively subsonic region.   

In the relatively supersonic region, marked by relative Mach number
greater than unity, we show that the resolvent modes are centred
around the relative sonic line rather than the critical layer and that
the majority of the energy is carried by the thermodynamic
fluctuations. These response modes are consistent with acoustic Mach
waves propagating towards the freestream and the idea of eddy
shocklets, where the instantaneous supersonic events cause local
shock-like structures in the boundary layer. Additionally, the modes
in this region are also shown to follow a modal amplification
mechanism.  The range of wave parameters corresponding to the
relatively supersonic region, where the compressibility effects are
concentrated, grows with Mach number, which might be an indicator of
why Morkovin's hypothesis fails for high Mach numbers. 

In the relatively subsonic region, we show that the principal response
modes are localised around the critical layer corresponding to the
mean velocity profile. Furthermore, with the semi-local scaling
proposed by \citet{trettel2016}, the mean velocity profiles
collapse for various Reynolds and Mach numbers.  This provides the
necessary condition for the resolvent modes to exhibit universality
and geometrical self-similarity when scaled with the semi-local
Reynolds number and wall-normal distance.  We show that the principal
response modes are indeed universal and self-similar and that they
follow the same scaling laws as the incompressible boundary layer when
normalised by the fluctuating kinetic energy. This validates the
notion of Morkovin's hypothesis for the relatively subsonic region on
a mode-by-mode basis.  Moreover, the velocity modes and the
temperature and density modes are qualitatively similar, consistent
with the strong Reynolds analogy. We also provide scaling laws for the
amplification factor of the principal response mode. 

Finally, we show that the energy distribution between the velocity
fluctuations and the thermodynamic fluctuations can be predicted from
the energy distribution of the most energetic response mode. Coupled
with the universality and self-similarity of the resolvent modes in
the relatively subsonic region, this has implications in modelling and
prediction of high-speed turbulence. As in the incompressible case,
the self-similar resolvent modes facilitate analytical developments in
the logarithmic region of the boundary layer. Additionally, this
allows prediction tools developed for resolvent analysis of
incompressible fluids to be applied to supersonic boundary layers. 

The results show that the main difference between the compressible
Navier-Stokes equations and the incompressible equations are due to
{\color{black}a different scaling law originating from} density
variations in the wall-normal direction and the acoustic contribution
in the relatively supersonic region. The full nonlinear closure then
propagates the deviation through triadic interactions, which results
in variations in the mean velocity profile. Future efforts will be
focused on studying the effect of the feedback loop by incorporating
limited self-interactions to estimate the forcing term as in
\citet{rosenberg2019} for the incompressible Navier-Stokes equations.
Also, further efforts are necessary to study the effects of higher
Mach numbers and different wall boundary conditions such as cooled
walls.

\vspace{0.5cm}
The authors acknowledge support from the Air Force Office of
Scientific Research grant FA9550-16-1-0232. The authors also thank Dr.
Adri\'an Lozano-Dur\'an and Dr. Minjeong Cho for their insightful
comments on the manuscript.

\appendix
\section{Modal amplification mechanism in the relatively supersonic
region}\label{sec:append:modal}

\begin{figure}
\vspace{0.25cm}
\centerline{
\subfloat[]{\includegraphics[width=0.47\textwidth]{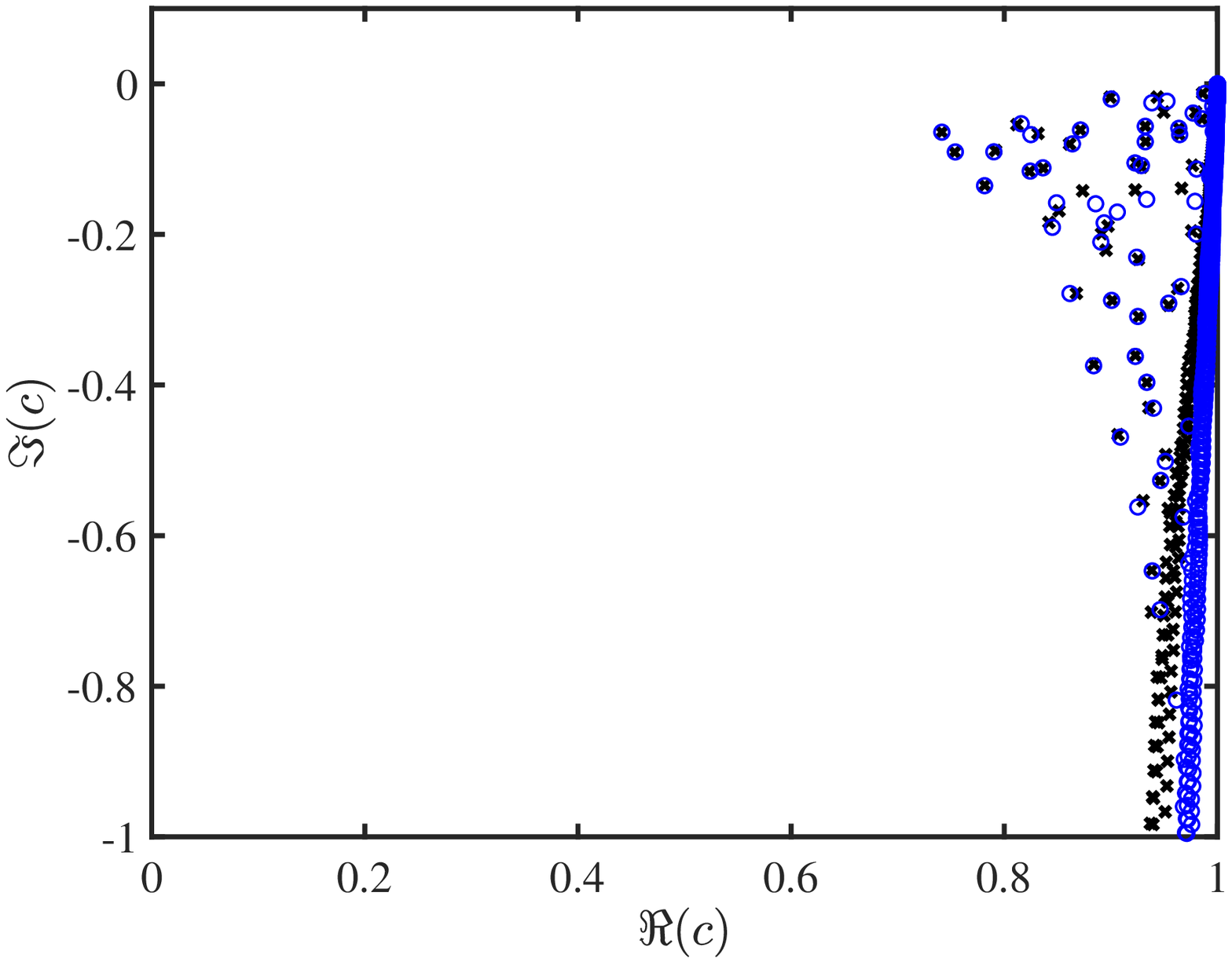}}
\hspace{0.2cm}
\subfloat[]{\includegraphics[width=0.47\textwidth]{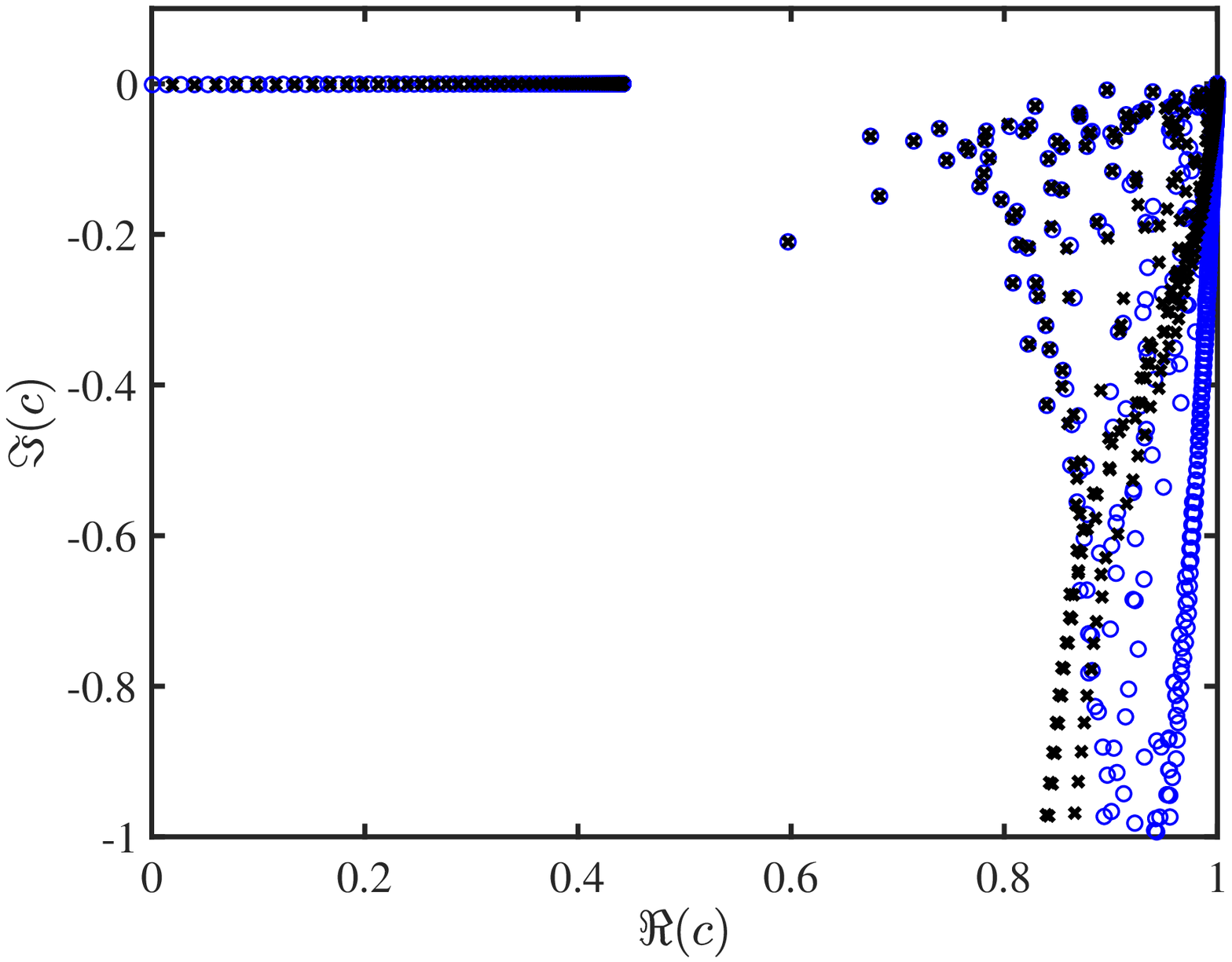}}
}
\caption{Eigenvalues for the linearised operator $\mathsfbi{L}$ for the
compressible ($M_\infty =4$) turbulent boundary layer for wave
parameters (a) $(\kappa_1\delta,\kappa_3\delta) = (0.2,2)$ and (b)
$(\kappa_1\delta,\kappa_3\delta) = (1,2)$ for $N_2 = 401$ ($\times$) and
$N_2 = 801$ ({\color{blue}\capcirc}).
\label{fig:eigen_vals}}
\end{figure}

The spectrum of the linear operator $\mathsfbi{L}$ is obtained from
solving the eigenvalue problem $\mathsfbi{L}\boldsymbol{q} =
\mi\omega\boldsymbol{q}$. Here, we plot the results for the $M_\infty
= 4$ case in terms of the complex wave speed $c = \Re(c) + \mi\Im(c)$
in figure \ref{fig:eigen_vals} for {\color{black}two different grid
resolutions in $x_2$, $N_2 = 401$ and $N_2 = 801$. We also tested
finite domain heights, compared to the semi-infinite domain, and the
resulting spectra are robust to this change (not shown). We see that
besides the discrete modes, the spectra consist of continuous branches
that are represented as discrete points which either converge towards
a vertical line, as expected for the viscous modes in both
incompressible and compressible boundary layers, or form a horizontal
line close to the real axis for the acoustic modes.} For the spectrum of
$(\kappa_1\delta,\kappa_3\delta) = (0.2,2)$, relative Mach number, 
$\overline{M}_\infty = 0.39$, is subsonic, and thus relatively supersonic modes are absent in
this case.  {\color{black} However, the spectrum for
$(\kappa_1\delta,\kappa_3\delta) = (1,2)$ (figure
\ref{fig:eigen_vals}(b)), where the relatively supersonic region is
present, shows an additional feature indicative of a continuous branch
of `acoustic' eigenmodes close to $\Im(c) = 0$. In the relatively
supersonic region, the resolvent norm shows peaks in both the leading
singular value and the energy contained in the principal resolvent
mode due to spectral amplification, i.e., the resonant effects from the
operator becoming close to normal as the wave speed approaches the
discrete acoustic eigenvalues. This shows that the irregular patterns
in figure \ref{fig:sv_energy}(b) are the consequence of
discretisation, and that the resonant effects will be universally
present in the continuous case.  }

\begin{figure}
\vspace{0.25cm}
\centerline{\includegraphics[width=0.6\textwidth]{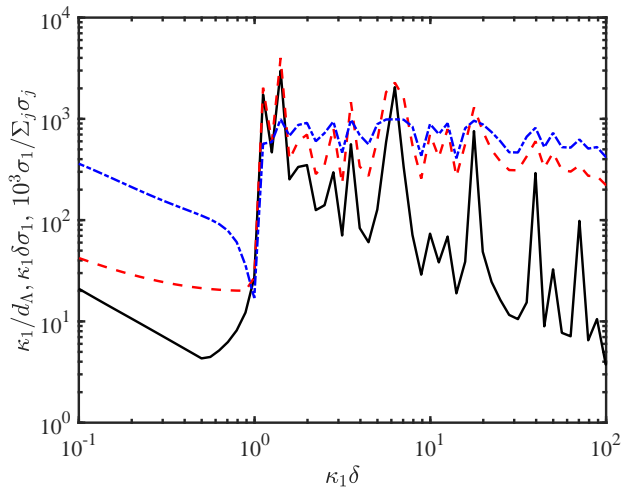}}
\caption{Inverse of the distance of the wave speed to the
closest acoustic eigenvalue with $\Im(c) = 0$, $\kappa_1/d_\Lambda$
(\solidline), compared with the premultiplied principal singular value 
$\kappa_1\delta\sigma_1$ ({\color{red}\dashline}) and energy contribution
from the leading singular value $\sigma_1^2/(\sum_j\sigma_j^2)$
({\color{blue}\dotdashline}) for $\kappa_3\delta = 2$.
\label{fig:eigen_dist}}
\end{figure}

{\color{black} For a given wave speed $c$, we can plot the distance
$d_\Lambda$, under the norm given in \eqref{eq:norm},
from $\omega = c\kappa_1$ to the closest discrete acoustic
eigenvalue as a function of wavenumbers $\kappa_1$ and $\kappa_3$.} In
figure \ref{fig:eigen_dist}, the inverse of the minimum eigenvalue
distance is shown to correlate well with the leading singular value,
which leads to correlation with the principal energy contribution
$\sigma_1^2/(\sum_j\sigma_j^2)$. As mentioned before, this is due to
the resonant amplification of the resolvent operator through modal
amplification mechanisms. The change in the discrete location of the
acoustic eigenvalues as a function of $\kappa_1$ and $\kappa_3$
explains the irregular patterns in figure \ref{fig:sv_energy}(b) in
the relatively supersonic region. However, the patterns are a purely
numerical phenomenon and are indicative of the fact that the resolvent
operator does not exist.


\section{Reynolds number scaling of the principal singular value}
\label{sec:append:operator}

In the inner region where the collapse of the mean velocity profile is
achieved through the semi-local scaling, the coordinates scale as
\begin{equation}
\kappa_1 \sim \Rey_\tau^*,\quad \kappa_3 \sim \Rey_\tau^*,\quad
\mathrm{d}/\mathrm{d}x_2 \sim \Rey_\tau^*.
\end{equation}
We then analyse the scaling of the linear operator $\mathsfbi{H}$ by
assessing the Reynolds number dependency of the terms in the linearised
operator $\mathsfbi{L}$. In this process, we assume that the Mach
number plays a secondary role, since we know $\Rey_\tau^*$ (or
$\Rey_\tau^\star$) reflects the Mach number dependency as seen in
\S \ref{sec:scaling:prm}. Doing so,
the linear operator $\mathsfbi{L}$ approximately scales as
\begin{equation}
\mathsfbi{L} \sim \left[\begin{array}{ccccc}
\Rey_\tau^* + \frac{\Rey_\tau^{*2}}{\Rey_\tau^\star} & 
\Rey_\tau^* + \frac{\Rey_\tau^{*2}}{\Rey_\tau^\star} & 
\Rey_\tau^* + \frac{\Rey_\tau^{*2}}{\Rey_\tau^\star} & 
\Rey_\tau^* &
\Rey_\tau^* + \frac{\Rey_\tau^{*2}}{\Rey_\tau^\star} \\
\frac{\Rey_\tau^{*2}}{\Rey_\tau^\star} & 
\Rey_\tau^* + \frac{\Rey_\tau^{*2}}{\Rey_\tau^\star} & 
\frac{\Rey_\tau^{*2}}{\Rey_\tau^\star} & 
\Rey_\tau^* &
\Rey_\tau^* + \frac{\Rey_\tau^{*2}}{\Rey_\tau^\star} \\
\frac{\Rey_\tau^{*2}}{\Rey_\tau^\star} & 
\frac{\Rey_\tau^{*2}}{\Rey_\tau^\star} & 
\Rey_\tau^* + \frac{\Rey_\tau^{*2}}{\Rey_\tau^\star} & 
\Rey_\tau^* &
\Rey_\tau^* \\ 
\Rey_\tau^* &
\Rey_\tau^* &
\Rey_\tau^* &
\Rey_\tau^* &
0 \\
\Rey_\tau^* + \frac{\Rey_\tau^{*2}}{\Rey_\tau^\star} & 
\Rey_\tau^* + \frac{\Rey_\tau^{*2}}{\Rey_\tau^\star} & 
\Rey_\tau^* &
0 &
\Rey_\tau^* + \frac{\Rey_\tau^{*2}}{\Rey_\tau^\star} 
\end{array}\right].
\end{equation}
As $\Rey_\tau^* \approx \Rey_\tau^\star$ in this region, the scaling
can be seen as $\mathsfbi{L}\sim \sqrt{\Rey_\tau^*\Rey_\tau^\star}$,
giving $\mathsfbi{H} = (\mi\omega\mathsfbi{I}+\mathsfbi{L})^{-1} \sim
1/\sqrt{\Rey_\tau^*\Rey_\tau^\star}$. Thus, the leading singular value is expected to be
proportional to $1/\sqrt{\Rey_\tau^*\Rey_\tau^\star}$. 

In the outer region, the coordinates scale as
\begin{equation}
\kappa_1 \sim 1/\Rey_\tau^*,\quad \kappa_3 \sim 1,\quad
\mathrm{d}/\mathrm{d}x_2 \sim 1.
\end{equation}
Additionally, if we assume that $\kappa_3^2$ dominates $\kappa_1^2$
for all values of $\Rey_\tau^*$, such that for the streamwise
wavenumber in the outer coordinate given by $\kappa_1^- =
\Rey_\tau^*\kappa_1$ we have $\kappa_3/\kappa_1^- >
\epsilon/\Rey_\tau^*$, the linear operator $\mathsfbi{L}$ scales as
\begin{equation}
\mathsfbi{L} \sim \left[\begin{array}{ccccc}
\frac{1}{\Rey_\tau^*} + \frac{1}{\Rey_\tau^\star} & 
\frac{1}{\Rey_\tau^*\Rey_\tau^\star} & 
\frac{1}{\Rey_\tau^*\Rey_\tau^\star} & 
\frac{1}{\Rey_\tau^*} & 
\frac{1}{\Rey_\tau^*} + \frac{1}{\Rey_\tau^\star} \\ 
\frac{1}{\Rey_\tau^*\Rey_\tau^\star} & 
\frac{1}{\Rey_\tau^*} + \frac{1}{\Rey_\tau^\star} & 
\frac{1}{\Rey_\tau^\star} & 
1 &
\frac{1}{\Rey_\tau^*\Rey_\tau^\star} \\ 
\frac{1}{\Rey_\tau^*\Rey_\tau^\star} & 
\frac{1}{\Rey_\tau^\star} & 
\frac{1}{\Rey_\tau^*} + \frac{1}{\Rey_\tau^\star} & 
1 &
1 \\
\frac{1}{\Rey_\tau^\star} &
1 &
1 &
\frac{1}{\Rey_\tau^\star} & 
0 \\
\frac{1}{\Rey_\tau^*} + \frac{1}{\Rey_\tau^\star} & 
\frac{1}{\Rey_\tau^*\Rey_\tau^\star} & 
1 &
0 &
\frac{1}{\Rey_\tau^*} + \frac{1}{\Rey_\tau^\star} \\ 
\end{array}\right].
\end{equation}
The resolvent $\mathsfbi{H}$ then scales as
$\Rey_\tau^*\Rey_\tau^\star$, which gives the scaling for the leading
singular value $\sigma_1$.

Finally, for the logarithmic region, the differential operators are
scaled as  
\begin{equation}
\kappa_1 \sim 1/(x_2^{c*}x_2^c),\quad \kappa_3 \sim 1/x_2^c,\quad
\mathrm{d}/\mathrm{d}x_2 \sim 1/x_2^c.
\end{equation}
We also assume that the spanwise coordinate dominates the spanwise
coordinate, i.e. $(\kappa_3/\kappa_1) > \epsilon$ with a conservative
estimation of $\epsilon \approx \sqrt{10}$, and arrive at 
\begin{equation}
\mathsfbi{L} \sim \left[\begin{array}{ccccc}
\frac{1}{x_2^{c*}x_2^c} + \frac{1}{x_2^{c\star} x_2^c} & 
\frac{1}{x_2^{c\star} x_2^c} & 
\frac{1}{x_2^{c*} x_2^{c\star}} & 
\frac{1}{x_2^{c*} x_2^c} & 
\frac{1}{x_2^{c*} x_2^c} + \frac{1}{x_2^{c\star} x_2^c} \\ 
\frac{1}{x_2^{c*} x_2^{c\star}} & 
\frac{1}{x_2^{c*} x_2^c} + \frac{1}{x_2^{c\star} x_2^c} & 
\frac{1}{x_2^{c\star} x_2^c} & 
\frac{1}{x_2^c} & 
\frac{1}{x_2^{c*} x_2^{c\star} x_2^c} \\ 
\frac{1}{x_2^{c*} x_2^{c\star} x_2^c} & 
\frac{1}{x_2^{c\star} x_2^c} & 
\frac{1}{x_2^{c*}x_2^c} + \frac{1}{x_2^{c\star} x_2^c} & 
\frac{1}{x_2^c} & 
\frac{1}{x_2^c} \\ 
\frac{1}{x_2^{c*} x_2^c} & 
\frac{1}{x_2^c} & 
\frac{1}{x_2^c} & 
\frac{1}{x_2^{c*} x_2^c} & 
0 \\
\frac{1}{x_2^{c*}x_2^c} + \frac{1}{x_2^{c\star} x_2^c} & 
\frac{1}{x_2^{c*} x_2^{c\star} x_2^c} & 
\frac{1}{x_2^c} & 
0 & 
\frac{1}{x_2^{c*}x_2^c} + \frac{1}{x_2^{c\star} x_2^c} 
\end{array}\right].
\end{equation}
And a similar analysis as the inner and outer regions reveals that the
singular values scale with $x_2^*x_2^\star x_2$.

\bibliography{comp_bl_res}
\bibliographystyle{jfm}

\end{document}